\documentclass[%
 aip,
 amsmath,amssymb,
 reprint,%
]{revtex4-1}

\usepackage{graphicx}
\usepackage{dcolumn}
\usepackage{bm}

\usepackage[utf8]{inputenc}
\usepackage[T1]{fontenc}
\usepackage{mathptmx}
\usepackage{latexsym}
\usepackage{amssymb}
\usepackage{amsthm}
\usepackage{enumitem}
\usepackage{amsmath}

\usepackage{xcolor}
\usepackage{color,soul}
\usepackage[normalem]{ulem}
\sethlcolor{yellow}

\begin{document}

\preprint{AIP/123-QED}

\title{Multispectral time-resolved energy-momentum microscopy using high-harmonic extreme ultraviolet radiation}



\author{Michael Heber}
\thanks{These authors contributed equally to this work.}
\affiliation{Deutsches Elektronen-Synchrotron DESY, 22607 Hamburg, Germany}
\author{Nils Wind}
\thanks{These authors contributed equally to this work.}
\affiliation{Institut f\"ur Experimentalphysik, Universit\"at Hamburg, 22761 Hamburg, Germany}
\affiliation{Ruprecht Haensel Laboratory, Deutsches Elektronen-Synchrotron DESY, 22607 Hamburg, Germany}
\author{Dmytro Kutnyakhov}
\email{dmytro.kutnyakhov@desy.de}
\author{Federico Pressacco}
\affiliation{Deutsches Elektronen-Synchrotron DESY, 22607 Hamburg, Germany}
\author{Tiberiu Arion}
\affiliation{Centre for Free-Electron Laser Science, Deutsches Elektronen-Synchrotron DESY, 22607 Hamburg, Germany}
\author{Friedrich Roth}
\affiliation{Institute of Experimental Physics, TU Bergakademie Freiberg, 09599 Freiberg, Germany}
\affiliation{Center for Efficient High Temperature Processes and Materials Conversion (ZeHS), TU Bergakademie Freiberg, 09599 Freiberg, Germany}
\author{Wolfgang Eberhardt}
\affiliation{Centre for Free-Electron Laser Science, Deutsches Elektronen-Synchrotron DESY, 22607 Hamburg, Germany}
\author{Kai Rossnagel}
\affiliation{Ruprecht Haensel Laboratory, Deutsches Elektronen-Synchrotron DESY, 22607 Hamburg, Germany}
\affiliation{Institut f\"ur Experimentelle und Angewandte Physik, Christian-Albrechts-Universit\"at zu Kiel, 24098 Kiel, Germany}

             
\begin{abstract}
A 790-nm-driven high-harmonic generation source with a repetition rate of 6\,kHz is combined with a toroidal-grating monochromator and a high-detection-efficiency photoelectron time-of-flight momentum microscope to enable time- and momentum-resolved photoemission spectroscopy over a spectral range of $23.6$--$45.5$\,eV with sub-100-fs time resolution. Three-dimensional (3D) Fermi surface mapping is demonstrated on graphene-covered Ir(111) with energy and momentum resolutions of $\lesssim$$100$\,meV and $\lesssim$$0.1$\,\AA$^{-1}$, respectively. The table-top experiment sets the stage for measuring the $k_z$-dependent ultrafast dynamics of 3D electronic structure, including band structure, Fermi surface, and carrier dynamics in 3D materials as well as 3D orbital dynamics in molecular layers.

\end{abstract}

\maketitle

\section{Introduction}

Angle-resolved photoemission spectroscopy (ARPES) is the standard method to determine how electrons behave at surfaces of solid materials.\cite{Damascelli2004,Lv2019,King2021,Sobota2021} Monochromatic photons having ultraviolet (UV) energies or higher eject electrons from a material’s surface, and the photocurrent is measured as a function of electron kinetic energy, emission direction, and photon energy. Direction is encoded in two emission angles or, equivalently, in the two components of the surface-parallel momentum $(k_x, k_y)$. Since the four measurement parameters can be straightforwardly related to the energy relative to the Fermi level ($E-E_\mathrm{F}$) and three-dimensional (3D) momentum $(k_x, k_y, k_z)$ of the electrons inside the material before photoexcitation,\cite{Chiang1980} the measured intensity distributions readily provide multidimensional images of the electronic structure in portions of four-dimensional (4D) energy-momentum space.\cite{Medjanik2017} Band structures and Fermi surfaces, but also momentum-dependent band renormalization and lifetime effects, can thus be accessed directly.\cite{Huefner1999,Valla1999,Krasovskii2007,Watson2019} Another intriguing application is orbital tomography, which can provide reconstructed real-space tomograms of molecular orbitals on solid surfaces.\cite{Puschnig2009,Graus2019} Depending on whether emission angles or surface-parallel momentum components are imaged onto the detector,\cite{Wannberg2009,Kotsugi2003,Schoenhense2015} the technique is referred to as ARPES or momentum microscopy, respectively.

In this energy-momentum imaging, the photon energy is an important parameter in at least three different ways. First, the photon energy determines the maximum detectable electron kinetic energy and 3D momentum, and thus the volume of the probed portion in energy-momentum space.\cite{Babenkov2019} 
Second, scanning the photon energy allows to retrieve the surface-perpendicular momentum component $k_z$, hence to perform full 4D energy-momentum imaging.\cite{Rossnagel2001,Nielsen2003,Medjanik2017,Watson2019} And third, tunability in the photon energy can also be useful to enhance the contrast of specific features in ARPES data by exploiting excitation resonances, escape-depth variation, or matrix-element effects.\cite{Strocov2014,Moser2017}

In principle, the same relevance of the probing photon energy and its tunability also applies to time-resolved ARPES (trARPES) or time-resolved momentum microscopy,\cite{Kutnyakhov2020,Maklar2020,Keunecke2020} in which time, or more precisely the time delay ($t_\mathrm{D}$) between femtosecond-scale pump and probe pulses, is added as a fifth measurement parameter. Over the past fifteen years, trARPES has evolved into a powerful ARPES modality providing direct dynamical information on electronic structure at the fundamental time scales of electronic and atomic motion, particularly on photoinduced transient changes of electronic states and their population.\cite{Perfetti2006,Schmitt2008,Rohwer2011,Smallwood2012,Gierz2013,Wang2013,Mahmood2016,Nicholson2018,Na2019,Wallauer2021} trARPES is now routinely performed using table-top laser sources based on fourth and higher harmonic generation (HHG) in solids\cite{Gauthier2020,Bao2022} and gases\cite{Dakovski2010,Frietsch2013,Eich2014,Cilento2016,Rohde2016,Corder2018,Puppin2019,Buss2019,Sie2019,Mills2019,Liu2020,Cucini2020,Lee2020,Keunecke2020,Peli2020,Guo2022}, respectively, as well as using free-electron lasers (FELs) based on the self-amplification of spontaneous emission (SASE) of free electrons in undulators.\cite{Kutnyakhov2020} The corresponding probe photon energies in trARPES range from the far UV to soft x-rays where a sweet spot currently is the intermediate extreme ultraviolet (XUV) regime.

\begin{figure*}
\centering
\includegraphics[width=\textwidth]{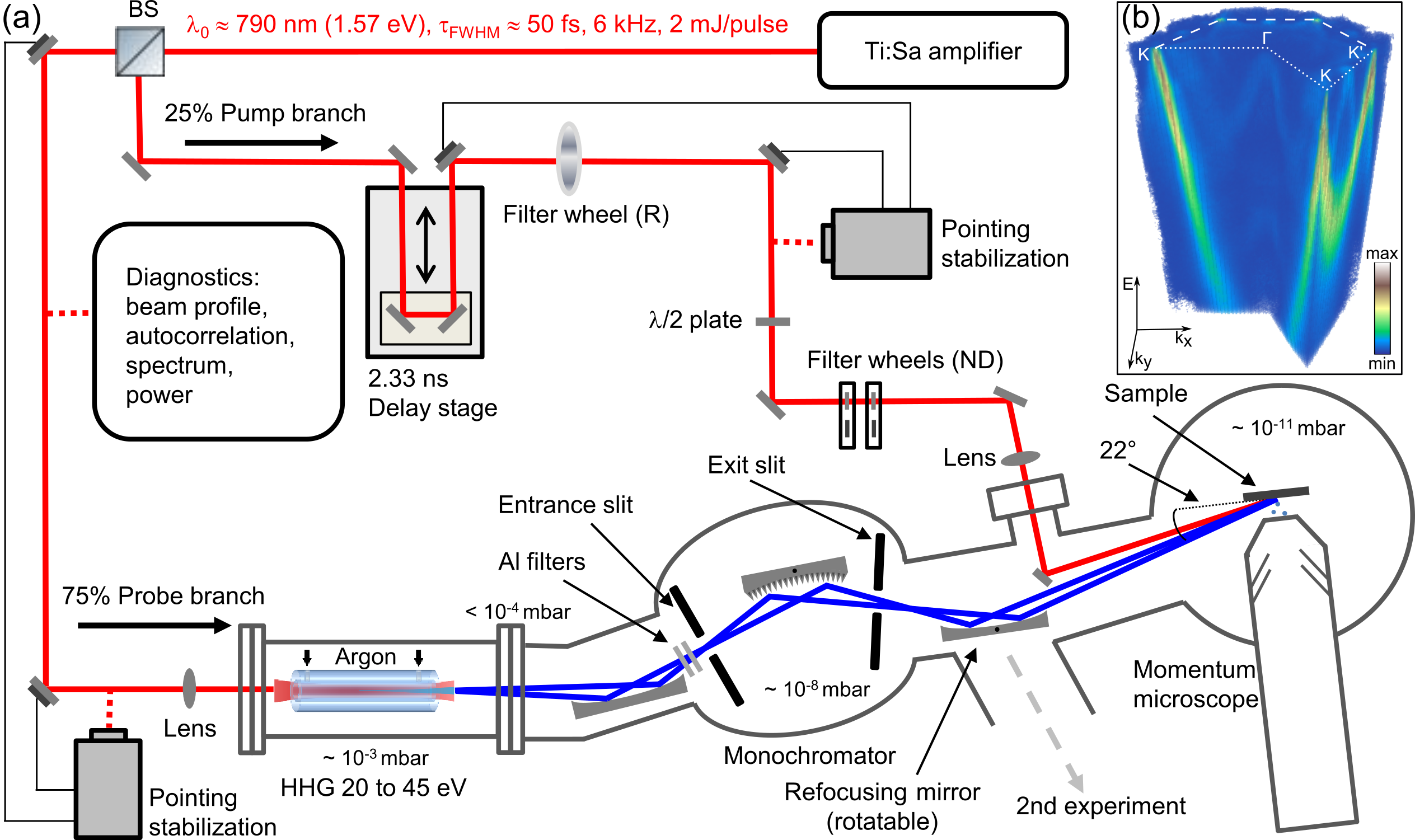}
\caption{(a) Schematic layout of the experimental setup for XUV multispectral time-resolved momentum microscopy, including the laser system, optics, and diagnostics, the Ar-filled capillary for HHG, the toroidal-grating monochromator, and the ToF momentum microscope for photoelectron detection. For details, see text. BS: beam splitter, R: reflective, ND: neutral density. 
(b) Exemplary 3D momentum microscopy data set representing photoemission intensity as a function of energy $E$ and surface-parallel momentum $(k_x, k_y)$. Data were measured with the $27^\mathrm{th}$ harmonic ($42.2$\,eV) from graphene/Ir(111) at room temperature and symmetrized by three-fold rotation. The hexagonal Brillouin zone of graphene is indicated.}
\label{fig:laserset}
\end{figure*}

In the XUV, HHG-based trARPES can optionally provide high time, energy, momentum, and spin resolution in conjunction with kHz-to-MHz repetition rates and a sufficiently wide detection window of the surface-parallel momentum $(k_x, k_y)$ to fully cover typical Brillouin-zone (BZ) dimensions.\cite{Dakovski2010,Frietsch2013,Eich2014,Cilento2016,Rohde2016,Ploetzing2016,Eich2017,Corder2018,Gort2018,Puppin2019,Buss2019,Sie2019,Mills2019,Liu2020,Cucini2020,Lee2020,Keunecke2020,Peli2020,Buehlmann2020,Guo2022} Moreover, HHG sources combined with monochromators enable multispectral measurements as they can deliver an useable, discretely tunable photon-energy range of about 8--40\,eV.\cite{Dakovski2010,Frietsch2013,Corder2018,Mills2019,Sie2019,Cucini2020} However, this multispectral capability has so far rarely been exploited in trARPES experiments, and recently several setups for HHG-based trARPES have even been optimized for operation at one specific photon energy only.\cite{Eich2014,Puppin2019,Lee2020,Keunecke2020} Here, contrary to this trend, we present the combination of a tunable, monochromatized, kHz-repetition-rate HHG source with a wide-momentum-acceptance time-of-flight (ToF) momentum microscope for efficient 4D energy-momentum mapping of ultrafast electronic structure dynamics. The overall system performance is demonstrated on bare and graphene-covered Ir(111).

\section{Experimental setup} \label{sec:method}

Our multispectral-HHG trARPES setup, as schematically illustrated in Fig.~\ref{fig:laserset}, measures a 5D photoemission data hypercube $I(E, k_x, k_y; k_z, t_\mathrm{D})$. The laser system gives the tuning of $k_z$ and $t_\mathrm{D}$, via adjustability of the probe photon energy and pump--probe delay, respectively, and the ToF momentum microscope provides the basic parallel 3D measurement of $I(E, k_x, k_y)$.

\subsection{Multispectral photon source}

The schematic layout of the laser system is shown in Fig.~\ref{fig:laserset}(a).\cite{Seo2016} A Ti:Sapphire laser amplifier (Wyvern 1000, KMLabs), operated at an output power of $\approx$12\,W and a repetition rate of 6\,kHz, delivers laser pulses at a center wavelength of $\approx$790\,nm, pulse duration of $\approx$50\,fs (FWHM, full width at half maximum), and pulse energy of $\approx$2\,mJ. A quarter of the amplifier output is coupled into the pump branch, propagated through a delay stage, and focused onto the sample in the ultrahigh vacuum (UHV) photoemission chamber. The spot size of the pump beam on the sample is typically about $(100\times400)$\,$\mu$m$^2$. The available pump-pulse energy of 500\,$\mu$J can be used for frequency conversion or attenuated to the mid-nJ to low-$\mu$J level for 790-nm excitation of the sample below the space-charge limit.\cite{Oloff2016,Schoenhense2021}

\begin{figure}
\centering
\includegraphics[width=0.5\textwidth]{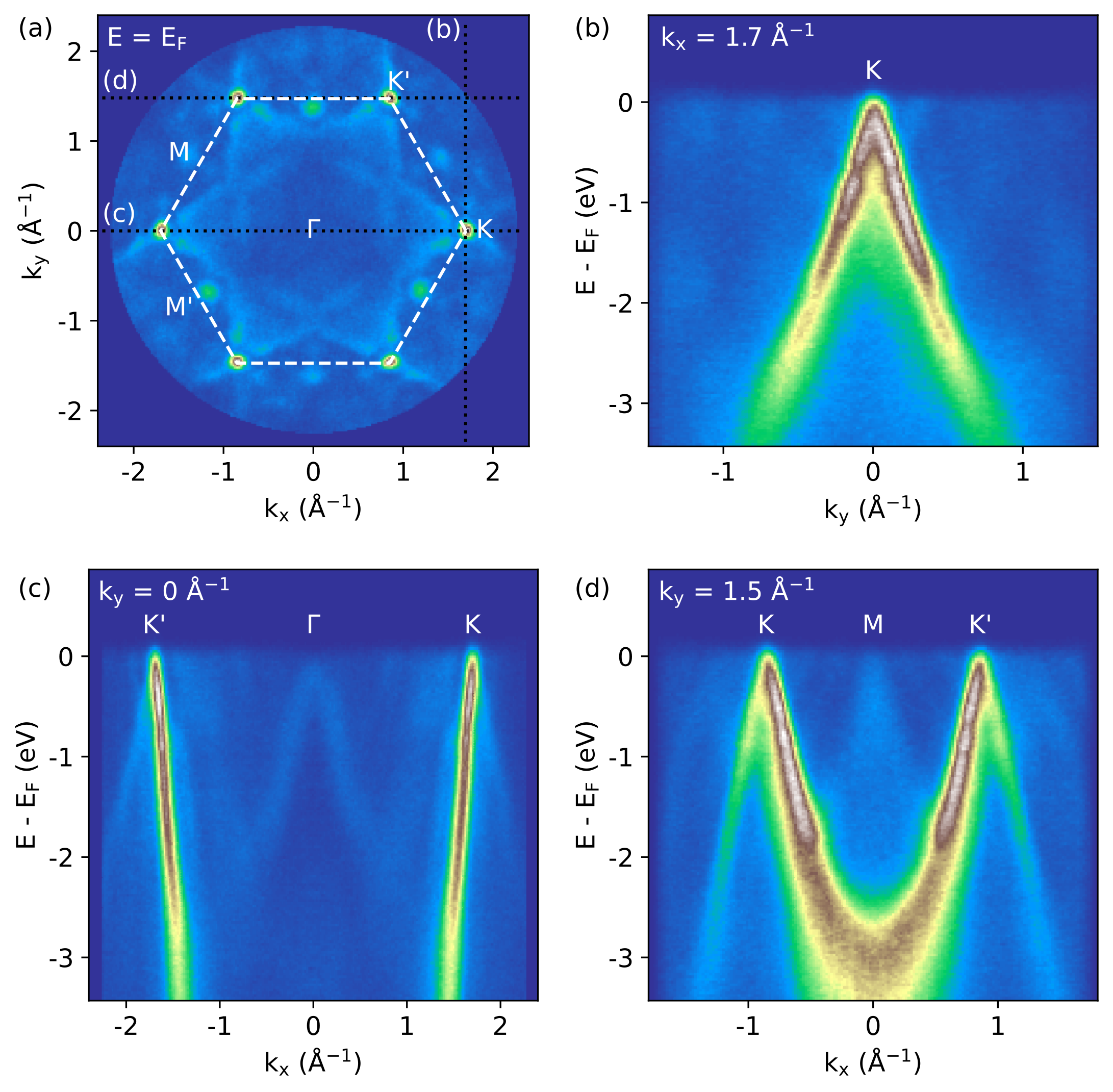}
\caption{Momentum microscopy data of graphene/Ir(111) recorded at room temperature with a photon energy of $42.4$\,eV ($27^\mathrm{th}$ harmonic). (a) $k_x$-versus-$k_y$ Fermi surface map (energy integration window: $\Delta E=\pm 0.1$\,eV). The hexagonal Brillouin zone of graphene and its high-symmetry points are indicated. (b)--(d) $E$-versus-$k$ band maps for different $k$ lines, as indicated by black dotted lines in (a) (momentum integration window: $\Delta k=\pm 0.07$\,\AA$^{-1}$).}
\label{fig:27th}
\end{figure}

The remaining 75\% ($\approx$50\,fs, $\approx$1.5\,mJ/pulse) of the amplifier output are coupled into the probe branch and focused with a lens (focal length $f=500$\,mm) into an Ar-filled waveguide capillary (XUUS, KMLabs), where higher harmonics in the XUV are generated. During operation, the Ar pressure inside the capillary is $\approx$60\,mbar, while the pressure outside stays below $\approx$5$\times10^{-3}$\,mbar. A ZrO$_2$-coated toroidal mirror ($f=307$\,mm) is used to focus the multispectral XUV light onto the entrance slit of the monochromator and to separate out much of the fundamental radiation. 
ZrO$_2$ specifically provides XUV reflectivity comparable to Au for photon energies around 40\,eV, and reduced reflectivity at $1.5$\,eV.\cite{Wood82,RefIndexZrO2,Henke93}
The remaining 790-nm light is blocked using few-100-nm thick Al filters. Differential pumping is used to maintain a 5-orders-of-magnitude pressure difference between the HHG source and the monochromator chamber. A rotatable Au-coated toroidal grating ($f=163$\,mm) with 550\,lines/mm (TGM300, Horiba Jobin Yvon SAS) images the entrance slit onto the exit slit with the selected higher harmonic light. 
For the 25$^{\rm{th}}$ harmonic, the footprint on the monochromator grating is estimated to be $(1 \times 1.7)$\,mm$^2$ translating into a nominal temporal probe-pulse broadening of 57\,fs. 
A second toroidal mirror ($f=520$\,mm) focuses the selected and monochromatized XUV light onto the sample at an angle of 22$^\circ$ with respect to the surface. This mirror is rotatable and can alternatively direct the beam to a gas-phase experiment with combined ion- and electron-ToF spectroscopy.\cite{PHDGerken2014,Gerken2014} Behind the refocusing mirror, the pump and probe beams propagate almost collinearly to the sample. At the sample, the XUV beam has a spot size of $(70\times800)$\,$\mu$m$^2$ and a flux of $3\times10^8$\,photons/s in the brightest harmonics, which are usually the 23$^\textrm{rd}$ and 25$^\textrm{th}$ (see central panel of Fig.~\ref{HHG_Series}). The practically useable part of the harmonic spectrum contains all eight odd harmonics from the 15$^\textrm{th}$ to the 29$^\textrm{th}$ corresponding to a photon energy range of $23.6$--$45.5$\,eV. According to ray-tracing simulations, the temporal probe pulse broadening due to the monochromator lies in the range of 50--100\,fs for all harmonics, independent of the slit size. The spectral resolution of the monochromator and the beamline is in the range of 100--240\,meV under measurement conditions with slit sizes of 150 $\mu$m. In particular, for the higher harmonics, the monochromator selects only a given harmonic rather than clipping the harmonic's spectral bandwidth. The overall high stability of the HHG source enables pump-probe experiments over several days, with the temporal overlap (time zero) remaining within the experimental time resolution. Average drifts in HHG intensity stay within 30\% over a week of continuous operation (with the possibility to retune and restore reduced intensity).

\subsection{Electron momentum microscope} \label{sec:detect}

The ToF momentum microscope employed in our laboratory-based setup is the same instrument used for FEL-based photoemission spectroscopy at the PG2 beamline of FLASH (DESY, Hamburg).\cite{Kutnyakhov2020} The high photoelectron detection efficiency of the instrument, which compensates for the moderate repetition rate of the photon pulses, results from a combination of three separate capabilities: (i) direct 2D momentum imaging with a field of view of $\pm2.4$\,\AA$^{-1}$ in both surface-parallel momentum directions, (ii) simultaneous ToF energy recording in an energy window of 7\,eV, and (iii) multi-hit detection of up to 3 electrons per pulse. The underlying principle of slit-less 3D photoelectron energy-momentum detection is implemented as follows:\cite{Schoenhense2015} The photocurrent emitted from the surface is imaged into an achromatic surface-parallel momentum image at the back-focal plane of the cathode objective lens; this hyperspectral image is subsequently magnified and high-pass-filtered by two lens systems, before it is spectrally dispersed in a field-free drift tube and finally captured on a delay-line detector (DLD). The 8-segment DLD (DLD6060-8s, Surface Concept) used in the current setup consists of two stacked 4-quadrant DLDs rotated by 45$^\circ$ with respect to each other. This novel detector provides improved multi-hit detection capability compared to a single 1- or 4-quadrant DLD, as well as improved resolution of hits occurring near segment boundaries. The length of the drift tube (800\,mm) and temporal resolution of the detector ($\approx$150\,ps) translate into a nominal energy resolution of $<$40\,meV for typical electron drift energies of 10--30\,eV. The nominal momentum resolution is $<$0.01\,\AA$^{-1}$, as given by the momentum field of view, active detector area (60\,mm diameter), and the spatial resolution of the detector ($\approx$80\,$\mu$m).  

\begin{figure*}
\centering
\includegraphics[width=1\textwidth]{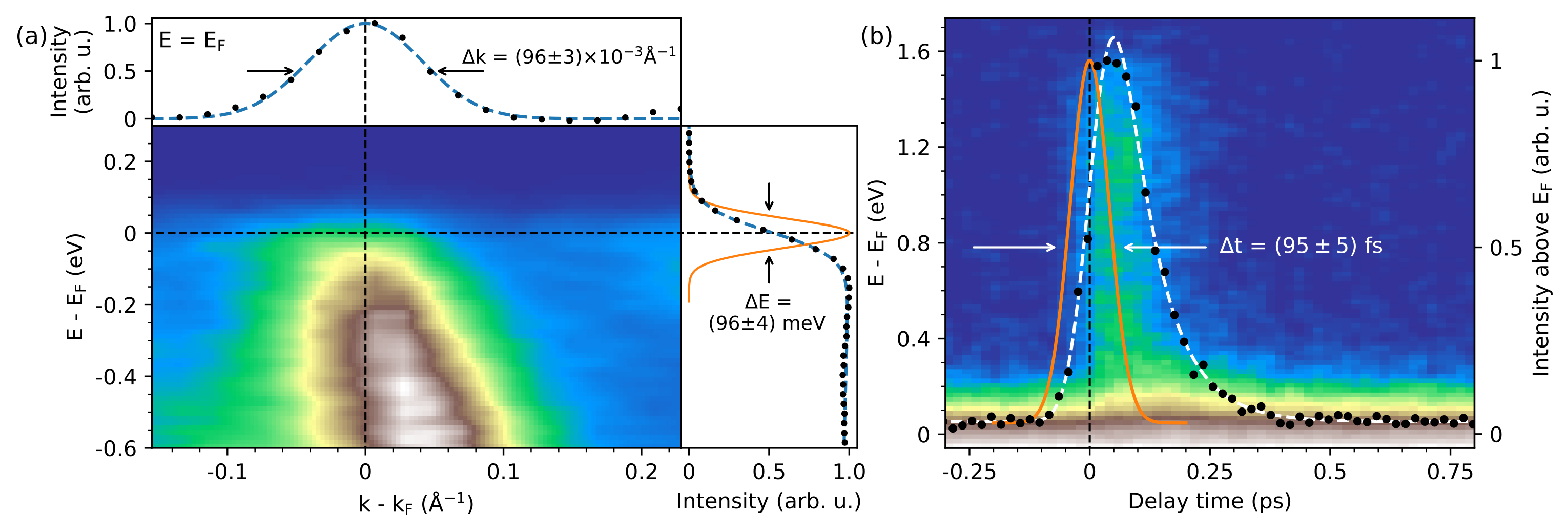}
\caption{(a) Energy-versus-momentum photoemission intensity map of graphene/Ir(111) recorded at room temperature near the $K$ point with a photon energy of $33.4$\,eV ($21^\mathrm{st}$ harmonic; momentum integration window: $\Delta k= 0.13$\,\AA$^{-1}$). Top panel: Momentum distribution curve (black dots) at $E_\mathrm{F}$ plus fit (blue dashed line). Right panel: Energy distribution curve (black dots) representing integrated MDC peak intensities plus fit (blue dashed line).
(b) Full momentum-integrated energy-versus-delay photoemission intensity map of Ir(111), recorded at room temperature with a photon energy of $36.1$\,eV ($23^\mathrm{rd}$ harmonic), and energy-integrated ($E \geq E_\mathrm{F}$) intensity transient (black dots) plus fit (white dashed line). Values for the Gaussian FWHM determined from the fits, corresponding to effective experimental resolutions, are indicated.}
\label{fig:resolution}
\end{figure*}

The 3D energy-momentum measurement system, based on single-event detection, fills in the photoemission data cube $I(E, k_x, k_y)$ over energy and momentum intervals with a characteristic width of 7\,eV and $4.8$\,\AA$^{-1}$, respectively, at a repetition rate of 6 kHz. When delay-time scanning is added, the resulting size of a typical 4D data hypercube is $\approx$100\,GB. An efficient data acquisition and data processing workflow is implemented using an open-source software package developed for high-throughput multidimensional photoemission spectroscopy experiments.\cite{Xian2020}
Figure~\ref{fig:laserset}(b) shows a 3D representation of an exemplary data set taken from graphene/Ir(111). Four different 2D cuts through this data set are shown in Fig.~\ref{fig:27th}, including the Fermi surface map $I(E_\mathrm{F}, k_x, k_y)$ [Fig.~\ref{fig:27th}(a)] and selected band maps $I(E, k_0, k_y)$ [Fig.~\ref{fig:27th}(b)] and $I(E, k_x, k_0)$ [Figs.~\ref{fig:27th}(c) and \ref{fig:27th}(d)] for different constant values of $k_0$ corresponding to lines passing through high-symmetry points of the graphene BZ, as indicated in Fig.~\ref{fig:27th}(a). In these maps, the strongest signal stems from the $\pi$-band of graphene, with its linear dispersion toward $E_\mathrm{F}$ and point-like Fermi surface at the $K$ and $K'$ points. The Ir~$5d$ bands appear as much weaker features. Their interaction with the $\pi$-band, however, leads to distinct kinks in the $\pi$-band dispersion.\cite{Tusche2016} The presented data vividly illustrate the efficiency and completeness of the ToF momentum microscopy approach to photoelectron detection.

\section{Performance}  \label{sec:performance}

We have characterized the performance of the experimental system by measuring the near-$E_\mathrm{F}$ electronic structure and the above-$E_\mathrm{F}$ carrier dynamics of graphene-covered and pristine Ir(111), respectively. Standard Ir(111) cleaning procedures and graphene growth recipes were applied.\cite{Diaye2008} Surface quality was checked by low-energy electron diffraction. All photoemission measurements were done at room temperature.

\subsection{Experimental resolutions}

We estimated the effective experimental energy and momentum resolutions from the Fermi-level crossing of the graphene $\pi$-band in the graphene/Ir(111) sample. Figure~\ref{fig:resolution}(a) shows a room-temperature $E$--$k$ photoemission intensity map, which was measured in the vicinity of the $K$ point with a photon energy of 33\,eV (21$^\textrm{st}$ harmonic) using probe-only photoemission. Also shown are the momentum distribution curve (MDC) extracted at $E_\mathrm{F}$ (top panel) and an energy distribution curve (EDC) obtained by the MDC method\cite{Ulstrup2014} (right panel), representing the energy distribution of fitted MDC peak areas.

The EDC was fitted with a room-temperature Fermi-Dirac distribution function convoluted with a Gaussian resolution function [Fig.~\ref{fig:resolution}(a), right panel]. The resulting Gaussian FWHM, corresponding to the total energy resolution, is $(96\pm4)$\,meV. This value includes a contribution of the photon source and monochromator of about 90\,meV. The contribution of the electron spectrometer including space-charge broadening is estimated to be 30\,meV. Over the entire usable spectral range of $\approx$24--46\,eV, the effective energy resolution varied between 80 and 135\,meV, where the photon-energy dependence of the grating resolution at fixed slit widths makes the dominant contribution. Note that these values are better than those given above for photon resolution because the field aperture of the momentum microscope typically acts as a virtual exit slit, selecting an effective field of view on the sample of $\approx$70\,$\mu$m.

\begin{figure*}
\centering
\includegraphics[width=1\textwidth]{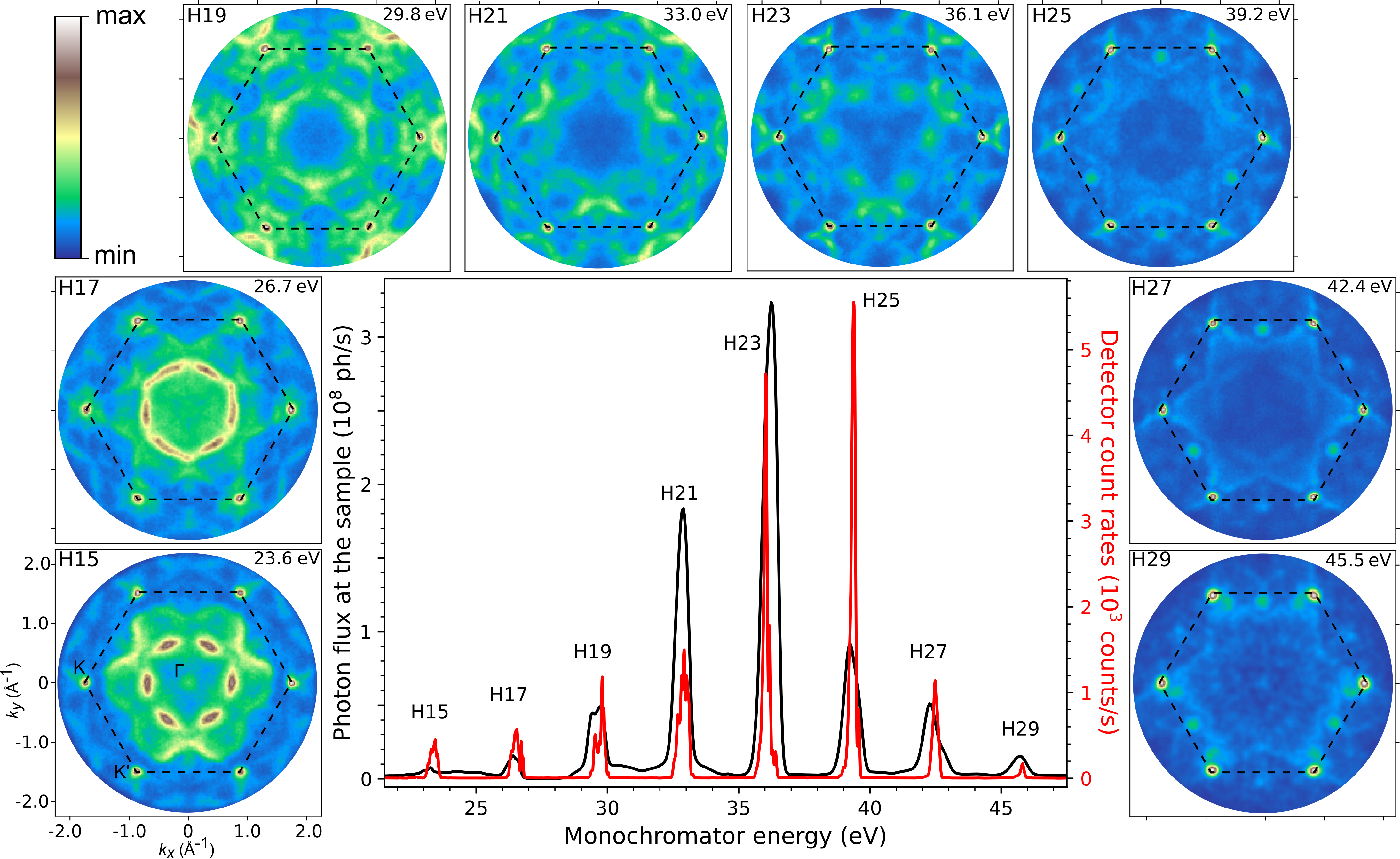}
\caption{Multispectral Fermi surface mapping of graphene/Ir(111) using higher-harmonic-generation-based time-of-flight momentum microscopy.
Central panel:
Higher harmonic spectra as given by photon flux (black line) and electron count rate (red line). Spectra were measured with an XUV-sensitive photodiode behind the exit slit of the monochromator and with the momentum microscope from a graphene-covered Ir(111) sample, respectively. The photon flux at the sample position was calculated, accounting for beamline transmission and beam attenuation. For the scans, the slit sizes of the monochromator were adjusted to a monochromator resolution of 150\,meV at the $23^\textrm{rd}$ harmonic.
Surrounding panels: Full series of Fermi-surface maps taken at room temperature over the photon-energy tuning range from $23.6$ to $45.5$\,eV ($15^\textrm{th}$ to $29^\textrm{th}$ harmonic; energy integration window: $\Delta E=\pm 0.1$\,eV). The hexagonal Brillouin zone of graphene is indicated.}
\label{HHG_Series}
\end{figure*}

The Gaussian FWHM determined from the momentum distribution curve at $E_\mathrm{F}$ is $(0.096\pm0.003)$\,\AA$^{-1}$ [Fig.~\ref{fig:resolution}(a), top panel]. After subtracting the intrinsic $\pi$-band momentum width of $0.031$\,\AA$^{-1}$,\cite{Kralj2011,Johannsen2013} the remaining effective momentum resolution is $(0.091\pm0.003)$\,\AA$^{-1}$.
We attribute the deterioration with respect to the nominal momentum resolution to less than optimal sample quality, electronic noise, and timing jitter in the position measurement on the detector.
With varying probe photon energy, no noticeable changes of the momentum resolution were detected.

To estimate the temporal cross-correlation between pump and probe pulses, we performed pump--probe photoemission measurements on pristine Ir(111) using pump and probe photon energies of $1.57$\,eV and $36.1$\,eV (23$^\textrm{rd}$ harmonic), respectively, and an incident pump fluence of $2.42$\,mJ/cm$^2$. Figure~\ref{fig:resolution}(b) shows a momentum-integrated $E$--$t_\mathrm{D}$ intensity map depicting the transient generation and relaxation of hot electrons above $E_\mathrm{F}$. A corresponding intensity transient (black data points), obtained by integrating over energies larger than $E_\mathrm{F}$, is overlaid. This signal was fitted with a step function multiplied by an exponential decay and convoluted with a Gaussian function. The resulting Gaussian FWHM is $(95\pm5)$\,fs, giving an estimate of the temporal system response function. Based on this value and with a modeled pump-pulse duration of $(67\pm5)$\,fs FWHM at the sample position (obtained by using an autocorrelation measurement in combination with a modeling of the additional optics), we estimate the duration of the probe pulse to $(67\pm5)$\,fs FWHM, assuming uncorrelated Gaussian-shaped pulses. 

\subsection{Multispectral energy-momentum mapping} \label{sec:FermiSurface}

The key novel characteristic of our experimental setup is the combination of highly efficient 3D photoemission intensity $I(E, k_x, k_y)$ imaging with a discrete tunability of the probe photon energy, thus making the energy-momentum mapping in trARPES $k_z$-dependent and 4D. Figure~\ref{HHG_Series} illustrates this experimental advance.

\begin{figure}
\centering
\includegraphics[width=0.5\textwidth]{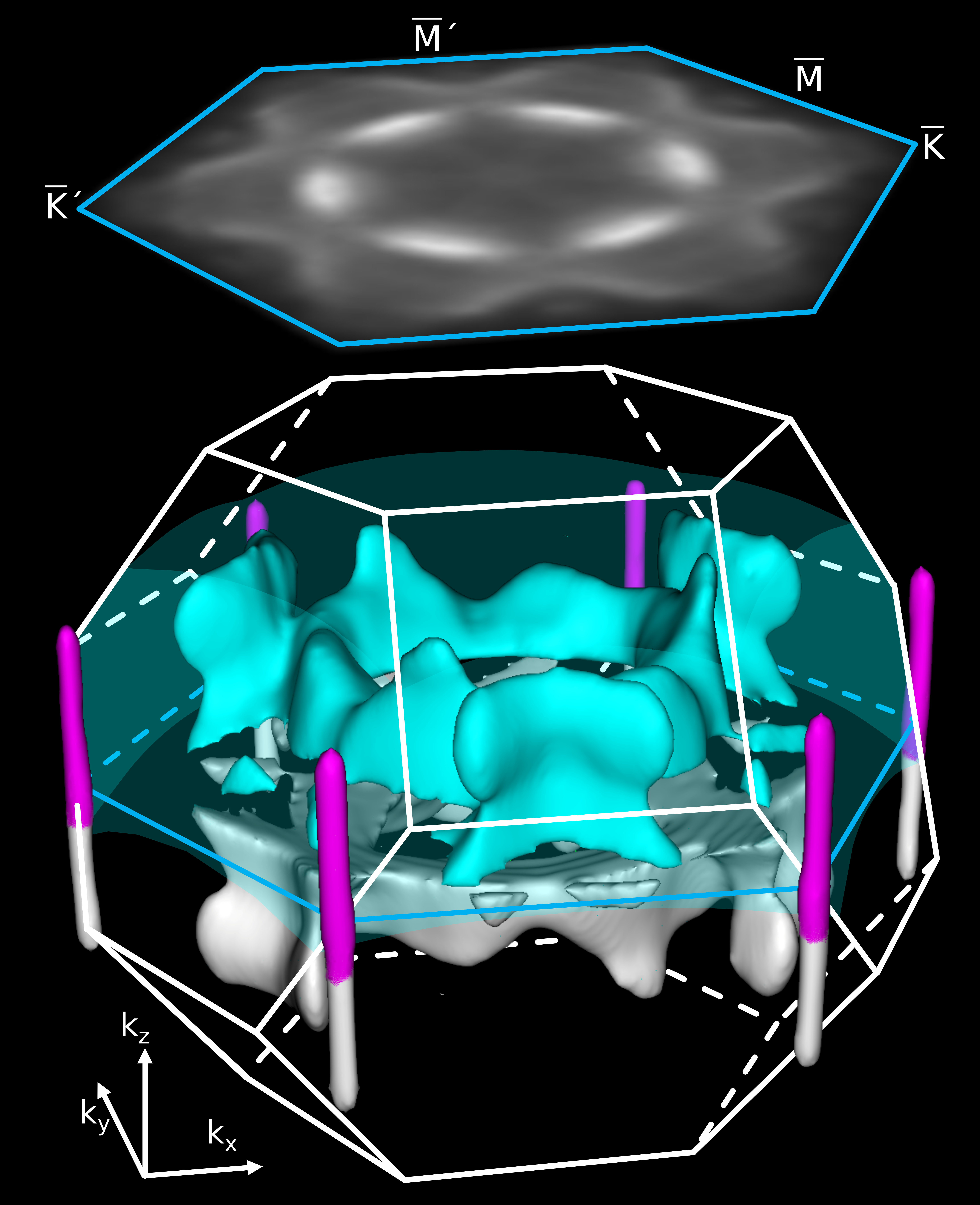}
\caption{Lower part: 3D Fermi surface tomogram of graphene/Ir(111), as reconstructed from the Fermi surface maps shown in Fig.~\ref{HHG_Series}. The 2D and 3D Brillouin zones of graphene and Ir(111) are indicated by blue and white lines, respectively. Fermi-surface sheets obtained from originally measured data are highlighted by pink and cyan color for graphene and Ir(111), respectively. Corresponding Fermi surface sheets obtained by point reflection about the $\Gamma$-point are shown in gray. Top part: Interpolated Fermi surface map through $\Gamma$.}
\label{fig:3DIr}
\end{figure}

The central panel of Fig.~\ref{HHG_Series} displays two typical XUV spectra as a function of the monochromator energy. The spectrum indicated by the black line gives the calculated photon flux at the sample position. This signal was measured by a calibrated XUV-sensitive Si photodiode behind the exit slit of the monochromator and corrected for beamline transmission, including Al-filter attenuation to prevent photocurrent saturation. The second spectrum (red line) represents the electron count rate at the detector, as measured from an electrically biased graphene/Ir(111) sample by the ToF momentum microscope under otherwise typical measurement settings. The practically useable photon-energy tuning range is $23.6$--$45.5$\,eV, at a spacing of $\approx$$3.1$\,eV, corresponding to the odd harmonic orders 15 to 29. The cutoff at $\approx$48\,eV is typical for Ar gas as a generating medium.\cite{Krause1992}
 
The panels surrounding the XUV spectra in Fig.~\ref{HHG_Series} show Fermi surface maps taken from graphene/Ir(111) with all eight harmonics. These 2D intensity maps are extracted from the full 3D data cubes that were originally measured. The data acquisition times varied from 2 to 12\,h. The multispectral Fermi surface maps reflect a superposition of the point-like 2D Fermi surface of graphene (centered at the corners of the indicated hexagonal graphene BZ) and the complex multi-sheet 3D Fermi surface of Ir(111). There is no photon energy-dependent change in the shape of the graphene Fermi points, whereas the variation in the shape of the Ir~$5d$ intensity pattern, i.e., $k_z$ dispersion, is pronounced. Another observation is that the relative contribution of the Ir~$5d$ signal to the total photoemission intensity at $E_\mathrm{F}$ is continuously suppressed upon increasing the photon energy from $23.6$\,eV to $45.5$\,eV. We attribute this effect to a decrease in the electron escape depth upon approaching the minimum of the universal curve of the inelastic mean free path around a kinetic energy of $\approx$50\,eV:\cite{Tanuma2011} With increasing surface sensitivity, less photoemission signal is obtained from Ir(111), which is covered by a graphene monolayer.

Under the assumption of free-electron-like final states within the direct-transition model of ARPES, constant-energy maps, such as the ones displayed in Fig.~\ref{HHG_Series}, map onto a spherical surface in 3D momentum space. The kinematic equation relating the surface-perpendicular momentum component to the other measurement parameters is:\cite{Chiang1980}
\begin{equation}
k_z = \sqrt{\frac{2m^*}{\hbar^2}\left[(E-E_\mathrm{F}) + h\nu + V_0^*\right] - \left(k_x^2 + k_y^2 \right)},
\label{eq:kzEq}
\end{equation}
where $h\nu$ is the photon energy and $m^*$ and $V_0^*$ are the effective electron mass and inner potential (referenced to $E_\mathrm{F}$) of the nearly-free-electron final-state parabola, respectively. The empirical parameters for Ir(111) are $m^* = 1.07\,m_\mathrm{e}$ and $V_0^* = 10$\,eV.\cite{Elmers2017} Using equation~(\ref{eq:kzEq}) and exploiting point symmetry about the center of the BZ ($\Gamma$ point), a tomogram of the 3D Fermi surface can be reconstructed from the stack of Fermi surface maps shown in Fig.~\ref{HHG_Series}.

Figure~\ref{fig:3DIr} displays the reconstructed portion of the 3D Fermi surface for graphene/Ir(111). Three sets of isosurfaces can be identified: the non-$k_z$-dispersive graphene Fermi rods centered at the corners of the hexagonal graphene BZ as well as an inner hexagon-shaped Fermi surface sheet and an outer star-shaped Fermi surface sheet derived from Ir~$5d$ bulk states. The reconstructed Ir~$5d$ Fermi surface sheets are in good agreement with a 3D Fermi surface tomogram obtained from pristine Ir(111) by soft x-ray momentum microscopy.\cite{Elmers2017} The available photon-energy range translates into a finite $k_z$ probing interval of $\Delta k_z = 0.9$\,\AA$^{-1}$, smaller than the characteristic $k_z$ dimension ($\overline{\Gamma L} = 1.42$\,\AA$^{-1}$) of the fcc BZ of Ir(111). Thus, our tomographic data cover $\approx$39\% of the BZ volume. For crystalline materials with smaller $k_z$ dimensions, particularly layered 3D electron materials,\cite{Rossnagel2001,Krasovskii2007,Watson2019,Bao2022} larger portions of the bulk BZ or entire bulk BZs can be scanned. Similarly, for layers of 3D molecules,\cite{Haag2020} 3D orbital momentum tomograms can be recorded.

\section{Conclusions} \label{sec:conclusion}

In conclusion, by combining a 790-nm-driven kHz-repetition-rate HHG source with a toroidal-grating monochromator and a high-detection-efficiency ToF momentum microscope, we have realized an experimental setup for probe photon energy-dependent time-resolved XUV-ARPES with good data collection efficiency and sub-100-fs time resolution. The photon energy tuning range is $\approx$24--46\,eV, sufficient to map band structures and Fermi surfaces as well as molecular orbital densities over an $k_z$ interval of $\approx$1\,\AA$^{-1}$. The system thus specifically enables $k_z$-selective probing of ultrafast electronic structure dynamics in 3D materials as well as ultrafast 3D orbital tomography of molecular layers.\cite{Wallauer2021,Baumgaertner2022} Moreover, the setup complements time-resolved momentum microscopy in the XUV to soft x-ray regime at the FEL FLASH, for which the same ToF momentum microscope is used.\cite{Kutnyakhov2020} The two complementary probe photon sources particularly enable a unique merging of trARPES with time-resolved x-ray photoelectron spectroscopy\cite{Hellmann2010,Dendzik2020,Pressacco2021} and diffraction\cite{Curcio2021} for combined investigations of ultrafast electronic, chemical, and geometric structure dynamics.

\begin{acknowledgments}
This work is dedicated to the late Wilfried Wurth. The work was supported by the Deutsche Forschungsgemeinschaft (DFG, German Research Foundation)---Project ID 491245950 and via the Collaborative Research Center (CRC) 925---Project ID 170620586 (project B2). We thank Holger Meyer and Sven Gieschen for support with the instrumentation. We also thank former members of Wolfgang Eberhardt's group, particularly Daniel Ramm, for help with the design and initial installation of the HHG source.
\end{acknowledgments}

\section*{Data Availability Statement}
The data that support the findings of this study are available from the corresponding author upon reasonable request.

\bibliography{HHG_HEXTOF_Revision}

\begin{thebibliography}{77}%
\makeatletter
\providecommand \@ifxundefined [1]{%
 \@ifx{#1\undefined}
}%
\providecommand \@ifnum [1]{%
 \ifnum #1\expandafter \@firstoftwo
 \else \expandafter \@secondoftwo
 \fi
}%
\providecommand \@ifx [1]{%
 \ifx #1\expandafter \@firstoftwo
 \else \expandafter \@secondoftwo
 \fi
}%
\providecommand \natexlab [1]{#1}%
\providecommand \enquote  [1]{``#1''}%
\providecommand \bibnamefont  [1]{#1}%
\providecommand \bibfnamefont [1]{#1}%
\providecommand \citenamefont [1]{#1}%
\providecommand \href@noop [0]{\@secondoftwo}%
\providecommand \href [0]{\begingroup \@sanitize@url \@href}%
\providecommand \@href[1]{\@@startlink{#1}\@@href}%
\providecommand \@@href[1]{\endgroup#1\@@endlink}%
\providecommand \@sanitize@url [0]{\catcode `\\12\catcode `\$12\catcode
  `\&12\catcode `\#12\catcode `\^12\catcode `\_12\catcode `\%12\relax}%
\providecommand \@@startlink[1]{}%
\providecommand \@@endlink[0]{}%
\providecommand \url  [0]{\begingroup\@sanitize@url \@url }%
\providecommand \@url [1]{\endgroup\@href {#1}{\urlprefix }}%
\providecommand \urlprefix  [0]{URL }%
\providecommand \Eprint [0]{\href }%
\providecommand \doibase [0]{http://dx.doi.org/}%
\providecommand \selectlanguage [0]{\@gobble}%
\providecommand \bibinfo  [0]{\@secondoftwo}%
\providecommand \bibfield  [0]{\@secondoftwo}%
\providecommand \translation [1]{[#1]}%
\providecommand \BibitemOpen [0]{}%
\providecommand \bibitemStop [0]{}%
\providecommand \bibitemNoStop [0]{.\EOS\space}%
\providecommand \EOS [0]{\spacefactor3000\relax}%
\providecommand \BibitemShut  [1]{\csname bibitem#1\endcsname}%
\let\auto@bib@innerbib\@empty
\bibitem [{\citenamefont {Damascelli}(2004)}]{Damascelli2004}%
  \BibitemOpen
  \bibfield  {author} {\bibinfo {author} {\bibfnamefont {A.}~\bibnamefont
  {Damascelli}},\ }\bibfield  {title} {\enquote {\bibinfo {title} {Probing the
  electronic structure of complex systems by {ARPES}},}\ }\href {\doibase
  10.1238/physica.topical.109a00061} {\bibfield  {journal} {\bibinfo  {journal}
  {Physica Scripta}\ }\textbf {\bibinfo {volume} {T109}},\ \bibinfo {pages}
  {61} (\bibinfo {year} {2004})}\BibitemShut {NoStop}%
\bibitem [{\citenamefont {Lv}, \citenamefont {Qian},\ and\ \citenamefont
  {Ding}(2019)}]{Lv2019}%
  \BibitemOpen
  \bibfield  {author} {\bibinfo {author} {\bibfnamefont {B.}~\bibnamefont
  {Lv}}, \bibinfo {author} {\bibfnamefont {T.}~\bibnamefont {Qian}}, \ and\
  \bibinfo {author} {\bibfnamefont {H.}~\bibnamefont {Ding}},\ }\bibfield
  {title} {\enquote {\bibinfo {title} {Angle-resolved photoemission
  spectroscopy and its application to topological materials},}\ }\href
  {\doibase 10.1038/s42254-019-0088-5} {\bibfield  {journal} {\bibinfo
  {journal} {Nature Reviews Physics}\ }\textbf {\bibinfo {volume} {1}},\
  \bibinfo {pages} {609--626} (\bibinfo {year} {2019})}\BibitemShut {NoStop}%
\bibitem [{\citenamefont {King}\ \emph {et~al.}(2021)\citenamefont {King},
  \citenamefont {Picozzi}, \citenamefont {Egdell},\ and\ \citenamefont
  {Panaccione}}]{King2021}%
  \BibitemOpen
  \bibfield  {author} {\bibinfo {author} {\bibfnamefont {P.~D.~C.}\
  \bibnamefont {King}}, \bibinfo {author} {\bibfnamefont {S.}~\bibnamefont
  {Picozzi}}, \bibinfo {author} {\bibfnamefont {R.~G.}\ \bibnamefont {Egdell}},
  \ and\ \bibinfo {author} {\bibfnamefont {G.}~\bibnamefont {Panaccione}},\
  }\bibfield  {title} {\enquote {\bibinfo {title} {Angle, spin, and depth
  resolved photoelectron spectroscopy on quantum materials},}\ }\href {\doibase
  10.1021/acs.chemrev.0c00616} {\bibfield  {journal} {\bibinfo  {journal}
  {Chemical Reviews}\ }\textbf {\bibinfo {volume} {121}},\ \bibinfo {pages}
  {2816--2856} (\bibinfo {year} {2021})},\ \bibinfo {note} {pMID: 33346644},\
  \Eprint {http://arxiv.org/abs/https://doi.org/10.1021/acs.chemrev.0c00616}
  {https://doi.org/10.1021/acs.chemrev.0c00616} \BibitemShut {NoStop}%
\bibitem [{\citenamefont {Sobota}, \citenamefont {He},\ and\ \citenamefont
  {Shen}(2021)}]{Sobota2021}%
  \BibitemOpen
  \bibfield  {author} {\bibinfo {author} {\bibfnamefont {J.~A.}\ \bibnamefont
  {Sobota}}, \bibinfo {author} {\bibfnamefont {Y.}~\bibnamefont {He}}, \ and\
  \bibinfo {author} {\bibfnamefont {Z.-X.}\ \bibnamefont {Shen}},\ }\bibfield
  {title} {\enquote {\bibinfo {title} {Angle-resolved photoemission studies of
  quantum materials},}\ }\href {\doibase 10.1103/RevModPhys.93.025006}
  {\bibfield  {journal} {\bibinfo  {journal} {Rev. Mod. Phys.}\ }\textbf
  {\bibinfo {volume} {93}},\ \bibinfo {pages} {025006} (\bibinfo {year}
  {2021})}\BibitemShut {NoStop}%
\bibitem [{\citenamefont {Chiang}\ \emph {et~al.}(1980)\citenamefont {Chiang},
  \citenamefont {Knapp}, \citenamefont {Aono},\ and\ \citenamefont
  {Eastman}}]{Chiang1980}%
  \BibitemOpen
  \bibfield  {author} {\bibinfo {author} {\bibfnamefont {T.~C.}\ \bibnamefont
  {Chiang}}, \bibinfo {author} {\bibfnamefont {J.~A.}\ \bibnamefont {Knapp}},
  \bibinfo {author} {\bibfnamefont {M.}~\bibnamefont {Aono}}, \ and\ \bibinfo
  {author} {\bibfnamefont {D.~E.}\ \bibnamefont {Eastman}},\ }\bibfield
  {title} {\enquote {\bibinfo {title} {Angle-resolved photoemission,
  valence-band dispersions
  $e(\stackrel{\ensuremath{\rightarrow}}{\mathrm{k}})$, and electron and hole
  lifetimes for gaas},}\ }\href {\doibase 10.1103/PhysRevB.21.3513} {\bibfield
  {journal} {\bibinfo  {journal} {Phys. Rev. B}\ }\textbf {\bibinfo {volume}
  {21}},\ \bibinfo {pages} {3513--3522} (\bibinfo {year} {1980})}\BibitemShut
  {NoStop}%
\bibitem [{\citenamefont {Medjanik}\ \emph {et~al.}(2017)\citenamefont
  {Medjanik}, \citenamefont {Fedchenko}, \citenamefont {Chernov}, \citenamefont
  {Kutnyakhov}, \citenamefont {Ellguth}, \citenamefont {Oelsner}, \citenamefont
  {Schönhense}, \citenamefont {Peixoto}, \citenamefont {Lutz}, \citenamefont
  {Min}, \citenamefont {Reinert}, \citenamefont {Däster}, \citenamefont
  {Acremann}, \citenamefont {Viefhaus}, \citenamefont {Wurth}, \citenamefont
  {Elmers},\ and\ \citenamefont {Schönhense}}]{Medjanik2017}%
  \BibitemOpen
  \bibfield  {author} {\bibinfo {author} {\bibfnamefont {K.}~\bibnamefont
  {Medjanik}}, \bibinfo {author} {\bibfnamefont {O.}~\bibnamefont {Fedchenko}},
  \bibinfo {author} {\bibfnamefont {S.}~\bibnamefont {Chernov}}, \bibinfo
  {author} {\bibfnamefont {D.}~\bibnamefont {Kutnyakhov}}, \bibinfo {author}
  {\bibfnamefont {M.}~\bibnamefont {Ellguth}}, \bibinfo {author} {\bibfnamefont
  {A.}~\bibnamefont {Oelsner}}, \bibinfo {author} {\bibfnamefont
  {B.}~\bibnamefont {Schönhense}}, \bibinfo {author} {\bibfnamefont
  {T.~R.~F.}\ \bibnamefont {Peixoto}}, \bibinfo {author} {\bibfnamefont
  {P.}~\bibnamefont {Lutz}}, \bibinfo {author} {\bibfnamefont {C.-H.}\
  \bibnamefont {Min}}, \bibinfo {author} {\bibfnamefont {F.}~\bibnamefont
  {Reinert}}, \bibinfo {author} {\bibfnamefont {S.}~\bibnamefont {Däster}},
  \bibinfo {author} {\bibfnamefont {Y.}~\bibnamefont {Acremann}}, \bibinfo
  {author} {\bibfnamefont {J.}~\bibnamefont {Viefhaus}}, \bibinfo {author}
  {\bibfnamefont {W.}~\bibnamefont {Wurth}}, \bibinfo {author} {\bibfnamefont
  {H.~J.}\ \bibnamefont {Elmers}}, \ and\ \bibinfo {author} {\bibfnamefont
  {G.}~\bibnamefont {Schönhense}},\ }\bibfield  {title} {\enquote {\bibinfo
  {title} {Direct 3d mapping of the fermi surface and fermi velocity},}\
  }\href {\doibase 10.1038/nmat4875} {\bibfield  {journal} {\bibinfo  {journal}
  {Nature Materials}\ }\textbf {\bibinfo {volume} {16}},\ \bibinfo {pages}
  {615--621} (\bibinfo {year} {2017})}\BibitemShut {NoStop}%
\bibitem [{\citenamefont {Hüfner}\ \emph {et~al.}(1999)\citenamefont
  {Hüfner}, \citenamefont {Claessen}, \citenamefont {Reinert}, \citenamefont
  {Straub}, \citenamefont {Strocov},\ and\ \citenamefont
  {Steiner}}]{Huefner1999}%
  \BibitemOpen
  \bibfield  {author} {\bibinfo {author} {\bibfnamefont {S.}~\bibnamefont
  {Hüfner}}, \bibinfo {author} {\bibfnamefont {R.}~\bibnamefont {Claessen}},
  \bibinfo {author} {\bibfnamefont {F.}~\bibnamefont {Reinert}}, \bibinfo
  {author} {\bibfnamefont {T.}~\bibnamefont {Straub}}, \bibinfo {author}
  {\bibfnamefont {V.}~\bibnamefont {Strocov}}, \ and\ \bibinfo {author}
  {\bibfnamefont {P.}~\bibnamefont {Steiner}},\ }\bibfield  {title} {\enquote
  {\bibinfo {title} {Photoemission spectroscopy in metals:: band
  structure-fermi surface–spectral function},}\ }\href {\doibase
  https://doi.org/10.1016/S0368-2048(99)00047-X} {\bibfield  {journal}
  {\bibinfo  {journal} {Journal of Electron Spectroscopy and Related
  Phenomena}\ }\textbf {\bibinfo {volume} {100}},\ \bibinfo {pages} {191--213}
  (\bibinfo {year} {1999})}\BibitemShut {NoStop}%
\bibitem [{\citenamefont {Valla}\ \emph {et~al.}(1999)\citenamefont {Valla},
  \citenamefont {Fedorov}, \citenamefont {Johnson},\ and\ \citenamefont
  {Hulbert}}]{Valla1999}%
  \BibitemOpen
  \bibfield  {author} {\bibinfo {author} {\bibfnamefont {T.}~\bibnamefont
  {Valla}}, \bibinfo {author} {\bibfnamefont {A.~V.}\ \bibnamefont {Fedorov}},
  \bibinfo {author} {\bibfnamefont {P.~D.}\ \bibnamefont {Johnson}}, \ and\
  \bibinfo {author} {\bibfnamefont {S.~L.}\ \bibnamefont {Hulbert}},\
  }\bibfield  {title} {\enquote {\bibinfo {title} {Many-body effects in
  angle-resolved photoemission: Quasiparticle energy and lifetime of a mo(110)
  surface state},}\ }\href {\doibase 10.1103/PhysRevLett.83.2085} {\bibfield
  {journal} {\bibinfo  {journal} {Phys. Rev. Lett.}\ }\textbf {\bibinfo
  {volume} {83}},\ \bibinfo {pages} {2085--2088} (\bibinfo {year}
  {1999})}\BibitemShut {NoStop}%
\bibitem [{\citenamefont {Krasovskii}\ \emph {et~al.}(2007)\citenamefont
  {Krasovskii}, \citenamefont {Rossnagel}, \citenamefont {Fedorov},
  \citenamefont {Schattke},\ and\ \citenamefont {Kipp}}]{Krasovskii2007}%
  \BibitemOpen
  \bibfield  {author} {\bibinfo {author} {\bibfnamefont {E.~E.}\ \bibnamefont
  {Krasovskii}}, \bibinfo {author} {\bibfnamefont {K.}~\bibnamefont
  {Rossnagel}}, \bibinfo {author} {\bibfnamefont {A.}~\bibnamefont {Fedorov}},
  \bibinfo {author} {\bibfnamefont {W.}~\bibnamefont {Schattke}}, \ and\
  \bibinfo {author} {\bibfnamefont {L.}~\bibnamefont {Kipp}},\ }\bibfield
  {title} {\enquote {\bibinfo {title} {Determination of the hole lifetime from
  photoemission: Ti $3d$ states in ${\mathrm{tite}}_{2}$},}\ }\href {\doibase
  10.1103/PhysRevLett.98.217604} {\bibfield  {journal} {\bibinfo  {journal}
  {Phys. Rev. Lett.}\ }\textbf {\bibinfo {volume} {98}},\ \bibinfo {pages}
  {217604} (\bibinfo {year} {2007})}\BibitemShut {NoStop}%
\bibitem [{\citenamefont {Watson}\ \emph {et~al.}(2019)\citenamefont {Watson},
  \citenamefont {Clark}, \citenamefont {Mazzola}, \citenamefont
  {Markovi\ifmmode~\acute{c}\else \'{c}\fi{}}, \citenamefont {Sunko},
  \citenamefont {Kim}, \citenamefont {Rossnagel},\ and\ \citenamefont
  {King}}]{Watson2019}%
  \BibitemOpen
  \bibfield  {author} {\bibinfo {author} {\bibfnamefont {M.~D.}\ \bibnamefont
  {Watson}}, \bibinfo {author} {\bibfnamefont {O.~J.}\ \bibnamefont {Clark}},
  \bibinfo {author} {\bibfnamefont {F.}~\bibnamefont {Mazzola}}, \bibinfo
  {author} {\bibfnamefont {I.}~\bibnamefont {Markovi\ifmmode~\acute{c}\else
  \'{c}\fi{}}}, \bibinfo {author} {\bibfnamefont {V.}~\bibnamefont {Sunko}},
  \bibinfo {author} {\bibfnamefont {T.~K.}\ \bibnamefont {Kim}}, \bibinfo
  {author} {\bibfnamefont {K.}~\bibnamefont {Rossnagel}}, \ and\ \bibinfo
  {author} {\bibfnamefont {P.~D.~C.}\ \bibnamefont {King}},\ }\bibfield
  {title} {\enquote {\bibinfo {title} {Orbital- and ${k}_{z}$-selective
  hybridization of se $4p$ and ti $3d$ states in the charge density wave phase
  of ${\mathrm{tise}}_{2}$},}\ }\href {\doibase 10.1103/PhysRevLett.122.076404}
  {\bibfield  {journal} {\bibinfo  {journal} {Phys. Rev. Lett.}\ }\textbf
  {\bibinfo {volume} {122}},\ \bibinfo {pages} {076404} (\bibinfo {year}
  {2019})}\BibitemShut {NoStop}%
\bibitem [{\citenamefont {Puschnig}\ \emph {et~al.}(2009)\citenamefont
  {Puschnig}, \citenamefont {Berkebile}, \citenamefont {Fleming}, \citenamefont
  {Koller}, \citenamefont {Emtsev}, \citenamefont {Seyller}, \citenamefont
  {Riley}, \citenamefont {Ambrosch-Draxl}, \citenamefont {Netzer},\ and\
  \citenamefont {Ramsey}}]{Puschnig2009}%
  \BibitemOpen
  \bibfield  {author} {\bibinfo {author} {\bibfnamefont {P.}~\bibnamefont
  {Puschnig}}, \bibinfo {author} {\bibfnamefont {S.}~\bibnamefont {Berkebile}},
  \bibinfo {author} {\bibfnamefont {A.~J.}\ \bibnamefont {Fleming}}, \bibinfo
  {author} {\bibfnamefont {G.}~\bibnamefont {Koller}}, \bibinfo {author}
  {\bibfnamefont {K.}~\bibnamefont {Emtsev}}, \bibinfo {author} {\bibfnamefont
  {T.}~\bibnamefont {Seyller}}, \bibinfo {author} {\bibfnamefont {J.~D.}\
  \bibnamefont {Riley}}, \bibinfo {author} {\bibfnamefont {C.}~\bibnamefont
  {Ambrosch-Draxl}}, \bibinfo {author} {\bibfnamefont {F.~P.}\ \bibnamefont
  {Netzer}}, \ and\ \bibinfo {author} {\bibfnamefont {M.~G.}\ \bibnamefont
  {Ramsey}},\ }\bibfield  {title} {\enquote {\bibinfo {title} {Reconstruction
  of molecular orbital densities from photoemission data},}\ }\href {\doibase
  10.1126/science.1176105} {\bibfield  {journal} {\bibinfo  {journal}
  {Science}\ }\textbf {\bibinfo {volume} {326}},\ \bibinfo {pages} {702--706}
  (\bibinfo {year} {2009})},\ \Eprint
  {http://arxiv.org/abs/https://www.science.org/doi/pdf/10.1126/science.1176105}
  {https://www.science.org/doi/pdf/10.1126/science.1176105} \BibitemShut
  {NoStop}%
\bibitem [{\citenamefont {Graus}\ \emph {et~al.}(2019)\citenamefont {Graus},
  \citenamefont {Metzger}, \citenamefont {Grimm}, \citenamefont {Nigge},
  \citenamefont {Feyer}, \citenamefont {Schöll},\ and\ \citenamefont
  {Reinert}}]{Graus2019}%
  \BibitemOpen
  \bibfield  {author} {\bibinfo {author} {\bibfnamefont {M.}~\bibnamefont
  {Graus}}, \bibinfo {author} {\bibfnamefont {C.}~\bibnamefont {Metzger}},
  \bibinfo {author} {\bibfnamefont {M.}~\bibnamefont {Grimm}}, \bibinfo
  {author} {\bibfnamefont {P.}~\bibnamefont {Nigge}}, \bibinfo {author}
  {\bibfnamefont {V.}~\bibnamefont {Feyer}}, \bibinfo {author} {\bibfnamefont
  {A.}~\bibnamefont {Schöll}}, \ and\ \bibinfo {author} {\bibfnamefont
  {F.}~\bibnamefont {Reinert}},\ }\bibfield  {title} {\enquote {\bibinfo
  {title} {Three-dimensional tomographic imaging of molecular orbitals by
  photoelectron momentum microscopy},}\ }\href {\doibase
  10.1140/epjb/e2019-100015-x} {\bibfield  {journal} {\bibinfo  {journal} {The
  European Physical Journal B}\ }\textbf {\bibinfo {volume} {92}},\ \bibinfo
  {pages} {80} (\bibinfo {year} {2019})}\BibitemShut {NoStop}%
\bibitem [{\citenamefont {Wannberg}(2009)}]{Wannberg2009}%
  \BibitemOpen
  \bibfield  {author} {\bibinfo {author} {\bibfnamefont {B.}~\bibnamefont
  {Wannberg}},\ }\bibfield  {title} {\enquote {\bibinfo {title} {Electron
  optics development for photo-electron spectrometers},}\ }\href {\doibase
  https://doi.org/10.1016/j.nima.2008.12.156} {\bibfield  {journal} {\bibinfo
  {journal} {Nuclear Instruments and Methods in Physics Research Section A:
  Accelerators, Spectrometers, Detectors and Associated Equipment}\ }\textbf
  {\bibinfo {volume} {601}},\ \bibinfo {pages} {182--194} (\bibinfo {year}
  {2009})},\ \bibinfo {note} {special issue in honour of Prof. Kai
  Siegbahn}\BibitemShut {NoStop}%
\bibitem [{\citenamefont {Kotsugi}\ \emph {et~al.}(2003)\citenamefont
  {Kotsugi}, \citenamefont {Kuch}, \citenamefont {Offi}, \citenamefont
  {Chelaru},\ and\ \citenamefont {Kirschner}}]{Kotsugi2003}%
  \BibitemOpen
  \bibfield  {author} {\bibinfo {author} {\bibfnamefont {M.}~\bibnamefont
  {Kotsugi}}, \bibinfo {author} {\bibfnamefont {W.}~\bibnamefont {Kuch}},
  \bibinfo {author} {\bibfnamefont {F.}~\bibnamefont {Offi}}, \bibinfo {author}
  {\bibfnamefont {L.~I.}\ \bibnamefont {Chelaru}}, \ and\ \bibinfo {author}
  {\bibfnamefont {J.}~\bibnamefont {Kirschner}},\ }\bibfield  {title} {\enquote
  {\bibinfo {title} {Microspectroscopic two-dimensional fermi surface mapping
  using a photoelectron emission microscope},}\ }\href {\doibase
  10.1063/1.1569404} {\bibfield  {journal} {\bibinfo  {journal} {Review of
  Scientific Instruments}\ }\textbf {\bibinfo {volume} {74}},\ \bibinfo {pages}
  {2754--2758} (\bibinfo {year} {2003})},\ \Eprint
  {http://arxiv.org/abs/https://doi.org/10.1063/1.1569404}
  {https://doi.org/10.1063/1.1569404} \BibitemShut {NoStop}%
\bibitem [{\citenamefont {Sch\"onhense}, \citenamefont {Medjanik},\ and\
  \citenamefont {Elmers}(2015)}]{Schoenhense2015}%
  \BibitemOpen
  \bibfield  {author} {\bibinfo {author} {\bibfnamefont {G.}~\bibnamefont
  {Sch\"onhense}}, \bibinfo {author} {\bibfnamefont {K.}~\bibnamefont
  {Medjanik}}, \ and\ \bibinfo {author} {\bibfnamefont {H.-J.}\ \bibnamefont
  {Elmers}},\ }\bibfield  {title} {\enquote {\bibinfo {title} {Space-, time-
  and spin-resolved photoemission},}\ }\href {\doibase
  http://dx.doi.org/10.1016/j.elspec.2015.05.016} {\bibfield  {journal}
  {\bibinfo  {journal} {Journal of Electron Spectroscopy and Related
  Phenomena}\ }\textbf {\bibinfo {volume} {200}},\ \bibinfo {pages} {94 -- 118}
  (\bibinfo {year} {2015})},\ \bibinfo {note} {special Anniversary Issue:
  Volume 200}\BibitemShut {NoStop}%
\bibitem [{\citenamefont {Babenkov}\ \emph {et~al.}(2019)\citenamefont
  {Babenkov}, \citenamefont {Medjanik}, \citenamefont {Vasilyev}, \citenamefont
  {Chernov}, \citenamefont {Schlueter}, \citenamefont {Gloskovskii},
  \citenamefont {Matveyev}, \citenamefont {Drube}, \citenamefont
  {Sch{\"o}nhense}, \citenamefont {Rossnagel}, \citenamefont {Elmers},\ and\
  \citenamefont {Sch{\"o}nhense}}]{Babenkov2019}%
  \BibitemOpen
  \bibfield  {author} {\bibinfo {author} {\bibfnamefont {S.}~\bibnamefont
  {Babenkov}}, \bibinfo {author} {\bibfnamefont {K.}~\bibnamefont {Medjanik}},
  \bibinfo {author} {\bibfnamefont {D.}~\bibnamefont {Vasilyev}}, \bibinfo
  {author} {\bibfnamefont {S.}~\bibnamefont {Chernov}}, \bibinfo {author}
  {\bibfnamefont {C.}~\bibnamefont {Schlueter}}, \bibinfo {author}
  {\bibfnamefont {A.}~\bibnamefont {Gloskovskii}}, \bibinfo {author}
  {\bibfnamefont {Y.}~\bibnamefont {Matveyev}}, \bibinfo {author}
  {\bibfnamefont {W.}~\bibnamefont {Drube}}, \bibinfo {author} {\bibfnamefont
  {B.}~\bibnamefont {Sch{\"o}nhense}}, \bibinfo {author} {\bibfnamefont
  {K.}~\bibnamefont {Rossnagel}}, \bibinfo {author} {\bibfnamefont {H.-J.}\
  \bibnamefont {Elmers}}, \ and\ \bibinfo {author} {\bibfnamefont
  {G.}~\bibnamefont {Sch{\"o}nhense}},\ }\bibfield  {title} {\enquote {\bibinfo
  {title} {High-accuracy bulk electronic bandmapping with eliminated
  diffraction effects using hard x-ray photoelectron momentum microscopy},}\
  }\href {\doibase 10.1038/s42005-019-0208-7} {\bibfield  {journal} {\bibinfo
  {journal} {Communications Physics}\ }\textbf {\bibinfo {volume} {2}},\
  \bibinfo {pages} {107} (\bibinfo {year} {2019})}\BibitemShut {NoStop}%
\bibitem [{\citenamefont {Rossnagel}\ \emph {et~al.}(2001)\citenamefont
  {Rossnagel}, \citenamefont {Kipp}, \citenamefont {Skibowski}, \citenamefont
  {Solterbeck}, \citenamefont {Strasser}, \citenamefont {Schattke},
  \citenamefont {Vo\ss{}}, \citenamefont {Kr\"uger}, \citenamefont {Mazur},\
  and\ \citenamefont {Pollmann}}]{Rossnagel2001}%
  \BibitemOpen
  \bibfield  {author} {\bibinfo {author} {\bibfnamefont {K.}~\bibnamefont
  {Rossnagel}}, \bibinfo {author} {\bibfnamefont {L.}~\bibnamefont {Kipp}},
  \bibinfo {author} {\bibfnamefont {M.}~\bibnamefont {Skibowski}}, \bibinfo
  {author} {\bibfnamefont {C.}~\bibnamefont {Solterbeck}}, \bibinfo {author}
  {\bibfnamefont {T.}~\bibnamefont {Strasser}}, \bibinfo {author}
  {\bibfnamefont {W.}~\bibnamefont {Schattke}}, \bibinfo {author}
  {\bibfnamefont {D.}~\bibnamefont {Vo\ss{}}}, \bibinfo {author} {\bibfnamefont
  {P.}~\bibnamefont {Kr\"uger}}, \bibinfo {author} {\bibfnamefont
  {A.}~\bibnamefont {Mazur}}, \ and\ \bibinfo {author} {\bibfnamefont
  {J.}~\bibnamefont {Pollmann}},\ }\bibfield  {title} {\enquote {\bibinfo
  {title} {Three-dimensional fermi surface determination by angle-resolved
  photoelectron spectroscopy},}\ }\href {\doibase 10.1103/PhysRevB.63.125104}
  {\bibfield  {journal} {\bibinfo  {journal} {Phys. Rev. B}\ }\textbf {\bibinfo
  {volume} {63}},\ \bibinfo {pages} {125104} (\bibinfo {year}
  {2001})}\BibitemShut {NoStop}%
\bibitem [{\citenamefont {Nielsen}\ \emph {et~al.}(2003)\citenamefont
  {Nielsen}, \citenamefont {Li}, \citenamefont {Lizzit}, \citenamefont
  {Goldoni},\ and\ \citenamefont {Hofmann}}]{Nielsen2003}%
  \BibitemOpen
  \bibfield  {author} {\bibinfo {author} {\bibfnamefont {M.~B.}\ \bibnamefont
  {Nielsen}}, \bibinfo {author} {\bibfnamefont {Z.}~\bibnamefont {Li}},
  \bibinfo {author} {\bibfnamefont {S.}~\bibnamefont {Lizzit}}, \bibinfo
  {author} {\bibfnamefont {A.}~\bibnamefont {Goldoni}}, \ and\ \bibinfo
  {author} {\bibfnamefont {P.}~\bibnamefont {Hofmann}},\ }\bibfield  {title}
  {\enquote {\bibinfo {title} {Bulk fermi surface mapping with high-energy
  angle-resolved photoemission},}\ }\href {\doibase
  10.1088/0953-8984/15/41/002} {\bibfield  {journal} {\bibinfo  {journal}
  {Journal of Physics: Condensed Matter}\ }\textbf {\bibinfo {volume} {15}},\
  \bibinfo {pages} {6919--6930} (\bibinfo {year} {2003})}\BibitemShut {NoStop}%
\bibitem [{\citenamefont {Strocov}\ \emph {et~al.}(2014)\citenamefont
  {Strocov}, \citenamefont {Kobayashi}, \citenamefont {Wang}, \citenamefont
  {Lev}, \citenamefont {Krempasky}, \citenamefont {Rogalev}, \citenamefont
  {Schmitt}, \citenamefont {Cancellieri},\ and\ \citenamefont
  {Reinle-Schmitt}}]{Strocov2014}%
  \BibitemOpen
  \bibfield  {author} {\bibinfo {author} {\bibfnamefont {V.~N.}\ \bibnamefont
  {Strocov}}, \bibinfo {author} {\bibfnamefont {M.}~\bibnamefont {Kobayashi}},
  \bibinfo {author} {\bibfnamefont {X.}~\bibnamefont {Wang}}, \bibinfo {author}
  {\bibfnamefont {L.~L.}\ \bibnamefont {Lev}}, \bibinfo {author} {\bibfnamefont
  {J.}~\bibnamefont {Krempasky}}, \bibinfo {author} {\bibfnamefont {V.~V.}\
  \bibnamefont {Rogalev}}, \bibinfo {author} {\bibfnamefont {T.}~\bibnamefont
  {Schmitt}}, \bibinfo {author} {\bibfnamefont {C.}~\bibnamefont
  {Cancellieri}}, \ and\ \bibinfo {author} {\bibfnamefont {M.~L.}\ \bibnamefont
  {Reinle-Schmitt}},\ }\bibfield  {title} {\enquote {\bibinfo {title}
  {Soft-x-ray arpes at the swiss light source: From 3d materials to buried
  interfaces and impurities},}\ }\href {\doibase 10.1080/08940886.2014.889550}
  {\bibfield  {journal} {\bibinfo  {journal} {Synchrotron Radiation News}\
  }\textbf {\bibinfo {volume} {27}},\ \bibinfo {pages} {31--40} (\bibinfo
  {year} {2014})},\ \Eprint
  {http://arxiv.org/abs/https://doi.org/10.1080/08940886.2014.889550}
  {https://doi.org/10.1080/08940886.2014.889550} \BibitemShut {NoStop}%
\bibitem [{\citenamefont {Moser}(2017)}]{Moser2017}%
  \BibitemOpen
  \bibfield  {author} {\bibinfo {author} {\bibfnamefont {S.}~\bibnamefont
  {Moser}},\ }\bibfield  {title} {\enquote {\bibinfo {title} {An
  experimentalist's guide to the matrix element in angle resolved
  photoemission},}\ }\href {\doibase
  https://doi.org/10.1016/j.elspec.2016.11.007} {\bibfield  {journal} {\bibinfo
   {journal} {Journal of Electron Spectroscopy and Related Phenomena}\ }\textbf
  {\bibinfo {volume} {214}},\ \bibinfo {pages} {29--52} (\bibinfo {year}
  {2017})}\BibitemShut {NoStop}%
\bibitem [{\citenamefont {Kutnyakhov}\ \emph {et~al.}(2020)\citenamefont
  {Kutnyakhov}, \citenamefont {Xian}, \citenamefont {Dendzik}, \citenamefont
  {Heber}, \citenamefont {Pressacco}, \citenamefont {Agustsson}, \citenamefont
  {Wenthaus}, \citenamefont {Meyer}, \citenamefont {Gieschen}, \citenamefont
  {Mercurio}, \citenamefont {Benz}, \citenamefont {B\"uhlman}, \citenamefont
  {D\"aster}, \citenamefont {Gort}, \citenamefont {Curcio}, \citenamefont
  {Volckaert}, \citenamefont {Bianchi}, \citenamefont {Sanders}, \citenamefont
  {Miwa}, \citenamefont {Ulstrup}, \citenamefont {Oelsner}, \citenamefont
  {Tusche}, \citenamefont {Chen}, \citenamefont {Vasilyev}, \citenamefont
  {Medjanik}, \citenamefont {Brenner}, \citenamefont {Dziarzhytski},
  \citenamefont {Redlin}, \citenamefont {Manschwetus}, \citenamefont {Dong},
  \citenamefont {Hauer}, \citenamefont {Rettig}, \citenamefont {Diekmann},
  \citenamefont {Rossnagel}, \citenamefont {Demsar}, \citenamefont {Elmers},
  \citenamefont {Hofmann}, \citenamefont {Ernstorfer}, \citenamefont
  {Sch\"onhense}, \citenamefont {Acremann},\ and\ \citenamefont
  {Wurth}}]{Kutnyakhov2020}%
  \BibitemOpen
  \bibfield  {author} {\bibinfo {author} {\bibfnamefont {D.}~\bibnamefont
  {Kutnyakhov}}, \bibinfo {author} {\bibfnamefont {R.~P.}\ \bibnamefont
  {Xian}}, \bibinfo {author} {\bibfnamefont {M.}~\bibnamefont {Dendzik}},
  \bibinfo {author} {\bibfnamefont {M.}~\bibnamefont {Heber}}, \bibinfo
  {author} {\bibfnamefont {F.}~\bibnamefont {Pressacco}}, \bibinfo {author}
  {\bibfnamefont {S.~Y.}\ \bibnamefont {Agustsson}}, \bibinfo {author}
  {\bibfnamefont {L.}~\bibnamefont {Wenthaus}}, \bibinfo {author}
  {\bibfnamefont {H.}~\bibnamefont {Meyer}}, \bibinfo {author} {\bibfnamefont
  {S.}~\bibnamefont {Gieschen}}, \bibinfo {author} {\bibfnamefont
  {G.}~\bibnamefont {Mercurio}}, \bibinfo {author} {\bibfnamefont
  {A.}~\bibnamefont {Benz}}, \bibinfo {author} {\bibfnamefont {K.}~\bibnamefont
  {B\"uhlman}}, \bibinfo {author} {\bibfnamefont {S.}~\bibnamefont {D\"aster}},
  \bibinfo {author} {\bibfnamefont {R.}~\bibnamefont {Gort}}, \bibinfo {author}
  {\bibfnamefont {D.}~\bibnamefont {Curcio}}, \bibinfo {author} {\bibfnamefont
  {K.}~\bibnamefont {Volckaert}}, \bibinfo {author} {\bibfnamefont
  {M.}~\bibnamefont {Bianchi}}, \bibinfo {author} {\bibfnamefont
  {C.}~\bibnamefont {Sanders}}, \bibinfo {author} {\bibfnamefont {J.~A.}\
  \bibnamefont {Miwa}}, \bibinfo {author} {\bibfnamefont {S.}~\bibnamefont
  {Ulstrup}}, \bibinfo {author} {\bibfnamefont {A.}~\bibnamefont {Oelsner}},
  \bibinfo {author} {\bibfnamefont {C.}~\bibnamefont {Tusche}}, \bibinfo
  {author} {\bibfnamefont {Y.-J.}\ \bibnamefont {Chen}}, \bibinfo {author}
  {\bibfnamefont {D.}~\bibnamefont {Vasilyev}}, \bibinfo {author}
  {\bibfnamefont {K.}~\bibnamefont {Medjanik}}, \bibinfo {author}
  {\bibfnamefont {G.}~\bibnamefont {Brenner}}, \bibinfo {author} {\bibfnamefont
  {S.}~\bibnamefont {Dziarzhytski}}, \bibinfo {author} {\bibfnamefont
  {H.}~\bibnamefont {Redlin}}, \bibinfo {author} {\bibfnamefont
  {B.}~\bibnamefont {Manschwetus}}, \bibinfo {author} {\bibfnamefont
  {S.}~\bibnamefont {Dong}}, \bibinfo {author} {\bibfnamefont {J.}~\bibnamefont
  {Hauer}}, \bibinfo {author} {\bibfnamefont {L.}~\bibnamefont {Rettig}},
  \bibinfo {author} {\bibfnamefont {F.}~\bibnamefont {Diekmann}}, \bibinfo
  {author} {\bibfnamefont {K.}~\bibnamefont {Rossnagel}}, \bibinfo {author}
  {\bibfnamefont {J.}~\bibnamefont {Demsar}}, \bibinfo {author} {\bibfnamefont
  {H.-J.}\ \bibnamefont {Elmers}}, \bibinfo {author} {\bibfnamefont
  {P.}~\bibnamefont {Hofmann}}, \bibinfo {author} {\bibfnamefont
  {R.}~\bibnamefont {Ernstorfer}}, \bibinfo {author} {\bibfnamefont
  {G.}~\bibnamefont {Sch\"onhense}}, \bibinfo {author} {\bibfnamefont
  {Y.}~\bibnamefont {Acremann}}, \ and\ \bibinfo {author} {\bibfnamefont
  {W.}~\bibnamefont {Wurth}},\ }\bibfield  {title} {\enquote {\bibinfo {title}
  {Time- and momentum-resolved photoemission studies using time-of-flight
  momentum microscopy at a free-electron laser},}\ }\href {\doibase
  10.1063/1.5118777} {\bibfield  {journal} {\bibinfo  {journal} {Review of
  Scientific Instruments}\ }\textbf {\bibinfo {volume} {91}},\ \bibinfo {pages}
  {013109} (\bibinfo {year} {2020})},\ \Eprint
  {http://arxiv.org/abs/https://doi.org/10.1063/1.5118777}
  {https://doi.org/10.1063/1.5118777} \BibitemShut {NoStop}%
\bibitem [{\citenamefont {Maklar}\ \emph {et~al.}(2020)\citenamefont {Maklar},
  \citenamefont {Dong}, \citenamefont {Beaulieu}, \citenamefont {Pincelli},
  \citenamefont {Dendzik}, \citenamefont {Windsor}, \citenamefont {Xian},
  \citenamefont {Wolf}, \citenamefont {Ernstorfer},\ and\ \citenamefont
  {Rettig}}]{Maklar2020}%
  \BibitemOpen
  \bibfield  {author} {\bibinfo {author} {\bibfnamefont {J.}~\bibnamefont
  {Maklar}}, \bibinfo {author} {\bibfnamefont {S.}~\bibnamefont {Dong}},
  \bibinfo {author} {\bibfnamefont {S.}~\bibnamefont {Beaulieu}}, \bibinfo
  {author} {\bibfnamefont {T.}~\bibnamefont {Pincelli}}, \bibinfo {author}
  {\bibfnamefont {M.}~\bibnamefont {Dendzik}}, \bibinfo {author} {\bibfnamefont
  {Y.~W.}\ \bibnamefont {Windsor}}, \bibinfo {author} {\bibfnamefont {R.~P.}\
  \bibnamefont {Xian}}, \bibinfo {author} {\bibfnamefont {M.}~\bibnamefont
  {Wolf}}, \bibinfo {author} {\bibfnamefont {R.}~\bibnamefont {Ernstorfer}}, \
  and\ \bibinfo {author} {\bibfnamefont {L.}~\bibnamefont {Rettig}},\
  }\bibfield  {title} {\enquote {\bibinfo {title} {A quantitative comparison of
  time-of-flight momentum microscopes and hemispherical analyzers for time- and
  angle-resolved photoemission spectroscopy experiments},}\ }\href {\doibase
  10.1063/5.0024493} {\bibfield  {journal} {\bibinfo  {journal} {Review of
  Scientific Instruments}\ }\textbf {\bibinfo {volume} {91}},\ \bibinfo {pages}
  {123112} (\bibinfo {year} {2020})},\ \Eprint
  {http://arxiv.org/abs/https://doi.org/10.1063/5.0024493}
  {https://doi.org/10.1063/5.0024493} \BibitemShut {NoStop}%
\bibitem [{\citenamefont {Keunecke}\ \emph {et~al.}(2020)\citenamefont
  {Keunecke}, \citenamefont {Möller}, \citenamefont {Schmitt}, \citenamefont
  {Nolte}, \citenamefont {Jansen}, \citenamefont {Reutzel}, \citenamefont
  {Gutberlet}, \citenamefont {Halasi}, \citenamefont {Steil}, \citenamefont
  {Steil},\ and\ \citenamefont {Mathias}}]{Keunecke2020}%
  \BibitemOpen
  \bibfield  {author} {\bibinfo {author} {\bibfnamefont {M.}~\bibnamefont
  {Keunecke}}, \bibinfo {author} {\bibfnamefont {C.}~\bibnamefont {Möller}},
  \bibinfo {author} {\bibfnamefont {D.}~\bibnamefont {Schmitt}}, \bibinfo
  {author} {\bibfnamefont {H.}~\bibnamefont {Nolte}}, \bibinfo {author}
  {\bibfnamefont {G.~S.~M.}\ \bibnamefont {Jansen}}, \bibinfo {author}
  {\bibfnamefont {M.}~\bibnamefont {Reutzel}}, \bibinfo {author} {\bibfnamefont
  {M.}~\bibnamefont {Gutberlet}}, \bibinfo {author} {\bibfnamefont
  {G.}~\bibnamefont {Halasi}}, \bibinfo {author} {\bibfnamefont
  {D.}~\bibnamefont {Steil}}, \bibinfo {author} {\bibfnamefont
  {S.}~\bibnamefont {Steil}}, \ and\ \bibinfo {author} {\bibfnamefont
  {S.}~\bibnamefont {Mathias}},\ }\bibfield  {title} {\enquote {\bibinfo
  {title} {Time-resolved momentum microscopy with a 1 mhz high-harmonic extreme
  ultraviolet beamline},}\ }\href {\doibase 10.1063/5.0006531} {\bibfield
  {journal} {\bibinfo  {journal} {Review of Scientific Instruments}\ }\textbf
  {\bibinfo {volume} {91}},\ \bibinfo {pages} {063905} (\bibinfo {year}
  {2020})},\ \Eprint {http://arxiv.org/abs/https://doi.org/10.1063/5.0006531}
  {https://doi.org/10.1063/5.0006531} \BibitemShut {NoStop}%
\bibitem [{\citenamefont {Perfetti}\ \emph {et~al.}(2006)\citenamefont
  {Perfetti}, \citenamefont {Loukakos}, \citenamefont {Lisowski}, \citenamefont
  {Bovensiepen}, \citenamefont {Berger}, \citenamefont {Biermann},
  \citenamefont {Cornaglia}, \citenamefont {Georges},\ and\ \citenamefont
  {Wolf}}]{Perfetti2006}%
  \BibitemOpen
  \bibfield  {author} {\bibinfo {author} {\bibfnamefont {L.}~\bibnamefont
  {Perfetti}}, \bibinfo {author} {\bibfnamefont {P.~A.}\ \bibnamefont
  {Loukakos}}, \bibinfo {author} {\bibfnamefont {M.}~\bibnamefont {Lisowski}},
  \bibinfo {author} {\bibfnamefont {U.}~\bibnamefont {Bovensiepen}}, \bibinfo
  {author} {\bibfnamefont {H.}~\bibnamefont {Berger}}, \bibinfo {author}
  {\bibfnamefont {S.}~\bibnamefont {Biermann}}, \bibinfo {author}
  {\bibfnamefont {P.~S.}\ \bibnamefont {Cornaglia}}, \bibinfo {author}
  {\bibfnamefont {A.}~\bibnamefont {Georges}}, \ and\ \bibinfo {author}
  {\bibfnamefont {M.}~\bibnamefont {Wolf}},\ }\bibfield  {title} {\enquote
  {\bibinfo {title} {Time evolution of the electronic structure of
  $1t\mathrm{\text{\ensuremath{-}}}{\mathrm{tas}}_{2}$ through the
  insulator-metal transition},}\ }\href {\doibase
  10.1103/PhysRevLett.97.067402} {\bibfield  {journal} {\bibinfo  {journal}
  {Phys. Rev. Lett.}\ }\textbf {\bibinfo {volume} {97}},\ \bibinfo {pages}
  {067402} (\bibinfo {year} {2006})}\BibitemShut {NoStop}%
\bibitem [{\citenamefont {Schmitt}\ \emph {et~al.}(2008)\citenamefont
  {Schmitt}, \citenamefont {Kirchmann}, \citenamefont {Bovensiepen},
  \citenamefont {Moore}, \citenamefont {Rettig}, \citenamefont {Krenz},
  \citenamefont {Chu}, \citenamefont {Ru}, \citenamefont {Perfetti},
  \citenamefont {Lu}, \citenamefont {Wolf}, \citenamefont {Fisher},\ and\
  \citenamefont {Shen}}]{Schmitt2008}%
  \BibitemOpen
  \bibfield  {author} {\bibinfo {author} {\bibfnamefont {F.}~\bibnamefont
  {Schmitt}}, \bibinfo {author} {\bibfnamefont {P.~S.}\ \bibnamefont
  {Kirchmann}}, \bibinfo {author} {\bibfnamefont {U.}~\bibnamefont
  {Bovensiepen}}, \bibinfo {author} {\bibfnamefont {R.~G.}\ \bibnamefont
  {Moore}}, \bibinfo {author} {\bibfnamefont {L.}~\bibnamefont {Rettig}},
  \bibinfo {author} {\bibfnamefont {M.}~\bibnamefont {Krenz}}, \bibinfo
  {author} {\bibfnamefont {J.-H.}\ \bibnamefont {Chu}}, \bibinfo {author}
  {\bibfnamefont {N.}~\bibnamefont {Ru}}, \bibinfo {author} {\bibfnamefont
  {L.}~\bibnamefont {Perfetti}}, \bibinfo {author} {\bibfnamefont {D.~H.}\
  \bibnamefont {Lu}}, \bibinfo {author} {\bibfnamefont {M.}~\bibnamefont
  {Wolf}}, \bibinfo {author} {\bibfnamefont {I.~R.}\ \bibnamefont {Fisher}}, \
  and\ \bibinfo {author} {\bibfnamefont {Z.-X.}\ \bibnamefont {Shen}},\
  }\bibfield  {title} {\enquote {\bibinfo {title} {Transient electronic
  structure and melting of a charge density wave in tbte<sub>3</sub>},}\ }\href
  {\doibase 10.1126/science.1160778} {\bibfield  {journal} {\bibinfo  {journal}
  {Science}\ }\textbf {\bibinfo {volume} {321}},\ \bibinfo {pages} {1649--1652}
  (\bibinfo {year} {2008})},\ \Eprint
  {http://arxiv.org/abs/https://www.science.org/doi/pdf/10.1126/science.1160778}
  {https://www.science.org/doi/pdf/10.1126/science.1160778} \BibitemShut
  {NoStop}%
\bibitem [{\citenamefont {Rohwer}\ \emph {et~al.}(2011)\citenamefont {Rohwer},
  \citenamefont {Hellmann}, \citenamefont {Wiesenmayer}, \citenamefont {Sohrt},
  \citenamefont {Stange}, \citenamefont {Slomski}, \citenamefont {Carr},
  \citenamefont {Liu}, \citenamefont {Avila}, \citenamefont {Kalläne},
  \citenamefont {Mathias}, \citenamefont {Kipp}, \citenamefont {Rossnagel},\
  and\ \citenamefont {Bauer}}]{Rohwer2011}%
  \BibitemOpen
  \bibfield  {author} {\bibinfo {author} {\bibfnamefont {T.}~\bibnamefont
  {Rohwer}}, \bibinfo {author} {\bibfnamefont {S.}~\bibnamefont {Hellmann}},
  \bibinfo {author} {\bibfnamefont {M.}~\bibnamefont {Wiesenmayer}}, \bibinfo
  {author} {\bibfnamefont {C.}~\bibnamefont {Sohrt}}, \bibinfo {author}
  {\bibfnamefont {A.}~\bibnamefont {Stange}}, \bibinfo {author} {\bibfnamefont
  {B.}~\bibnamefont {Slomski}}, \bibinfo {author} {\bibfnamefont
  {A.}~\bibnamefont {Carr}}, \bibinfo {author} {\bibfnamefont {Y.}~\bibnamefont
  {Liu}}, \bibinfo {author} {\bibfnamefont {L.~M.}\ \bibnamefont {Avila}},
  \bibinfo {author} {\bibfnamefont {M.}~\bibnamefont {Kalläne}}, \bibinfo
  {author} {\bibfnamefont {S.}~\bibnamefont {Mathias}}, \bibinfo {author}
  {\bibfnamefont {L.}~\bibnamefont {Kipp}}, \bibinfo {author} {\bibfnamefont
  {K.}~\bibnamefont {Rossnagel}}, \ and\ \bibinfo {author} {\bibfnamefont
  {M.}~\bibnamefont {Bauer}},\ }\bibfield  {title} {\enquote {\bibinfo {title}
  {Collapse of long-range charge order tracked by time-resolved photoemission
  at high momenta},}\ }\href {\doibase 10.1038/nature09829} {\bibfield
  {journal} {\bibinfo  {journal} {Nature}\ }\textbf {\bibinfo {volume} {471}},\
  \bibinfo {pages} {490--493} (\bibinfo {year} {2011})}\BibitemShut {NoStop}%
\bibitem [{\citenamefont {Smallwood}\ \emph {et~al.}(2012)\citenamefont
  {Smallwood}, \citenamefont {Hinton}, \citenamefont {Jozwiak}, \citenamefont
  {Zhang}, \citenamefont {Koralek}, \citenamefont {Eisaki}, \citenamefont
  {Lee}, \citenamefont {Orenstein},\ and\ \citenamefont
  {Lanzara}}]{Smallwood2012}%
  \BibitemOpen
  \bibfield  {author} {\bibinfo {author} {\bibfnamefont {C.~L.}\ \bibnamefont
  {Smallwood}}, \bibinfo {author} {\bibfnamefont {J.~P.}\ \bibnamefont
  {Hinton}}, \bibinfo {author} {\bibfnamefont {C.}~\bibnamefont {Jozwiak}},
  \bibinfo {author} {\bibfnamefont {W.}~\bibnamefont {Zhang}}, \bibinfo
  {author} {\bibfnamefont {J.~D.}\ \bibnamefont {Koralek}}, \bibinfo {author}
  {\bibfnamefont {H.}~\bibnamefont {Eisaki}}, \bibinfo {author} {\bibfnamefont
  {D.-H.}\ \bibnamefont {Lee}}, \bibinfo {author} {\bibfnamefont
  {J.}~\bibnamefont {Orenstein}}, \ and\ \bibinfo {author} {\bibfnamefont
  {A.}~\bibnamefont {Lanzara}},\ }\bibfield  {title} {\enquote {\bibinfo
  {title} {Tracking cooper pairs in a cuprate superconductor by ultrafast
  angle-resolved photoemission},}\ }\href {\doibase 10.1126/science.1217423}
  {\bibfield  {journal} {\bibinfo  {journal} {Science}\ }\textbf {\bibinfo
  {volume} {336}},\ \bibinfo {pages} {1137--1139} (\bibinfo {year} {2012})},\
  \Eprint
  {http://arxiv.org/abs/https://www.science.org/doi/pdf/10.1126/science.1217423}
  {https://www.science.org/doi/pdf/10.1126/science.1217423} \BibitemShut
  {NoStop}%
\bibitem [{\citenamefont {Gierz}\ \emph {et~al.}(2013)\citenamefont {Gierz},
  \citenamefont {Petersen}, \citenamefont {Mitrano}, \citenamefont {Cacho},
  \citenamefont {Turcu}, \citenamefont {Springate}, \citenamefont {Stöhr},
  \citenamefont {Köhler}, \citenamefont {Starke},\ and\ \citenamefont
  {Cavalleri}}]{Gierz2013}%
  \BibitemOpen
  \bibfield  {author} {\bibinfo {author} {\bibfnamefont {I.}~\bibnamefont
  {Gierz}}, \bibinfo {author} {\bibfnamefont {J.~C.}\ \bibnamefont {Petersen}},
  \bibinfo {author} {\bibfnamefont {M.}~\bibnamefont {Mitrano}}, \bibinfo
  {author} {\bibfnamefont {C.}~\bibnamefont {Cacho}}, \bibinfo {author}
  {\bibfnamefont {I.~C.~E.}\ \bibnamefont {Turcu}}, \bibinfo {author}
  {\bibfnamefont {E.}~\bibnamefont {Springate}}, \bibinfo {author}
  {\bibfnamefont {A.}~\bibnamefont {Stöhr}}, \bibinfo {author} {\bibfnamefont
  {A.}~\bibnamefont {Köhler}}, \bibinfo {author} {\bibfnamefont
  {U.}~\bibnamefont {Starke}}, \ and\ \bibinfo {author} {\bibfnamefont
  {A.}~\bibnamefont {Cavalleri}},\ }\bibfield  {title} {\enquote {\bibinfo
  {title} {Snapshots of non-equilibrium dirac carrier distributions in
  graphene},}\ }\href {\doibase 10.1038/nmat3757} {\bibfield  {journal}
  {\bibinfo  {journal} {Nature Materials}\ }\textbf {\bibinfo {volume} {12}},\
  \bibinfo {pages} {1119--1124} (\bibinfo {year} {2013})}\BibitemShut {NoStop}%
\bibitem [{\citenamefont {Wang}\ \emph {et~al.}(2013)\citenamefont {Wang},
  \citenamefont {Steinberg}, \citenamefont {Jarillo-Herrero},\ and\
  \citenamefont {Gedik}}]{Wang2013}%
  \BibitemOpen
  \bibfield  {author} {\bibinfo {author} {\bibfnamefont {Y.~H.}\ \bibnamefont
  {Wang}}, \bibinfo {author} {\bibfnamefont {H.}~\bibnamefont {Steinberg}},
  \bibinfo {author} {\bibfnamefont {P.}~\bibnamefont {Jarillo-Herrero}}, \ and\
  \bibinfo {author} {\bibfnamefont {N.}~\bibnamefont {Gedik}},\ }\bibfield
  {title} {\enquote {\bibinfo {title} {Observation of floquet-bloch states on
  the surface of a topological insulator},}\ }\href {\doibase
  10.1126/science.1239834} {\bibfield  {journal} {\bibinfo  {journal}
  {Science}\ }\textbf {\bibinfo {volume} {342}},\ \bibinfo {pages} {453--457}
  (\bibinfo {year} {2013})},\ \Eprint
  {http://arxiv.org/abs/https://www.science.org/doi/pdf/10.1126/science.1239834}
  {https://www.science.org/doi/pdf/10.1126/science.1239834} \BibitemShut
  {NoStop}%
\bibitem [{\citenamefont {Mahmood}\ \emph {et~al.}(2016)\citenamefont
  {Mahmood}, \citenamefont {Chan}, \citenamefont {Alpichshev}, \citenamefont
  {Gardner}, \citenamefont {Lee}, \citenamefont {Lee},\ and\ \citenamefont
  {Gedik}}]{Mahmood2016}%
  \BibitemOpen
  \bibfield  {author} {\bibinfo {author} {\bibfnamefont {F.}~\bibnamefont
  {Mahmood}}, \bibinfo {author} {\bibfnamefont {C.-K.}\ \bibnamefont {Chan}},
  \bibinfo {author} {\bibfnamefont {Z.}~\bibnamefont {Alpichshev}}, \bibinfo
  {author} {\bibfnamefont {D.}~\bibnamefont {Gardner}}, \bibinfo {author}
  {\bibfnamefont {Y.}~\bibnamefont {Lee}}, \bibinfo {author} {\bibfnamefont
  {P.~A.}\ \bibnamefont {Lee}}, \ and\ \bibinfo {author} {\bibfnamefont
  {N.}~\bibnamefont {Gedik}},\ }\bibfield  {title} {\enquote {\bibinfo {title}
  {Selective scattering between floquet–bloch and volkov states in a
  topological insulator},}\ }\href {\doibase 10.1038/nphys3609} {\bibfield
  {journal} {\bibinfo  {journal} {Nature Physics}\ }\textbf {\bibinfo {volume}
  {12}},\ \bibinfo {pages} {306--310} (\bibinfo {year} {2016})}\BibitemShut
  {NoStop}%
\bibitem [{\citenamefont {Nicholson}\ \emph {et~al.}(2018)\citenamefont
  {Nicholson}, \citenamefont {Lücke}, \citenamefont {Schmidt}, \citenamefont
  {Puppin}, \citenamefont {Rettig}, \citenamefont {Ernstorfer},\ and\
  \citenamefont {Wolf}}]{Nicholson2018}%
  \BibitemOpen
  \bibfield  {author} {\bibinfo {author} {\bibfnamefont {C.~W.}\ \bibnamefont
  {Nicholson}}, \bibinfo {author} {\bibfnamefont {A.}~\bibnamefont {Lücke}},
  \bibinfo {author} {\bibfnamefont {W.~G.}\ \bibnamefont {Schmidt}}, \bibinfo
  {author} {\bibfnamefont {M.}~\bibnamefont {Puppin}}, \bibinfo {author}
  {\bibfnamefont {L.}~\bibnamefont {Rettig}}, \bibinfo {author} {\bibfnamefont
  {R.}~\bibnamefont {Ernstorfer}}, \ and\ \bibinfo {author} {\bibfnamefont
  {M.}~\bibnamefont {Wolf}},\ }\bibfield  {title} {\enquote {\bibinfo {title}
  {Beyond the molecular movie: Dynamics of bands and bonds during a
  photoinduced phase transition},}\ }\href {\doibase 10.1126/science.aar4183}
  {\bibfield  {journal} {\bibinfo  {journal} {Science}\ }\textbf {\bibinfo
  {volume} {362}},\ \bibinfo {pages} {821--825} (\bibinfo {year} {2018})},\
  \Eprint
  {http://arxiv.org/abs/https://www.science.org/doi/pdf/10.1126/science.aar4183}
  {https://www.science.org/doi/pdf/10.1126/science.aar4183} \BibitemShut
  {NoStop}%
\bibitem [{\citenamefont {Na}\ \emph {et~al.}(2019)\citenamefont {Na},
  \citenamefont {Mills}, \citenamefont {Boschini}, \citenamefont {Michiardi},
  \citenamefont {Nosarzewski}, \citenamefont {Day}, \citenamefont {Razzoli},
  \citenamefont {Sheyerman}, \citenamefont {Schneider}, \citenamefont {Levy},
  \citenamefont {Zhdanovich}, \citenamefont {Devereaux}, \citenamefont
  {Kemper}, \citenamefont {Jones},\ and\ \citenamefont {Damascelli}}]{Na2019}%
  \BibitemOpen
  \bibfield  {author} {\bibinfo {author} {\bibfnamefont {M.~X.}\ \bibnamefont
  {Na}}, \bibinfo {author} {\bibfnamefont {A.~K.}\ \bibnamefont {Mills}},
  \bibinfo {author} {\bibfnamefont {F.}~\bibnamefont {Boschini}}, \bibinfo
  {author} {\bibfnamefont {M.}~\bibnamefont {Michiardi}}, \bibinfo {author}
  {\bibfnamefont {B.}~\bibnamefont {Nosarzewski}}, \bibinfo {author}
  {\bibfnamefont {R.~P.}\ \bibnamefont {Day}}, \bibinfo {author} {\bibfnamefont
  {E.}~\bibnamefont {Razzoli}}, \bibinfo {author} {\bibfnamefont
  {A.}~\bibnamefont {Sheyerman}}, \bibinfo {author} {\bibfnamefont
  {M.}~\bibnamefont {Schneider}}, \bibinfo {author} {\bibfnamefont
  {G.}~\bibnamefont {Levy}}, \bibinfo {author} {\bibfnamefont {S.}~\bibnamefont
  {Zhdanovich}}, \bibinfo {author} {\bibfnamefont {T.~P.}\ \bibnamefont
  {Devereaux}}, \bibinfo {author} {\bibfnamefont {A.~F.}\ \bibnamefont
  {Kemper}}, \bibinfo {author} {\bibfnamefont {D.~J.}\ \bibnamefont {Jones}}, \
  and\ \bibinfo {author} {\bibfnamefont {A.}~\bibnamefont {Damascelli}},\
  }\bibfield  {title} {\enquote {\bibinfo {title} {Direct determination of
  mode-projected electron-phonon coupling in the time domain},}\ }\href
  {\doibase 10.1126/science.aaw1662} {\bibfield  {journal} {\bibinfo  {journal}
  {Science}\ }\textbf {\bibinfo {volume} {366}},\ \bibinfo {pages} {1231--1236}
  (\bibinfo {year} {2019})},\ \Eprint
  {http://arxiv.org/abs/https://www.science.org/doi/pdf/10.1126/science.aaw1662}
  {https://www.science.org/doi/pdf/10.1126/science.aaw1662} \BibitemShut
  {NoStop}%
\bibitem [{\citenamefont {Wallauer}\ \emph {et~al.}(2021)\citenamefont
  {Wallauer}, \citenamefont {Raths}, \citenamefont {Stallberg}, \citenamefont
  {M{\"u}nster}, \citenamefont {Brandstetter}, \citenamefont {Yang},
  \citenamefont {G{\"u}dde}, \citenamefont {Puschnig}, \citenamefont
  {Soubatch}, \citenamefont {Kumpf}, \citenamefont {Bocquet}, \citenamefont
  {Tautz},\ and\ \citenamefont {H{\"o}fer}}]{Wallauer2021}%
  \BibitemOpen
  \bibfield  {author} {\bibinfo {author} {\bibfnamefont {R.}~\bibnamefont
  {Wallauer}}, \bibinfo {author} {\bibfnamefont {M.}~\bibnamefont {Raths}},
  \bibinfo {author} {\bibfnamefont {K.}~\bibnamefont {Stallberg}}, \bibinfo
  {author} {\bibfnamefont {L.}~\bibnamefont {M{\"u}nster}}, \bibinfo {author}
  {\bibfnamefont {D.}~\bibnamefont {Brandstetter}}, \bibinfo {author}
  {\bibfnamefont {X.}~\bibnamefont {Yang}}, \bibinfo {author} {\bibfnamefont
  {J.}~\bibnamefont {G{\"u}dde}}, \bibinfo {author} {\bibfnamefont
  {P.}~\bibnamefont {Puschnig}}, \bibinfo {author} {\bibfnamefont
  {S.}~\bibnamefont {Soubatch}}, \bibinfo {author} {\bibfnamefont
  {C.}~\bibnamefont {Kumpf}}, \bibinfo {author} {\bibfnamefont {F.~C.}\
  \bibnamefont {Bocquet}}, \bibinfo {author} {\bibfnamefont {F.~S.}\
  \bibnamefont {Tautz}}, \ and\ \bibinfo {author} {\bibfnamefont
  {U.}~\bibnamefont {H{\"o}fer}},\ }\bibfield  {title} {\enquote {\bibinfo
  {title} {Tracing orbital images on ultrafast time scales},}\ }\href {\doibase
  10.1126/science.abf3286} {\bibfield  {journal} {\bibinfo  {journal}
  {Science}\ } (\bibinfo {year} {2021}),\ 10.1126/science.abf3286}\BibitemShut
  {NoStop}%
\bibitem [{\citenamefont {Gauthier}\ \emph {et~al.}(2020)\citenamefont
  {Gauthier}, \citenamefont {Sobota}, \citenamefont {Gauthier}, \citenamefont
  {Xu}, \citenamefont {Pfau}, \citenamefont {Rotundu}, \citenamefont {Shen},\
  and\ \citenamefont {Kirchmann}}]{Gauthier2020}%
  \BibitemOpen
  \bibfield  {author} {\bibinfo {author} {\bibfnamefont {A.}~\bibnamefont
  {Gauthier}}, \bibinfo {author} {\bibfnamefont {J.~A.}\ \bibnamefont
  {Sobota}}, \bibinfo {author} {\bibfnamefont {N.}~\bibnamefont {Gauthier}},
  \bibinfo {author} {\bibfnamefont {K.-J.}\ \bibnamefont {Xu}}, \bibinfo
  {author} {\bibfnamefont {H.}~\bibnamefont {Pfau}}, \bibinfo {author}
  {\bibfnamefont {C.~R.}\ \bibnamefont {Rotundu}}, \bibinfo {author}
  {\bibfnamefont {Z.-X.}\ \bibnamefont {Shen}}, \ and\ \bibinfo {author}
  {\bibfnamefont {P.~S.}\ \bibnamefont {Kirchmann}},\ }\bibfield  {title}
  {\enquote {\bibinfo {title} {Tuning time and energy resolution in
  time-resolved photoemission spectroscopy with nonlinear crystals},}\ }\href
  {\doibase 10.1063/5.0018834} {\bibfield  {journal} {\bibinfo  {journal}
  {Journal of Applied Physics}\ }\textbf {\bibinfo {volume} {128}},\ \bibinfo
  {pages} {093101} (\bibinfo {year} {2020})},\ \Eprint
  {http://arxiv.org/abs/https://doi.org/10.1063/5.0018834}
  {https://doi.org/10.1063/5.0018834} \BibitemShut {NoStop}%
\bibitem [{\citenamefont {Bao}\ \emph {et~al.}(2022)\citenamefont {Bao},
  \citenamefont {Zhong}, \citenamefont {Zhou}, \citenamefont {Feng},
  \citenamefont {Wang},\ and\ \citenamefont {Zhou}}]{Bao2022}%
  \BibitemOpen
  \bibfield  {author} {\bibinfo {author} {\bibfnamefont {C.}~\bibnamefont
  {Bao}}, \bibinfo {author} {\bibfnamefont {H.}~\bibnamefont {Zhong}}, \bibinfo
  {author} {\bibfnamefont {S.}~\bibnamefont {Zhou}}, \bibinfo {author}
  {\bibfnamefont {R.}~\bibnamefont {Feng}}, \bibinfo {author} {\bibfnamefont
  {Y.}~\bibnamefont {Wang}}, \ and\ \bibinfo {author} {\bibfnamefont
  {S.}~\bibnamefont {Zhou}},\ }\bibfield  {title} {\enquote {\bibinfo {title}
  {Ultrafast time- and angle-resolved photoemission spectroscopy with widely
  tunable probe photon energy of 5.3–7.0 ev for investigating dynamics of
  three-dimensional materials},}\ }\href {\doibase 10.1063/5.0070004}
  {\bibfield  {journal} {\bibinfo  {journal} {Review of Scientific
  Instruments}\ }\textbf {\bibinfo {volume} {93}},\ \bibinfo {pages} {013902}
  (\bibinfo {year} {2022})},\ \Eprint
  {http://arxiv.org/abs/https://doi.org/10.1063/5.0070004}
  {https://doi.org/10.1063/5.0070004} \BibitemShut {NoStop}%
\bibitem [{\citenamefont {Dakovski}\ \emph {et~al.}(2010)\citenamefont
  {Dakovski}, \citenamefont {Li}, \citenamefont {Durakiewicz},\ and\
  \citenamefont {Rodriguez}}]{Dakovski2010}%
  \BibitemOpen
  \bibfield  {author} {\bibinfo {author} {\bibfnamefont {G.~L.}\ \bibnamefont
  {Dakovski}}, \bibinfo {author} {\bibfnamefont {Y.}~\bibnamefont {Li}},
  \bibinfo {author} {\bibfnamefont {T.}~\bibnamefont {Durakiewicz}}, \ and\
  \bibinfo {author} {\bibfnamefont {G.}~\bibnamefont {Rodriguez}},\ }\bibfield
  {title} {\enquote {\bibinfo {title} {Tunable ultrafast extreme ultraviolet
  source for time- and angle-resolved photoemission spectroscopy},}\ }\href
  {\doibase 10.1063/1.3460267} {\bibfield  {journal} {\bibinfo  {journal}
  {Review of Scientific Instruments}\ }\textbf {\bibinfo {volume} {81}},\
  \bibinfo {pages} {073108} (\bibinfo {year} {2010})},\ \Eprint
  {http://arxiv.org/abs/https://doi.org/10.1063/1.3460267}
  {https://doi.org/10.1063/1.3460267} \BibitemShut {NoStop}%
\bibitem [{\citenamefont {Frietsch}\ \emph {et~al.}(2013)\citenamefont
  {Frietsch}, \citenamefont {Carley}, \citenamefont {Döbrich}, \citenamefont
  {Gahl}, \citenamefont {Teichmann}, \citenamefont {Schwarzkopf}, \citenamefont
  {Wernet},\ and\ \citenamefont {Weinelt}}]{Frietsch2013}%
  \BibitemOpen
  \bibfield  {author} {\bibinfo {author} {\bibfnamefont {B.}~\bibnamefont
  {Frietsch}}, \bibinfo {author} {\bibfnamefont {R.}~\bibnamefont {Carley}},
  \bibinfo {author} {\bibfnamefont {K.}~\bibnamefont {Döbrich}}, \bibinfo
  {author} {\bibfnamefont {C.}~\bibnamefont {Gahl}}, \bibinfo {author}
  {\bibfnamefont {M.}~\bibnamefont {Teichmann}}, \bibinfo {author}
  {\bibfnamefont {O.}~\bibnamefont {Schwarzkopf}}, \bibinfo {author}
  {\bibfnamefont {P.}~\bibnamefont {Wernet}}, \ and\ \bibinfo {author}
  {\bibfnamefont {M.}~\bibnamefont {Weinelt}},\ }\bibfield  {title} {\enquote
  {\bibinfo {title} {A high-order harmonic generation apparatus for time- and
  angle-resolved photoelectron spectroscopy},}\ }\href {\doibase
  10.1063/1.4812992} {\bibfield  {journal} {\bibinfo  {journal} {Review of
  Scientific Instruments}\ }\textbf {\bibinfo {volume} {84}},\ \bibinfo {pages}
  {075106} (\bibinfo {year} {2013})},\ \Eprint
  {http://arxiv.org/abs/https://doi.org/10.1063/1.4812992}
  {https://doi.org/10.1063/1.4812992} \BibitemShut {NoStop}%
\bibitem [{\citenamefont {Eich}\ \emph {et~al.}(2014)\citenamefont {Eich},
  \citenamefont {Stange}, \citenamefont {Carr}, \citenamefont {Urbancic},
  \citenamefont {Popmintchev}, \citenamefont {Wiesenmayer}, \citenamefont
  {Jansen}, \citenamefont {Ruffing}, \citenamefont {Jakobs}, \citenamefont
  {Rohwer}, \citenamefont {Hellmann}, \citenamefont {Chen}, \citenamefont
  {Matyba}, \citenamefont {Kipp}, \citenamefont {Rossnagel}, \citenamefont
  {Bauer}, \citenamefont {Murnane}, \citenamefont {Kapteyn}, \citenamefont
  {Mathias},\ and\ \citenamefont {Aeschlimann}}]{Eich2014}%
  \BibitemOpen
  \bibfield  {author} {\bibinfo {author} {\bibfnamefont {S.}~\bibnamefont
  {Eich}}, \bibinfo {author} {\bibfnamefont {A.}~\bibnamefont {Stange}},
  \bibinfo {author} {\bibfnamefont {A.}~\bibnamefont {Carr}}, \bibinfo {author}
  {\bibfnamefont {J.}~\bibnamefont {Urbancic}}, \bibinfo {author}
  {\bibfnamefont {T.}~\bibnamefont {Popmintchev}}, \bibinfo {author}
  {\bibfnamefont {M.}~\bibnamefont {Wiesenmayer}}, \bibinfo {author}
  {\bibfnamefont {K.}~\bibnamefont {Jansen}}, \bibinfo {author} {\bibfnamefont
  {A.}~\bibnamefont {Ruffing}}, \bibinfo {author} {\bibfnamefont
  {S.}~\bibnamefont {Jakobs}}, \bibinfo {author} {\bibfnamefont
  {T.}~\bibnamefont {Rohwer}}, \bibinfo {author} {\bibfnamefont
  {S.}~\bibnamefont {Hellmann}}, \bibinfo {author} {\bibfnamefont
  {C.}~\bibnamefont {Chen}}, \bibinfo {author} {\bibfnamefont {P.}~\bibnamefont
  {Matyba}}, \bibinfo {author} {\bibfnamefont {L.}~\bibnamefont {Kipp}},
  \bibinfo {author} {\bibfnamefont {K.}~\bibnamefont {Rossnagel}}, \bibinfo
  {author} {\bibfnamefont {M.}~\bibnamefont {Bauer}}, \bibinfo {author}
  {\bibfnamefont {M.}~\bibnamefont {Murnane}}, \bibinfo {author} {\bibfnamefont
  {H.}~\bibnamefont {Kapteyn}}, \bibinfo {author} {\bibfnamefont
  {S.}~\bibnamefont {Mathias}}, \ and\ \bibinfo {author} {\bibfnamefont
  {M.}~\bibnamefont {Aeschlimann}},\ }\bibfield  {title} {\enquote {\bibinfo
  {title} {Time- and angle-resolved photoemission spectroscopy with optimized
  high-harmonic pulses using frequency-doubled ti:sapphire lasers},}\ }\href
  {\doibase https://doi.org/10.1016/j.elspec.2014.04.013} {\bibfield  {journal}
  {\bibinfo  {journal} {Journal of Electron Spectroscopy and Related
  Phenomena}\ }\textbf {\bibinfo {volume} {195}},\ \bibinfo {pages} {231--236}
  (\bibinfo {year} {2014})}\BibitemShut {NoStop}%
\bibitem [{\citenamefont {Cilento}\ \emph {et~al.}(2016)\citenamefont
  {Cilento}, \citenamefont {Crepaldi}, \citenamefont {Manzoni}, \citenamefont
  {Sterzi}, \citenamefont {Zacchigna}, \citenamefont {Bugnon}, \citenamefont
  {Berger},\ and\ \citenamefont {Parmigiani}}]{Cilento2016}%
  \BibitemOpen
  \bibfield  {author} {\bibinfo {author} {\bibfnamefont {F.}~\bibnamefont
  {Cilento}}, \bibinfo {author} {\bibfnamefont {A.}~\bibnamefont {Crepaldi}},
  \bibinfo {author} {\bibfnamefont {G.}~\bibnamefont {Manzoni}}, \bibinfo
  {author} {\bibfnamefont {A.}~\bibnamefont {Sterzi}}, \bibinfo {author}
  {\bibfnamefont {M.}~\bibnamefont {Zacchigna}}, \bibinfo {author}
  {\bibfnamefont {P.}~\bibnamefont {Bugnon}}, \bibinfo {author} {\bibfnamefont
  {H.}~\bibnamefont {Berger}}, \ and\ \bibinfo {author} {\bibfnamefont
  {F.}~\bibnamefont {Parmigiani}},\ }\bibfield  {title} {\enquote {\bibinfo
  {title} {Advancing non-equilibrium arpes experiments by a 9.3ev coherent
  ultrafast photon source},}\ }\href {\doibase
  https://doi.org/10.1016/j.elspec.2015.12.002} {\bibfield  {journal} {\bibinfo
   {journal} {Journal of Electron Spectroscopy and Related Phenomena}\ }\textbf
  {\bibinfo {volume} {207}},\ \bibinfo {pages} {7--13} (\bibinfo {year}
  {2016})}\BibitemShut {NoStop}%
\bibitem [{\citenamefont {Rohde}\ \emph {et~al.}(2016)\citenamefont {Rohde},
  \citenamefont {Hendel}, \citenamefont {Stange}, \citenamefont {Hanff},
  \citenamefont {Oloff}, \citenamefont {Yang}, \citenamefont {Rossnagel},\ and\
  \citenamefont {Bauer}}]{Rohde2016}%
  \BibitemOpen
  \bibfield  {author} {\bibinfo {author} {\bibfnamefont {G.}~\bibnamefont
  {Rohde}}, \bibinfo {author} {\bibfnamefont {A.}~\bibnamefont {Hendel}},
  \bibinfo {author} {\bibfnamefont {A.}~\bibnamefont {Stange}}, \bibinfo
  {author} {\bibfnamefont {K.}~\bibnamefont {Hanff}}, \bibinfo {author}
  {\bibfnamefont {L.-P.}\ \bibnamefont {Oloff}}, \bibinfo {author}
  {\bibfnamefont {L.~X.}\ \bibnamefont {Yang}}, \bibinfo {author}
  {\bibfnamefont {K.}~\bibnamefont {Rossnagel}}, \ and\ \bibinfo {author}
  {\bibfnamefont {M.}~\bibnamefont {Bauer}},\ }\bibfield  {title} {\enquote
  {\bibinfo {title} {Time-resolved arpes with sub-15 fs temporal and near
  fourier-limited spectral resolution},}\ }\href {\doibase 10.1063/1.4963668}
  {\bibfield  {journal} {\bibinfo  {journal} {Review of Scientific
  Instruments}\ }\textbf {\bibinfo {volume} {87}},\ \bibinfo {pages} {103102}
  (\bibinfo {year} {2016})},\ \Eprint
  {http://arxiv.org/abs/https://doi.org/10.1063/1.4963668}
  {https://doi.org/10.1063/1.4963668} \BibitemShut {NoStop}%
\bibitem [{\citenamefont {Corder}\ \emph {et~al.}(2018)\citenamefont {Corder},
  \citenamefont {Zhao}, \citenamefont {Bakalis}, \citenamefont {Li},
  \citenamefont {Kershis}, \citenamefont {Muraca}, \citenamefont {White},\ and\
  \citenamefont {Allison}}]{Corder2018}%
  \BibitemOpen
  \bibfield  {author} {\bibinfo {author} {\bibfnamefont {C.}~\bibnamefont
  {Corder}}, \bibinfo {author} {\bibfnamefont {P.}~\bibnamefont {Zhao}},
  \bibinfo {author} {\bibfnamefont {J.}~\bibnamefont {Bakalis}}, \bibinfo
  {author} {\bibfnamefont {X.}~\bibnamefont {Li}}, \bibinfo {author}
  {\bibfnamefont {M.~D.}\ \bibnamefont {Kershis}}, \bibinfo {author}
  {\bibfnamefont {A.~R.}\ \bibnamefont {Muraca}}, \bibinfo {author}
  {\bibfnamefont {M.~G.}\ \bibnamefont {White}}, \ and\ \bibinfo {author}
  {\bibfnamefont {T.~K.}\ \bibnamefont {Allison}},\ }\bibfield  {title}
  {\enquote {\bibinfo {title} {Ultrafast extreme ultraviolet photoemission
  without space charge},}\ }\href {\doibase 10.1063/1.5045578} {\bibfield
  {journal} {\bibinfo  {journal} {Structural Dynamics}\ }\textbf {\bibinfo
  {volume} {5}},\ \bibinfo {pages} {054301} (\bibinfo {year} {2018})},\ \Eprint
  {http://arxiv.org/abs/https://doi.org/10.1063/1.5045578}
  {https://doi.org/10.1063/1.5045578} \BibitemShut {NoStop}%
\bibitem [{\citenamefont {Puppin}\ \emph {et~al.}(2019)\citenamefont {Puppin},
  \citenamefont {Deng}, \citenamefont {Nicholson}, \citenamefont {Feldl},
  \citenamefont {Schr\"oter}, \citenamefont {Vita}, \citenamefont {Kirchmann},
  \citenamefont {Monney}, \citenamefont {Rettig}, \citenamefont {Wolf},\ and\
  \citenamefont {Ernstorfer}}]{Puppin2019}%
  \BibitemOpen
  \bibfield  {author} {\bibinfo {author} {\bibfnamefont {M.}~\bibnamefont
  {Puppin}}, \bibinfo {author} {\bibfnamefont {Y.}~\bibnamefont {Deng}},
  \bibinfo {author} {\bibfnamefont {C.~W.}\ \bibnamefont {Nicholson}}, \bibinfo
  {author} {\bibfnamefont {J.}~\bibnamefont {Feldl}}, \bibinfo {author}
  {\bibfnamefont {N.~B.~M.}\ \bibnamefont {Schr\"oter}}, \bibinfo {author}
  {\bibfnamefont {H.}~\bibnamefont {Vita}}, \bibinfo {author} {\bibfnamefont
  {P.~S.}\ \bibnamefont {Kirchmann}}, \bibinfo {author} {\bibfnamefont
  {C.}~\bibnamefont {Monney}}, \bibinfo {author} {\bibfnamefont
  {L.}~\bibnamefont {Rettig}}, \bibinfo {author} {\bibfnamefont
  {M.}~\bibnamefont {Wolf}}, \ and\ \bibinfo {author} {\bibfnamefont
  {R.}~\bibnamefont {Ernstorfer}},\ }\bibfield  {title} {\enquote {\bibinfo
  {title} {Time- and angle-resolved photoemission spectroscopy of solids in the
  extreme ultraviolet at 500 khz repetition rate},}\ }\href {\doibase
  10.1063/1.5081938} {\bibfield  {journal} {\bibinfo  {journal} {Review of
  Scientific Instruments}\ }\textbf {\bibinfo {volume} {90}},\ \bibinfo {pages}
  {023104} (\bibinfo {year} {2019})},\ \Eprint
  {http://arxiv.org/abs/https://doi.org/10.1063/1.5081938}
  {https://doi.org/10.1063/1.5081938} \BibitemShut {NoStop}%
\bibitem [{\citenamefont {Buss}\ \emph {et~al.}(2019)\citenamefont {Buss},
  \citenamefont {Wang}, \citenamefont {Xu}, \citenamefont {Maklar},
  \citenamefont {Joucken}, \citenamefont {Zeng}, \citenamefont {Stoll},
  \citenamefont {Jozwiak}, \citenamefont {Pepper}, \citenamefont {Chuang},
  \citenamefont {Denlinger}, \citenamefont {Hussain}, \citenamefont {Lanzara},\
  and\ \citenamefont {Kaindl}}]{Buss2019}%
  \BibitemOpen
  \bibfield  {author} {\bibinfo {author} {\bibfnamefont {J.~H.}\ \bibnamefont
  {Buss}}, \bibinfo {author} {\bibfnamefont {H.}~\bibnamefont {Wang}}, \bibinfo
  {author} {\bibfnamefont {Y.}~\bibnamefont {Xu}}, \bibinfo {author}
  {\bibfnamefont {J.}~\bibnamefont {Maklar}}, \bibinfo {author} {\bibfnamefont
  {F.}~\bibnamefont {Joucken}}, \bibinfo {author} {\bibfnamefont
  {L.}~\bibnamefont {Zeng}}, \bibinfo {author} {\bibfnamefont {S.}~\bibnamefont
  {Stoll}}, \bibinfo {author} {\bibfnamefont {C.}~\bibnamefont {Jozwiak}},
  \bibinfo {author} {\bibfnamefont {J.}~\bibnamefont {Pepper}}, \bibinfo
  {author} {\bibfnamefont {Y.-D.}\ \bibnamefont {Chuang}}, \bibinfo {author}
  {\bibfnamefont {J.~D.}\ \bibnamefont {Denlinger}}, \bibinfo {author}
  {\bibfnamefont {Z.}~\bibnamefont {Hussain}}, \bibinfo {author} {\bibfnamefont
  {A.}~\bibnamefont {Lanzara}}, \ and\ \bibinfo {author} {\bibfnamefont
  {R.~A.}\ \bibnamefont {Kaindl}},\ }\bibfield  {title} {\enquote {\bibinfo
  {title} {A setup for extreme-ultraviolet ultrafast angle-resolved
  photoelectron spectroscopy at 50-khz repetition rate},}\ }\href {\doibase
  10.1063/1.5079677} {\bibfield  {journal} {\bibinfo  {journal} {Review of
  Scientific Instruments}\ }\textbf {\bibinfo {volume} {90}},\ \bibinfo {pages}
  {023105} (\bibinfo {year} {2019})},\ \Eprint
  {http://arxiv.org/abs/https://doi.org/10.1063/1.5079677}
  {https://doi.org/10.1063/1.5079677} \BibitemShut {NoStop}%
\bibitem [{\citenamefont {Sie}\ \emph {et~al.}(2019)\citenamefont {Sie},
  \citenamefont {Rohwer}, \citenamefont {Lee},\ and\ \citenamefont
  {Gedik}}]{Sie2019}%
  \BibitemOpen
  \bibfield  {author} {\bibinfo {author} {\bibfnamefont {E.~J.}\ \bibnamefont
  {Sie}}, \bibinfo {author} {\bibfnamefont {T.}~\bibnamefont {Rohwer}},
  \bibinfo {author} {\bibfnamefont {C.}~\bibnamefont {Lee}}, \ and\ \bibinfo
  {author} {\bibfnamefont {N.}~\bibnamefont {Gedik}},\ }\bibfield  {title}
  {\enquote {\bibinfo {title} {Time-resolved {XUV} arpes with tunable 24-33 ev
  laser pulses at 30 mev resolution},}\ }\href
  {https://doi.org/10.1038/s41467-019-11492-3} {\bibfield  {journal} {\bibinfo
  {journal} {Nature Communications}\ }\textbf {\bibinfo {volume} {10}},\
  \bibinfo {pages} {3535--} (\bibinfo {year} {2019})}\BibitemShut {NoStop}%
\bibitem [{\citenamefont {Mills}\ \emph {et~al.}(2019)\citenamefont {Mills},
  \citenamefont {Zhdanovich}, \citenamefont {Na}, \citenamefont {Boschini},
  \citenamefont {Razzoli}, \citenamefont {Michiardi}, \citenamefont
  {Sheyerman}, \citenamefont {Schneider}, \citenamefont {Hammond},
  \citenamefont {Süss}, \citenamefont {Felser}, \citenamefont {Damascelli},\
  and\ \citenamefont {Jones}}]{Mills2019}%
  \BibitemOpen
  \bibfield  {author} {\bibinfo {author} {\bibfnamefont {A.~K.}\ \bibnamefont
  {Mills}}, \bibinfo {author} {\bibfnamefont {S.}~\bibnamefont {Zhdanovich}},
  \bibinfo {author} {\bibfnamefont {M.~X.}\ \bibnamefont {Na}}, \bibinfo
  {author} {\bibfnamefont {F.}~\bibnamefont {Boschini}}, \bibinfo {author}
  {\bibfnamefont {E.}~\bibnamefont {Razzoli}}, \bibinfo {author} {\bibfnamefont
  {M.}~\bibnamefont {Michiardi}}, \bibinfo {author} {\bibfnamefont
  {A.}~\bibnamefont {Sheyerman}}, \bibinfo {author} {\bibfnamefont
  {M.}~\bibnamefont {Schneider}}, \bibinfo {author} {\bibfnamefont {T.~J.}\
  \bibnamefont {Hammond}}, \bibinfo {author} {\bibfnamefont {V.}~\bibnamefont
  {Süss}}, \bibinfo {author} {\bibfnamefont {C.}~\bibnamefont {Felser}},
  \bibinfo {author} {\bibfnamefont {A.}~\bibnamefont {Damascelli}}, \ and\
  \bibinfo {author} {\bibfnamefont {D.~J.}\ \bibnamefont {Jones}},\ }\bibfield
  {title} {\enquote {\bibinfo {title} {Cavity-enhanced high harmonic generation
  for extreme ultraviolet time- and angle-resolved photoemission
  spectroscopy},}\ }\href {\doibase 10.1063/1.5090507} {\bibfield  {journal}
  {\bibinfo  {journal} {Review of Scientific Instruments}\ }\textbf {\bibinfo
  {volume} {90}},\ \bibinfo {pages} {083001} (\bibinfo {year} {2019})},\
  \Eprint {http://arxiv.org/abs/https://doi.org/10.1063/1.5090507}
  {https://doi.org/10.1063/1.5090507} \BibitemShut {NoStop}%
\bibitem [{\citenamefont {Liu}\ \emph {et~al.}(2020)\citenamefont {Liu},
  \citenamefont {Beetar}, \citenamefont {Hosen}, \citenamefont {Dhakal},
  \citenamefont {Sims}, \citenamefont {Kabir}, \citenamefont {Etienne},
  \citenamefont {Dimitri}, \citenamefont {Regmi}, \citenamefont {Liu},
  \citenamefont {Pathak}, \citenamefont {Kaczorowski}, \citenamefont
  {Neupane},\ and\ \citenamefont {Chini}}]{Liu2020}%
  \BibitemOpen
  \bibfield  {author} {\bibinfo {author} {\bibfnamefont {Y.}~\bibnamefont
  {Liu}}, \bibinfo {author} {\bibfnamefont {J.~E.}\ \bibnamefont {Beetar}},
  \bibinfo {author} {\bibfnamefont {M.~M.}\ \bibnamefont {Hosen}}, \bibinfo
  {author} {\bibfnamefont {G.}~\bibnamefont {Dhakal}}, \bibinfo {author}
  {\bibfnamefont {C.}~\bibnamefont {Sims}}, \bibinfo {author} {\bibfnamefont
  {F.}~\bibnamefont {Kabir}}, \bibinfo {author} {\bibfnamefont {M.~B.}\
  \bibnamefont {Etienne}}, \bibinfo {author} {\bibfnamefont {K.}~\bibnamefont
  {Dimitri}}, \bibinfo {author} {\bibfnamefont {S.}~\bibnamefont {Regmi}},
  \bibinfo {author} {\bibfnamefont {Y.}~\bibnamefont {Liu}}, \bibinfo {author}
  {\bibfnamefont {A.~K.}\ \bibnamefont {Pathak}}, \bibinfo {author}
  {\bibfnamefont {D.}~\bibnamefont {Kaczorowski}}, \bibinfo {author}
  {\bibfnamefont {M.}~\bibnamefont {Neupane}}, \ and\ \bibinfo {author}
  {\bibfnamefont {M.}~\bibnamefont {Chini}},\ }\bibfield  {title} {\enquote
  {\bibinfo {title} {Extreme ultraviolet time- and angle-resolved photoemission
  setup with 21.5 mev resolution using high-order harmonic generation from a
  turn-key yb:kgw amplifier},}\ }\href {\doibase 10.1063/1.5121425} {\bibfield
  {journal} {\bibinfo  {journal} {Review of Scientific Instruments}\ }\textbf
  {\bibinfo {volume} {91}},\ \bibinfo {pages} {013102} (\bibinfo {year}
  {2020})},\ \Eprint {http://arxiv.org/abs/https://doi.org/10.1063/1.5121425}
  {https://doi.org/10.1063/1.5121425} \BibitemShut {NoStop}%
\bibitem [{\citenamefont {Cucini}\ \emph {et~al.}(2020)\citenamefont {Cucini},
  \citenamefont {Pincelli}, \citenamefont {Panaccione}, \citenamefont {Kopic},
  \citenamefont {Frassetto}, \citenamefont {Miotti}, \citenamefont
  {Pierantozzi}, \citenamefont {Peli}, \citenamefont {Fondacaro}, \citenamefont
  {De~Luisa}, \citenamefont {De~Vita}, \citenamefont {Carrara}, \citenamefont
  {Krizmancic}, \citenamefont {Payne}, \citenamefont {Salvador}, \citenamefont
  {Sterzi}, \citenamefont {Poletto}, \citenamefont {Parmigiani}, \citenamefont
  {Rossi},\ and\ \citenamefont {Cilento}}]{Cucini2020}%
  \BibitemOpen
  \bibfield  {author} {\bibinfo {author} {\bibfnamefont {R.}~\bibnamefont
  {Cucini}}, \bibinfo {author} {\bibfnamefont {T.}~\bibnamefont {Pincelli}},
  \bibinfo {author} {\bibfnamefont {G.}~\bibnamefont {Panaccione}}, \bibinfo
  {author} {\bibfnamefont {D.}~\bibnamefont {Kopic}}, \bibinfo {author}
  {\bibfnamefont {F.}~\bibnamefont {Frassetto}}, \bibinfo {author}
  {\bibfnamefont {P.}~\bibnamefont {Miotti}}, \bibinfo {author} {\bibfnamefont
  {G.~M.}\ \bibnamefont {Pierantozzi}}, \bibinfo {author} {\bibfnamefont
  {S.}~\bibnamefont {Peli}}, \bibinfo {author} {\bibfnamefont {A.}~\bibnamefont
  {Fondacaro}}, \bibinfo {author} {\bibfnamefont {A.}~\bibnamefont {De~Luisa}},
  \bibinfo {author} {\bibfnamefont {A.}~\bibnamefont {De~Vita}}, \bibinfo
  {author} {\bibfnamefont {P.}~\bibnamefont {Carrara}}, \bibinfo {author}
  {\bibfnamefont {D.}~\bibnamefont {Krizmancic}}, \bibinfo {author}
  {\bibfnamefont {D.~T.}\ \bibnamefont {Payne}}, \bibinfo {author}
  {\bibfnamefont {F.}~\bibnamefont {Salvador}}, \bibinfo {author}
  {\bibfnamefont {A.}~\bibnamefont {Sterzi}}, \bibinfo {author} {\bibfnamefont
  {L.}~\bibnamefont {Poletto}}, \bibinfo {author} {\bibfnamefont
  {F.}~\bibnamefont {Parmigiani}}, \bibinfo {author} {\bibfnamefont
  {G.}~\bibnamefont {Rossi}}, \ and\ \bibinfo {author} {\bibfnamefont
  {F.}~\bibnamefont {Cilento}},\ }\bibfield  {title} {\enquote {\bibinfo
  {title} {Coherent narrowband light source for ultrafast photoelectron
  spectroscopy in the 17–31 ev photon energy range},}\ }\href {\doibase
  10.1063/1.5131216} {\bibfield  {journal} {\bibinfo  {journal} {Structural
  Dynamics}\ }\textbf {\bibinfo {volume} {7}},\ \bibinfo {pages} {014303}
  (\bibinfo {year} {2020})},\ \Eprint
  {http://arxiv.org/abs/https://doi.org/10.1063/1.5131216}
  {https://doi.org/10.1063/1.5131216} \BibitemShut {NoStop}%
\bibitem [{\citenamefont {Lee}\ \emph {et~al.}(2020)\citenamefont {Lee},
  \citenamefont {Rohwer}, \citenamefont {Sie}, \citenamefont {Zong},
  \citenamefont {Baldini}, \citenamefont {Straquadine}, \citenamefont
  {Walmsley}, \citenamefont {Gardner}, \citenamefont {Lee}, \citenamefont
  {Fisher},\ and\ \citenamefont {Gedik}}]{Lee2020}%
  \BibitemOpen
  \bibfield  {author} {\bibinfo {author} {\bibfnamefont {C.}~\bibnamefont
  {Lee}}, \bibinfo {author} {\bibfnamefont {T.}~\bibnamefont {Rohwer}},
  \bibinfo {author} {\bibfnamefont {E.~J.}\ \bibnamefont {Sie}}, \bibinfo
  {author} {\bibfnamefont {A.}~\bibnamefont {Zong}}, \bibinfo {author}
  {\bibfnamefont {E.}~\bibnamefont {Baldini}}, \bibinfo {author} {\bibfnamefont
  {J.}~\bibnamefont {Straquadine}}, \bibinfo {author} {\bibfnamefont
  {P.}~\bibnamefont {Walmsley}}, \bibinfo {author} {\bibfnamefont
  {D.}~\bibnamefont {Gardner}}, \bibinfo {author} {\bibfnamefont {Y.~S.}\
  \bibnamefont {Lee}}, \bibinfo {author} {\bibfnamefont {I.~R.}\ \bibnamefont
  {Fisher}}, \ and\ \bibinfo {author} {\bibfnamefont {N.}~\bibnamefont
  {Gedik}},\ }\bibfield  {title} {\enquote {\bibinfo {title} {High resolution
  time- and angle-resolved photoemission spectroscopy with 11 ev laser
  pulses},}\ }\href {\doibase 10.1063/1.5139556} {\bibfield  {journal}
  {\bibinfo  {journal} {Review of Scientific Instruments}\ }\textbf {\bibinfo
  {volume} {91}},\ \bibinfo {pages} {043102} (\bibinfo {year} {2020})},\
  \Eprint {http://arxiv.org/abs/https://doi.org/10.1063/1.5139556}
  {https://doi.org/10.1063/1.5139556} \BibitemShut {NoStop}%
\bibitem [{\citenamefont {Peli}\ \emph {et~al.}(2020)\citenamefont {Peli},
  \citenamefont {Puntel}, \citenamefont {Kopic}, \citenamefont {Sockol},
  \citenamefont {Parmigiani},\ and\ \citenamefont {Cilento}}]{Peli2020}%
  \BibitemOpen
  \bibfield  {author} {\bibinfo {author} {\bibfnamefont {S.}~\bibnamefont
  {Peli}}, \bibinfo {author} {\bibfnamefont {D.}~\bibnamefont {Puntel}},
  \bibinfo {author} {\bibfnamefont {D.}~\bibnamefont {Kopic}}, \bibinfo
  {author} {\bibfnamefont {B.}~\bibnamefont {Sockol}}, \bibinfo {author}
  {\bibfnamefont {F.}~\bibnamefont {Parmigiani}}, \ and\ \bibinfo {author}
  {\bibfnamefont {F.}~\bibnamefont {Cilento}},\ }\bibfield  {title} {\enquote
  {\bibinfo {title} {Time-resolved vuv arpes at 10.8 ev photon energy and mhz
  repetition rate},}\ }\href {\doibase
  https://doi.org/10.1016/j.elspec.2020.146978} {\bibfield  {journal} {\bibinfo
   {journal} {Journal of Electron Spectroscopy and Related Phenomena}\ }\textbf
  {\bibinfo {volume} {243}},\ \bibinfo {pages} {146978} (\bibinfo {year}
  {2020})}\BibitemShut {NoStop}%
\bibitem [{\citenamefont {Guo}\ \emph {et~al.}(2022)\citenamefont {Guo},
  \citenamefont {Dendzik}, \citenamefont {Grubišić-Čabo}, \citenamefont
  {Berntsen}, \citenamefont {Li}, \citenamefont {Chen}, \citenamefont {Matta},
  \citenamefont {Starke}, \citenamefont {Hessmo}, \citenamefont
  {Weissenrieder},\ and\ \citenamefont {Tjernberg}}]{Guo2022}%
  \BibitemOpen
  \bibfield  {author} {\bibinfo {author} {\bibfnamefont {Q.}~\bibnamefont
  {Guo}}, \bibinfo {author} {\bibfnamefont {M.}~\bibnamefont {Dendzik}},
  \bibinfo {author} {\bibfnamefont {A.}~\bibnamefont {Grubišić-Čabo}},
  \bibinfo {author} {\bibfnamefont {M.~H.}\ \bibnamefont {Berntsen}}, \bibinfo
  {author} {\bibfnamefont {C.}~\bibnamefont {Li}}, \bibinfo {author}
  {\bibfnamefont {W.}~\bibnamefont {Chen}}, \bibinfo {author} {\bibfnamefont
  {B.}~\bibnamefont {Matta}}, \bibinfo {author} {\bibfnamefont
  {U.}~\bibnamefont {Starke}}, \bibinfo {author} {\bibfnamefont
  {B.}~\bibnamefont {Hessmo}}, \bibinfo {author} {\bibfnamefont
  {J.}~\bibnamefont {Weissenrieder}}, \ and\ \bibinfo {author} {\bibfnamefont
  {O.}~\bibnamefont {Tjernberg}},\ }\href@noop {} {\enquote {\bibinfo {title}
  {A narrow bandwidth extreme ultra-violet light source for time- and
  angle-resolved photoemission spectroscopy},}\ } (\bibinfo {year} {2022}),\
  \Eprint {http://arxiv.org/abs/2201.12456} {arXiv:2201.12456
  [cond-mat.str-el]} \BibitemShut {NoStop}%
\bibitem [{\citenamefont {Pl\"otzing}\ \emph {et~al.}(2016)\citenamefont
  {Pl\"otzing}, \citenamefont {Adam}, \citenamefont {Weier}, \citenamefont
  {Plucinski}, \citenamefont {Eich}, \citenamefont {Emmerich}, \citenamefont
  {Rollinger}, \citenamefont {Aeschlimann}, \citenamefont {Mathias},\ and\
  \citenamefont {Schneider}}]{Ploetzing2016}%
  \BibitemOpen
  \bibfield  {author} {\bibinfo {author} {\bibfnamefont {M.}~\bibnamefont
  {Pl\"otzing}}, \bibinfo {author} {\bibfnamefont {R.}~\bibnamefont {Adam}},
  \bibinfo {author} {\bibfnamefont {C.}~\bibnamefont {Weier}}, \bibinfo
  {author} {\bibfnamefont {L.}~\bibnamefont {Plucinski}}, \bibinfo {author}
  {\bibfnamefont {S.}~\bibnamefont {Eich}}, \bibinfo {author} {\bibfnamefont
  {S.}~\bibnamefont {Emmerich}}, \bibinfo {author} {\bibfnamefont
  {M.}~\bibnamefont {Rollinger}}, \bibinfo {author} {\bibfnamefont
  {M.}~\bibnamefont {Aeschlimann}}, \bibinfo {author} {\bibfnamefont
  {S.}~\bibnamefont {Mathias}}, \ and\ \bibinfo {author} {\bibfnamefont
  {C.~M.}\ \bibnamefont {Schneider}},\ }\bibfield  {title} {\enquote {\bibinfo
  {title} {Spin-resolved photoelectron spectroscopy using femtosecond extreme
  ultraviolet light pulses from high-order harmonic generation},}\ }\href
  {\doibase 10.1063/1.4946782} {\bibfield  {journal} {\bibinfo  {journal}
  {Review of Scientific Instruments}\ }\textbf {\bibinfo {volume} {87}},\
  \bibinfo {pages} {043903} (\bibinfo {year} {2016})},\ \Eprint
  {http://arxiv.org/abs/https://doi.org/10.1063/1.4946782}
  {https://doi.org/10.1063/1.4946782} \BibitemShut {NoStop}%
\bibitem [{\citenamefont {Eich}\ \emph {et~al.}(2017)\citenamefont {Eich},
  \citenamefont {Pl\"otzing}, \citenamefont {Rollinger}, \citenamefont
  {Emmerich}, \citenamefont {Adam}, \citenamefont {Chen}, \citenamefont
  {Kapteyn}, \citenamefont {Murnane}, \citenamefont {Plucinski}, \citenamefont
  {Steil}, \citenamefont {Stadtm\"uller}, \citenamefont {Cinchetti},
  \citenamefont {Aeschlimann}, \citenamefont {Schneider},\ and\ \citenamefont
  {Mathias}}]{Eich2017}%
  \BibitemOpen
  \bibfield  {author} {\bibinfo {author} {\bibfnamefont {S.}~\bibnamefont
  {Eich}}, \bibinfo {author} {\bibfnamefont {M.}~\bibnamefont {Pl\"otzing}},
  \bibinfo {author} {\bibfnamefont {M.}~\bibnamefont {Rollinger}}, \bibinfo
  {author} {\bibfnamefont {S.}~\bibnamefont {Emmerich}}, \bibinfo {author}
  {\bibfnamefont {R.}~\bibnamefont {Adam}}, \bibinfo {author} {\bibfnamefont
  {C.}~\bibnamefont {Chen}}, \bibinfo {author} {\bibfnamefont {H.~C.}\
  \bibnamefont {Kapteyn}}, \bibinfo {author} {\bibfnamefont {M.~M.}\
  \bibnamefont {Murnane}}, \bibinfo {author} {\bibfnamefont {L.}~\bibnamefont
  {Plucinski}}, \bibinfo {author} {\bibfnamefont {D.}~\bibnamefont {Steil}},
  \bibinfo {author} {\bibfnamefont {B.}~\bibnamefont {Stadtm\"uller}}, \bibinfo
  {author} {\bibfnamefont {M.}~\bibnamefont {Cinchetti}}, \bibinfo {author}
  {\bibfnamefont {M.}~\bibnamefont {Aeschlimann}}, \bibinfo {author}
  {\bibfnamefont {C.~M.}\ \bibnamefont {Schneider}}, \ and\ \bibinfo {author}
  {\bibfnamefont {S.}~\bibnamefont {Mathias}},\ }\bibfield  {title} {\enquote
  {\bibinfo {title} {Band structure evolution during the ultrafast
  ferromagnetic-paramagnetic phase transition in cobalt},}\ }\href {\doibase
  10.1126/sciadv.1602094} {\bibfield  {journal} {\bibinfo  {journal} {Science
  Advances}\ }\textbf {\bibinfo {volume} {3}},\ \bibinfo {pages} {e1602094}
  (\bibinfo {year} {2017})},\ \Eprint
  {http://arxiv.org/abs/https://www.science.org/doi/pdf/10.1126/sciadv.1602094}
  {https://www.science.org/doi/pdf/10.1126/sciadv.1602094} \BibitemShut
  {NoStop}%
\bibitem [{\citenamefont {Gort}\ \emph {et~al.}(2018)\citenamefont {Gort},
  \citenamefont {B\"uhlmann}, \citenamefont {D\"aster}, \citenamefont
  {Salvatella}, \citenamefont {Hartmann}, \citenamefont {Zemp}, \citenamefont
  {Holenstein}, \citenamefont {Stieger}, \citenamefont {Fognini}, \citenamefont
  {Michlmayr}, \citenamefont {B\"ahler}, \citenamefont {Vaterlaus},\ and\
  \citenamefont {Acremann}}]{Gort2018}%
  \BibitemOpen
  \bibfield  {author} {\bibinfo {author} {\bibfnamefont {R.}~\bibnamefont
  {Gort}}, \bibinfo {author} {\bibfnamefont {K.}~\bibnamefont {B\"uhlmann}},
  \bibinfo {author} {\bibfnamefont {S.}~\bibnamefont {D\"aster}}, \bibinfo
  {author} {\bibfnamefont {G.}~\bibnamefont {Salvatella}}, \bibinfo {author}
  {\bibfnamefont {N.}~\bibnamefont {Hartmann}}, \bibinfo {author}
  {\bibfnamefont {Y.}~\bibnamefont {Zemp}}, \bibinfo {author} {\bibfnamefont
  {S.}~\bibnamefont {Holenstein}}, \bibinfo {author} {\bibfnamefont
  {C.}~\bibnamefont {Stieger}}, \bibinfo {author} {\bibfnamefont
  {A.}~\bibnamefont {Fognini}}, \bibinfo {author} {\bibfnamefont {T.~U.}\
  \bibnamefont {Michlmayr}}, \bibinfo {author} {\bibfnamefont {T.}~\bibnamefont
  {B\"ahler}}, \bibinfo {author} {\bibfnamefont {A.}~\bibnamefont {Vaterlaus}},
  \ and\ \bibinfo {author} {\bibfnamefont {Y.}~\bibnamefont {Acremann}},\
  }\bibfield  {title} {\enquote {\bibinfo {title} {Early stages of ultrafast
  spin dynamics in a $3d$ ferromagnet},}\ }\href {\doibase
  10.1103/PhysRevLett.121.087206} {\bibfield  {journal} {\bibinfo  {journal}
  {Phys. Rev. Lett.}\ }\textbf {\bibinfo {volume} {121}},\ \bibinfo {pages}
  {087206} (\bibinfo {year} {2018})}\BibitemShut {NoStop}%
\bibitem [{\citenamefont {B\"uhlmann}\ \emph {et~al.}(2020)\citenamefont
  {B\"uhlmann}, \citenamefont {Gort}, \citenamefont {Fognini}, \citenamefont
  {D\"aster}, \citenamefont {Holenstein}, \citenamefont {Hartmann},
  \citenamefont {Zemp}, \citenamefont {Salvatella}, \citenamefont {Michlmayr},
  \citenamefont {B\"ahler}, \citenamefont {Kutnyakhov}, \citenamefont
  {Medjanik}, \citenamefont {Sch\"onhense}, \citenamefont {Vaterlaus},\ and\
  \citenamefont {Acremann}}]{Buehlmann2020}%
  \BibitemOpen
  \bibfield  {author} {\bibinfo {author} {\bibfnamefont {K.}~\bibnamefont
  {B\"uhlmann}}, \bibinfo {author} {\bibfnamefont {R.}~\bibnamefont {Gort}},
  \bibinfo {author} {\bibfnamefont {A.}~\bibnamefont {Fognini}}, \bibinfo
  {author} {\bibfnamefont {S.}~\bibnamefont {D\"aster}}, \bibinfo {author}
  {\bibfnamefont {S.}~\bibnamefont {Holenstein}}, \bibinfo {author}
  {\bibfnamefont {N.}~\bibnamefont {Hartmann}}, \bibinfo {author}
  {\bibfnamefont {Y.}~\bibnamefont {Zemp}}, \bibinfo {author} {\bibfnamefont
  {G.}~\bibnamefont {Salvatella}}, \bibinfo {author} {\bibfnamefont {T.~U.}\
  \bibnamefont {Michlmayr}}, \bibinfo {author} {\bibfnamefont {T.}~\bibnamefont
  {B\"ahler}}, \bibinfo {author} {\bibfnamefont {D.}~\bibnamefont
  {Kutnyakhov}}, \bibinfo {author} {\bibfnamefont {K.}~\bibnamefont
  {Medjanik}}, \bibinfo {author} {\bibfnamefont {G.}~\bibnamefont
  {Sch\"onhense}}, \bibinfo {author} {\bibfnamefont {A.}~\bibnamefont
  {Vaterlaus}}, \ and\ \bibinfo {author} {\bibfnamefont {Y.}~\bibnamefont
  {Acremann}},\ }\bibfield  {title} {\enquote {\bibinfo {title} {Compact setup
  for spin-, time-, and angle-resolved photoemission spectroscopy},}\ }\href
  {\doibase 10.1063/5.0004861} {\bibfield  {journal} {\bibinfo  {journal}
  {Review of Scientific Instruments}\ }\textbf {\bibinfo {volume} {91}},\
  \bibinfo {pages} {063001} (\bibinfo {year} {2020})},\ \Eprint
  {http://arxiv.org/abs/https://doi.org/10.1063/5.0004861}
  {https://doi.org/10.1063/5.0004861} \BibitemShut {NoStop}%
\bibitem [{\citenamefont {Seo}\ \emph {et~al.}(2016)\citenamefont {Seo},
  \citenamefont {Arion}, \citenamefont {Roth}, \citenamefont {Ramm},
  \citenamefont {Lupulescu},\ and\ \citenamefont {Eberhardt}}]{Seo2016}%
  \BibitemOpen
  \bibfield  {author} {\bibinfo {author} {\bibfnamefont {H.~O.}\ \bibnamefont
  {Seo}}, \bibinfo {author} {\bibfnamefont {T.}~\bibnamefont {Arion}}, \bibinfo
  {author} {\bibfnamefont {F.}~\bibnamefont {Roth}}, \bibinfo {author}
  {\bibfnamefont {D.}~\bibnamefont {Ramm}}, \bibinfo {author} {\bibfnamefont
  {C.}~\bibnamefont {Lupulescu}}, \ and\ \bibinfo {author} {\bibfnamefont
  {W.}~\bibnamefont {Eberhardt}},\ }\bibfield  {title} {\enquote {\bibinfo
  {title} {Improving the efficiency of high harmonic generation (hhg) by
  ne-admixing into a pure ar gas medium},}\ }\href {\doibase
  10.1007/s00340-016-6347-6} {\bibfield  {journal} {\bibinfo  {journal}
  {Applied Physics B}\ }\textbf {\bibinfo {volume} {122}},\ \bibinfo {pages}
  {70} (\bibinfo {year} {2016})}\BibitemShut {NoStop}%
\bibitem [{\citenamefont {Oloff}\ \emph {et~al.}(2016)\citenamefont {Oloff},
  \citenamefont {Hanff}, \citenamefont {Stange}, \citenamefont {Rohde},
  \citenamefont {Diekmann}, \citenamefont {Bauer},\ and\ \citenamefont
  {Rossnagel}}]{Oloff2016}%
  \BibitemOpen
  \bibfield  {author} {\bibinfo {author} {\bibfnamefont {L.-P.}\ \bibnamefont
  {Oloff}}, \bibinfo {author} {\bibfnamefont {K.}~\bibnamefont {Hanff}},
  \bibinfo {author} {\bibfnamefont {A.}~\bibnamefont {Stange}}, \bibinfo
  {author} {\bibfnamefont {G.}~\bibnamefont {Rohde}}, \bibinfo {author}
  {\bibfnamefont {F.}~\bibnamefont {Diekmann}}, \bibinfo {author}
  {\bibfnamefont {M.}~\bibnamefont {Bauer}}, \ and\ \bibinfo {author}
  {\bibfnamefont {K.}~\bibnamefont {Rossnagel}},\ }\bibfield  {title} {\enquote
  {\bibinfo {title} {Pump laser-induced space-charge effects in hhg-driven
  time- and angle-resolved photoelectron spectroscopy},}\ }\href {\doibase
  10.1063/1.4953643} {\bibfield  {journal} {\bibinfo  {journal} {Journal of
  Applied Physics}\ }\textbf {\bibinfo {volume} {119}},\ \bibinfo {pages}
  {225106} (\bibinfo {year} {2016})},\ \Eprint
  {http://arxiv.org/abs/https://doi.org/10.1063/1.4953643}
  {https://doi.org/10.1063/1.4953643} \BibitemShut {NoStop}%
\bibitem [{\citenamefont {Sch\"onhense}\ \emph {et~al.}(2021)\citenamefont
  {Sch\"onhense}, \citenamefont {Kutnyakhov}, \citenamefont {Pressacco},
  \citenamefont {Heber}, \citenamefont {Wind}, \citenamefont {Agustsson},
  \citenamefont {Babenkov}, \citenamefont {Vasilyev}, \citenamefont
  {Fedchenko}, \citenamefont {Chernov}, \citenamefont {Rettig}, \citenamefont
  {Sch\"onhense}, \citenamefont {Wenthaus}, \citenamefont {Brenner},
  \citenamefont {Dziarzhytski}, \citenamefont {Palutke}, \citenamefont
  {Mahatha}, \citenamefont {Schirmel}, \citenamefont {Redlin}, \citenamefont
  {Manschwetus}, \citenamefont {Hartl}, \citenamefont {Matveyev}, \citenamefont
  {Gloskovskii}, \citenamefont {Schlueter}, \citenamefont {Shokeen},
  \citenamefont {Duerr}, \citenamefont {Allison}, \citenamefont {Beye},
  \citenamefont {Rossnagel}, \citenamefont {Elmers},\ and\ \citenamefont
  {Medjanik}}]{Schoenhense2021}%
  \BibitemOpen
  \bibfield  {author} {\bibinfo {author} {\bibfnamefont {G.}~\bibnamefont
  {Sch\"onhense}}, \bibinfo {author} {\bibfnamefont {D.}~\bibnamefont
  {Kutnyakhov}}, \bibinfo {author} {\bibfnamefont {F.}~\bibnamefont
  {Pressacco}}, \bibinfo {author} {\bibfnamefont {M.}~\bibnamefont {Heber}},
  \bibinfo {author} {\bibfnamefont {N.}~\bibnamefont {Wind}}, \bibinfo {author}
  {\bibfnamefont {S.~Y.}\ \bibnamefont {Agustsson}}, \bibinfo {author}
  {\bibfnamefont {S.}~\bibnamefont {Babenkov}}, \bibinfo {author}
  {\bibfnamefont {D.}~\bibnamefont {Vasilyev}}, \bibinfo {author}
  {\bibfnamefont {O.}~\bibnamefont {Fedchenko}}, \bibinfo {author}
  {\bibfnamefont {S.}~\bibnamefont {Chernov}}, \bibinfo {author} {\bibfnamefont
  {L.}~\bibnamefont {Rettig}}, \bibinfo {author} {\bibfnamefont
  {B.}~\bibnamefont {Sch\"onhense}}, \bibinfo {author} {\bibfnamefont
  {L.}~\bibnamefont {Wenthaus}}, \bibinfo {author} {\bibfnamefont
  {G.}~\bibnamefont {Brenner}}, \bibinfo {author} {\bibfnamefont
  {S.}~\bibnamefont {Dziarzhytski}}, \bibinfo {author} {\bibfnamefont
  {S.}~\bibnamefont {Palutke}}, \bibinfo {author} {\bibfnamefont {S.~K.}\
  \bibnamefont {Mahatha}}, \bibinfo {author} {\bibfnamefont {N.}~\bibnamefont
  {Schirmel}}, \bibinfo {author} {\bibfnamefont {H.}~\bibnamefont {Redlin}},
  \bibinfo {author} {\bibfnamefont {B.}~\bibnamefont {Manschwetus}}, \bibinfo
  {author} {\bibfnamefont {I.}~\bibnamefont {Hartl}}, \bibinfo {author}
  {\bibfnamefont {Y.}~\bibnamefont {Matveyev}}, \bibinfo {author}
  {\bibfnamefont {A.}~\bibnamefont {Gloskovskii}}, \bibinfo {author}
  {\bibfnamefont {C.}~\bibnamefont {Schlueter}}, \bibinfo {author}
  {\bibfnamefont {V.}~\bibnamefont {Shokeen}}, \bibinfo {author} {\bibfnamefont
  {H.}~\bibnamefont {Duerr}}, \bibinfo {author} {\bibfnamefont {T.~K.}\
  \bibnamefont {Allison}}, \bibinfo {author} {\bibfnamefont {M.}~\bibnamefont
  {Beye}}, \bibinfo {author} {\bibfnamefont {K.}~\bibnamefont {Rossnagel}},
  \bibinfo {author} {\bibfnamefont {H.~J.}\ \bibnamefont {Elmers}}, \ and\
  \bibinfo {author} {\bibfnamefont {K.}~\bibnamefont {Medjanik}},\ }\bibfield
  {title} {\enquote {\bibinfo {title} {Suppression of the vacuum space-charge
  effect in fs-photoemission by a retarding electrostatic front lens},}\ }\href
  {\doibase 10.1063/5.0046567} {\bibfield  {journal} {\bibinfo  {journal}
  {Review of Scientific Instruments}\ }\textbf {\bibinfo {volume} {92}},\
  \bibinfo {pages} {053703} (\bibinfo {year} {2021})},\ \Eprint
  {http://arxiv.org/abs/https://doi.org/10.1063/5.0046567}
  {https://doi.org/10.1063/5.0046567} \BibitemShut {NoStop}%
\bibitem [{\citenamefont {Wood}\ and\ \citenamefont {Nassau}(1982)}]{Wood82}%
  \BibitemOpen
  \bibfield  {author} {\bibinfo {author} {\bibfnamefont {D.~L.}\ \bibnamefont
  {Wood}}\ and\ \bibinfo {author} {\bibfnamefont {K.}~\bibnamefont {Nassau}},\
  }\bibfield  {title} {\enquote {\bibinfo {title} {Refractive index of cubic
  zirconia stabilized with yttria},}\ }\href {\doibase 10.1364/AO.21.002978}
  {\bibfield  {journal} {\bibinfo  {journal} {Appl. Opt.}\ }\textbf {\bibinfo
  {volume} {21}},\ \bibinfo {pages} {2978--2981} (\bibinfo {year}
  {1982})}\BibitemShut {NoStop}%
\bibitem [{\citenamefont {Polyanskiy}()}]{RefIndexZrO2}%
  \BibitemOpen
  \bibfield  {author} {\bibinfo {author} {\bibfnamefont {M.~N.}\ \bibnamefont
  {Polyanskiy}},\ }\href@noop {} {\enquote {\bibinfo {title} {Refractive index
  database},}\ }\bibinfo {howpublished}
  {\url{https://refractiveindex.info/?shelf=main&book=ZrO2&page=Wood}},\
  \bibinfo {note} {accessed on 2022-05-24}\BibitemShut {NoStop}%
\bibitem [{\citenamefont {Henke}, \citenamefont {Gullikson},\ and\
  \citenamefont {Davis}(1993)}]{Henke93}%
  \BibitemOpen
  \bibfield  {author} {\bibinfo {author} {\bibfnamefont {B.}~\bibnamefont
  {Henke}}, \bibinfo {author} {\bibfnamefont {E.}~\bibnamefont {Gullikson}}, \
  and\ \bibinfo {author} {\bibfnamefont {J.}~\bibnamefont {Davis}},\ }\bibfield
   {title} {\enquote {\bibinfo {title} {X-ray interactions: Photoabsorption,
  scattering, transmission, and reflection at e = 50-30,000 ev, z = 1-92},}\
  }\href {\doibase https://doi.org/10.1006/adnd.1993.1013} {\bibfield
  {journal} {\bibinfo  {journal} {Atomic Data and Nuclear Data Tables}\
  }\textbf {\bibinfo {volume} {54}},\ \bibinfo {pages} {181--342} (\bibinfo
  {year} {1993})}\BibitemShut {NoStop}%
\bibitem [{\citenamefont {Gerken}(2014)}]{PHDGerken2014}%
  \BibitemOpen
  \bibfield  {author} {\bibinfo {author} {\bibfnamefont {N.~C.}\ \bibnamefont
  {Gerken}},\ }\href@noop {} {\emph {\bibinfo {title} {XUV Multiphoton
  Excitation and Charged State Dynamics in Small Quantum Systems}}}\ (\bibinfo
  {publisher} {Verlag Dr. Hut},\ \bibinfo {year} {2014})\BibitemShut {NoStop}%
\bibitem [{\citenamefont {Gerken}\ \emph {et~al.}(2014)\citenamefont {Gerken},
  \citenamefont {Klumpp}, \citenamefont {Sorokin}, \citenamefont {Tiedtke},
  \citenamefont {Richter}, \citenamefont {B\"urk}, \citenamefont {Mertens},
  \citenamefont {Jurani\ifmmode~\acute{c}\else \'{c}\fi{}},\ and\ \citenamefont
  {Martins}}]{Gerken2014}%
  \BibitemOpen
  \bibfield  {author} {\bibinfo {author} {\bibfnamefont {N.}~\bibnamefont
  {Gerken}}, \bibinfo {author} {\bibfnamefont {S.}~\bibnamefont {Klumpp}},
  \bibinfo {author} {\bibfnamefont {A.~A.}\ \bibnamefont {Sorokin}}, \bibinfo
  {author} {\bibfnamefont {K.}~\bibnamefont {Tiedtke}}, \bibinfo {author}
  {\bibfnamefont {M.}~\bibnamefont {Richter}}, \bibinfo {author} {\bibfnamefont
  {V.}~\bibnamefont {B\"urk}}, \bibinfo {author} {\bibfnamefont
  {K.}~\bibnamefont {Mertens}}, \bibinfo {author} {\bibfnamefont
  {P.}~\bibnamefont {Jurani\ifmmode~\acute{c}\else \'{c}\fi{}}}, \ and\
  \bibinfo {author} {\bibfnamefont {M.}~\bibnamefont {Martins}},\ }\bibfield
  {title} {\enquote {\bibinfo {title} {Time-dependent multiphoton ionization of
  xenon in the soft-x-ray regime},}\ }\href {\doibase
  10.1103/PhysRevLett.112.213002} {\bibfield  {journal} {\bibinfo  {journal}
  {Phys. Rev. Lett.}\ }\textbf {\bibinfo {volume} {112}},\ \bibinfo {pages}
  {213002} (\bibinfo {year} {2014})}\BibitemShut {NoStop}%
\bibitem [{\citenamefont {Xian}\ \emph {et~al.}(2020)\citenamefont {Xian},
  \citenamefont {Acremann}, \citenamefont {Agustsson}, \citenamefont {Dendzik},
  \citenamefont {B{\"u}hlmann}, \citenamefont {Curcio}, \citenamefont
  {Kutnyakhov}, \citenamefont {Pressacco}, \citenamefont {Heber}, \citenamefont
  {Dong}, \citenamefont {Pincelli}, \citenamefont {Demsar}, \citenamefont
  {Wurth}, \citenamefont {Hofmann}, \citenamefont {Wolf}, \citenamefont
  {Scheidgen}, \citenamefont {Rettig},\ and\ \citenamefont
  {Ernstorfer}}]{Xian2020}%
  \BibitemOpen
  \bibfield  {author} {\bibinfo {author} {\bibfnamefont {R.~P.}\ \bibnamefont
  {Xian}}, \bibinfo {author} {\bibfnamefont {Y.}~\bibnamefont {Acremann}},
  \bibinfo {author} {\bibfnamefont {S.~Y.}\ \bibnamefont {Agustsson}}, \bibinfo
  {author} {\bibfnamefont {M.}~\bibnamefont {Dendzik}}, \bibinfo {author}
  {\bibfnamefont {K.}~\bibnamefont {B{\"u}hlmann}}, \bibinfo {author}
  {\bibfnamefont {D.}~\bibnamefont {Curcio}}, \bibinfo {author} {\bibfnamefont
  {D.}~\bibnamefont {Kutnyakhov}}, \bibinfo {author} {\bibfnamefont
  {F.}~\bibnamefont {Pressacco}}, \bibinfo {author} {\bibfnamefont
  {M.}~\bibnamefont {Heber}}, \bibinfo {author} {\bibfnamefont
  {S.}~\bibnamefont {Dong}}, \bibinfo {author} {\bibfnamefont {T.}~\bibnamefont
  {Pincelli}}, \bibinfo {author} {\bibfnamefont {J.}~\bibnamefont {Demsar}},
  \bibinfo {author} {\bibfnamefont {W.}~\bibnamefont {Wurth}}, \bibinfo
  {author} {\bibfnamefont {P.}~\bibnamefont {Hofmann}}, \bibinfo {author}
  {\bibfnamefont {M.}~\bibnamefont {Wolf}}, \bibinfo {author} {\bibfnamefont
  {M.}~\bibnamefont {Scheidgen}}, \bibinfo {author} {\bibfnamefont
  {L.}~\bibnamefont {Rettig}}, \ and\ \bibinfo {author} {\bibfnamefont
  {R.}~\bibnamefont {Ernstorfer}},\ }\bibfield  {title} {\enquote {\bibinfo
  {title} {An open-source, end-to-end workflow for multidimensional
  photoemission spectroscopy},}\ }\href {\doibase 10.1038/s41597-020-00769-8}
  {\bibfield  {journal} {\bibinfo  {journal} {Scientific Data}\ }\textbf
  {\bibinfo {volume} {7}},\ \bibinfo {pages} {442} (\bibinfo {year}
  {2020})}\BibitemShut {NoStop}%
\bibitem [{\citenamefont {Tusche}\ \emph {et~al.}(2016)\citenamefont {Tusche},
  \citenamefont {Goslawski}, \citenamefont {Kutnyakhov}, \citenamefont
  {Ellguth}, \citenamefont {Medjanik}, \citenamefont {Elmers}, \citenamefont
  {Chernov}, \citenamefont {Wallauer}, \citenamefont {Engel}, \citenamefont
  {Jankowiak},\ and\ \citenamefont {Schönhense}}]{Tusche2016}%
  \BibitemOpen
  \bibfield  {author} {\bibinfo {author} {\bibfnamefont {C.}~\bibnamefont
  {Tusche}}, \bibinfo {author} {\bibfnamefont {P.}~\bibnamefont {Goslawski}},
  \bibinfo {author} {\bibfnamefont {D.}~\bibnamefont {Kutnyakhov}}, \bibinfo
  {author} {\bibfnamefont {M.}~\bibnamefont {Ellguth}}, \bibinfo {author}
  {\bibfnamefont {K.}~\bibnamefont {Medjanik}}, \bibinfo {author}
  {\bibfnamefont {H.~J.}\ \bibnamefont {Elmers}}, \bibinfo {author}
  {\bibfnamefont {S.}~\bibnamefont {Chernov}}, \bibinfo {author} {\bibfnamefont
  {R.}~\bibnamefont {Wallauer}}, \bibinfo {author} {\bibfnamefont
  {D.}~\bibnamefont {Engel}}, \bibinfo {author} {\bibfnamefont
  {A.}~\bibnamefont {Jankowiak}}, \ and\ \bibinfo {author} {\bibfnamefont
  {G.}~\bibnamefont {Schönhense}},\ }\bibfield  {title} {\enquote {\bibinfo
  {title} {Multi-mhz time-of-flight electronic bandstructure imaging of
  graphene on ir(111)},}\ }\href {\doibase 10.1063/1.4955015} {\bibfield
  {journal} {\bibinfo  {journal} {Applied Physics Letters}\ }\textbf {\bibinfo
  {volume} {108}},\ \bibinfo {pages} {261602} (\bibinfo {year} {2016})},\
  \Eprint {http://arxiv.org/abs/https://doi.org/10.1063/1.4955015}
  {https://doi.org/10.1063/1.4955015} \BibitemShut {NoStop}%
\bibitem [{\citenamefont {N'Diaye}\ \emph {et~al.}(2008)\citenamefont
  {N'Diaye}, \citenamefont {Coraux}, \citenamefont {Plasa}, \citenamefont
  {Busse},\ and\ \citenamefont {Michely}}]{Diaye2008}%
  \BibitemOpen
  \bibfield  {author} {\bibinfo {author} {\bibfnamefont {A.~T.}\ \bibnamefont
  {N'Diaye}}, \bibinfo {author} {\bibfnamefont {J.}~\bibnamefont {Coraux}},
  \bibinfo {author} {\bibfnamefont {T.~N.}\ \bibnamefont {Plasa}}, \bibinfo
  {author} {\bibfnamefont {C.}~\bibnamefont {Busse}}, \ and\ \bibinfo {author}
  {\bibfnamefont {T.}~\bibnamefont {Michely}},\ }\bibfield  {title} {\enquote
  {\bibinfo {title} {Structure of epitaxial graphene on ir(111)},}\ }\href
  {\doibase 10.1088/1367-2630/10/4/043033} {\bibfield  {journal} {\bibinfo
  {journal} {New Journal of Physics}\ }\textbf {\bibinfo {volume} {10}},\
  \bibinfo {pages} {043033} (\bibinfo {year} {2008})}\BibitemShut {NoStop}%
\bibitem [{\citenamefont {Ulstrup}\ \emph {et~al.}(2014)\citenamefont
  {Ulstrup}, \citenamefont {Johannsen}, \citenamefont {Grioni},\ and\
  \citenamefont {Hofmann}}]{Ulstrup2014}%
  \BibitemOpen
  \bibfield  {author} {\bibinfo {author} {\bibfnamefont {S.}~\bibnamefont
  {Ulstrup}}, \bibinfo {author} {\bibfnamefont {J.~C.}\ \bibnamefont
  {Johannsen}}, \bibinfo {author} {\bibfnamefont {M.}~\bibnamefont {Grioni}}, \
  and\ \bibinfo {author} {\bibfnamefont {P.}~\bibnamefont {Hofmann}},\
  }\bibfield  {title} {\enquote {\bibinfo {title} {Extracting the temperature
  of hot carriers in time- and angle-resolved photoemission},}\ }\href
  {\doibase 10.1063/1.4863322} {\bibfield  {journal} {\bibinfo  {journal}
  {Review of Scientific Instruments}\ }\textbf {\bibinfo {volume} {85}},\
  \bibinfo {pages} {013907} (\bibinfo {year} {2014})},\ \Eprint
  {http://arxiv.org/abs/https://doi.org/10.1063/1.4863322}
  {https://doi.org/10.1063/1.4863322} \BibitemShut {NoStop}%
\bibitem [{\citenamefont {Kralj}\ \emph {et~al.}(2011)\citenamefont {Kralj},
  \citenamefont {Pletikosi\ifmmode~\acute{c}\else \'{c}\fi{}}, \citenamefont
  {Petrovi\ifmmode~\acute{c}\else \'{c}\fi{}}, \citenamefont {Pervan},
  \citenamefont {Milun}, \citenamefont {N'Diaye}, \citenamefont {Busse},
  \citenamefont {Michely}, \citenamefont {Fujii},\ and\ \citenamefont
  {Vobornik}}]{Kralj2011}%
  \BibitemOpen
  \bibfield  {author} {\bibinfo {author} {\bibfnamefont {M.}~\bibnamefont
  {Kralj}}, \bibinfo {author} {\bibfnamefont {I.}~\bibnamefont
  {Pletikosi\ifmmode~\acute{c}\else \'{c}\fi{}}}, \bibinfo {author}
  {\bibfnamefont {M.}~\bibnamefont {Petrovi\ifmmode~\acute{c}\else
  \'{c}\fi{}}}, \bibinfo {author} {\bibfnamefont {P.}~\bibnamefont {Pervan}},
  \bibinfo {author} {\bibfnamefont {M.}~\bibnamefont {Milun}}, \bibinfo
  {author} {\bibfnamefont {A.~T.}\ \bibnamefont {N'Diaye}}, \bibinfo {author}
  {\bibfnamefont {C.}~\bibnamefont {Busse}}, \bibinfo {author} {\bibfnamefont
  {T.}~\bibnamefont {Michely}}, \bibinfo {author} {\bibfnamefont
  {J.}~\bibnamefont {Fujii}}, \ and\ \bibinfo {author} {\bibfnamefont
  {I.}~\bibnamefont {Vobornik}},\ }\bibfield  {title} {\enquote {\bibinfo
  {title} {Graphene on ir(111) characterized by angle-resolved
  photoemission},}\ }\href {\doibase 10.1103/PhysRevB.84.075427} {\bibfield
  {journal} {\bibinfo  {journal} {Phys. Rev. B}\ }\textbf {\bibinfo {volume}
  {84}},\ \bibinfo {pages} {075427} (\bibinfo {year} {2011})}\BibitemShut
  {NoStop}%
\bibitem [{\citenamefont {Johannsen}\ \emph {et~al.}(2013)\citenamefont
  {Johannsen}, \citenamefont {Ulstrup}, \citenamefont {Bianchi}, \citenamefont
  {Hatch}, \citenamefont {Guan}, \citenamefont {Mazzola}, \citenamefont
  {Hornek{\ae}r}, \citenamefont {Fromm}, \citenamefont {Raidel}, \citenamefont
  {Seyller},\ and\ \citenamefont {Hofmann}}]{Johannsen2013}%
  \BibitemOpen
  \bibfield  {author} {\bibinfo {author} {\bibfnamefont {J.~C.}\ \bibnamefont
  {Johannsen}}, \bibinfo {author} {\bibfnamefont {S.}~\bibnamefont {Ulstrup}},
  \bibinfo {author} {\bibfnamefont {M.}~\bibnamefont {Bianchi}}, \bibinfo
  {author} {\bibfnamefont {R.}~\bibnamefont {Hatch}}, \bibinfo {author}
  {\bibfnamefont {D.}~\bibnamefont {Guan}}, \bibinfo {author} {\bibfnamefont
  {F.}~\bibnamefont {Mazzola}}, \bibinfo {author} {\bibfnamefont
  {L.}~\bibnamefont {Hornek{\ae}r}}, \bibinfo {author} {\bibfnamefont
  {F.}~\bibnamefont {Fromm}}, \bibinfo {author} {\bibfnamefont
  {C.}~\bibnamefont {Raidel}}, \bibinfo {author} {\bibfnamefont
  {T.}~\bibnamefont {Seyller}}, \ and\ \bibinfo {author} {\bibfnamefont
  {P.}~\bibnamefont {Hofmann}},\ }\bibfield  {title} {\enquote {\bibinfo
  {title} {Electron{\textendash}phonon coupling in quasi-free-standing
  graphene},}\ }\href {\doibase 10.1088/0953-8984/25/9/094001} {\bibfield
  {journal} {\bibinfo  {journal} {Journal of Physics: Condensed Matter}\
  }\textbf {\bibinfo {volume} {25}},\ \bibinfo {pages} {094001} (\bibinfo
  {year} {2013})}\BibitemShut {NoStop}%
\bibitem [{\citenamefont {Krause}, \citenamefont {Schafer},\ and\ \citenamefont
  {Kulander}(1992)}]{Krause1992}%
  \BibitemOpen
  \bibfield  {author} {\bibinfo {author} {\bibfnamefont {J.~L.}\ \bibnamefont
  {Krause}}, \bibinfo {author} {\bibfnamefont {K.~J.}\ \bibnamefont {Schafer}},
  \ and\ \bibinfo {author} {\bibfnamefont {K.~C.}\ \bibnamefont {Kulander}},\
  }\bibfield  {title} {\enquote {\bibinfo {title} {High-order harmonic
  generation from atoms and ions in the high intensity regime},}\ }\href
  {\doibase 10.1103/PhysRevLett.68.3535} {\bibfield  {journal} {\bibinfo
  {journal} {Phys. Rev. Lett.}\ }\textbf {\bibinfo {volume} {68}},\ \bibinfo
  {pages} {3535--3538} (\bibinfo {year} {1992})}\BibitemShut {NoStop}%
\bibitem [{\citenamefont {Tanuma}, \citenamefont {Powell},\ and\ \citenamefont
  {Penn}(2011)}]{Tanuma2011}%
  \BibitemOpen
  \bibfield  {author} {\bibinfo {author} {\bibfnamefont {S.}~\bibnamefont
  {Tanuma}}, \bibinfo {author} {\bibfnamefont {C.~J.}\ \bibnamefont {Powell}},
  \ and\ \bibinfo {author} {\bibfnamefont {D.~R.}\ \bibnamefont {Penn}},\
  }\bibfield  {title} {\enquote {\bibinfo {title} {Calculations of electron
  inelastic mean free paths. ix. data for 41 elemental solids over the 50 ev to
  30 kev range},}\ }\href {\doibase https://doi.org/10.1002/sia.3522}
  {\bibfield  {journal} {\bibinfo  {journal} {Surface and Interface Analysis}\
  }\textbf {\bibinfo {volume} {43}},\ \bibinfo {pages} {689--713} (\bibinfo
  {year} {2011})},\ \Eprint
  {http://arxiv.org/abs/https://onlinelibrary.wiley.com/doi/pdf/10.1002/sia.3522}
  {https://onlinelibrary.wiley.com/doi/pdf/10.1002/sia.3522} \BibitemShut
  {NoStop}%
\bibitem [{\citenamefont {Elmers}\ \emph {et~al.}(2017)\citenamefont {Elmers},
  \citenamefont {Kutnyakhov}, \citenamefont {Chernov}, \citenamefont
  {Medjanik}, \citenamefont {Fedchenko}, \citenamefont
  {Zaporozhchenko-Zymakova}, \citenamefont {Ellguth}, \citenamefont {Tusche},
  \citenamefont {Viefhaus},\ and\ \citenamefont {Schönhense}}]{Elmers2017}%
  \BibitemOpen
  \bibfield  {author} {\bibinfo {author} {\bibfnamefont {H.~J.}\ \bibnamefont
  {Elmers}}, \bibinfo {author} {\bibfnamefont {D.}~\bibnamefont {Kutnyakhov}},
  \bibinfo {author} {\bibfnamefont {S.~V.}\ \bibnamefont {Chernov}}, \bibinfo
  {author} {\bibfnamefont {K.}~\bibnamefont {Medjanik}}, \bibinfo {author}
  {\bibfnamefont {O.}~\bibnamefont {Fedchenko}}, \bibinfo {author}
  {\bibfnamefont {A.}~\bibnamefont {Zaporozhchenko-Zymakova}}, \bibinfo
  {author} {\bibfnamefont {M.}~\bibnamefont {Ellguth}}, \bibinfo {author}
  {\bibfnamefont {C.}~\bibnamefont {Tusche}}, \bibinfo {author} {\bibfnamefont
  {J.}~\bibnamefont {Viefhaus}}, \ and\ \bibinfo {author} {\bibfnamefont
  {G.}~\bibnamefont {Schönhense}},\ }\bibfield  {title} {\enquote {\bibinfo
  {title} {Hosting of surface states in spin{\textendash}orbit induced
  projected bulk band gaps of w(1{\hspace{0.167em}}1{\hspace{0.167em}}0) and
  ir(1{\hspace{0.167em}}1{\hspace{0.167em}}1)},}\ }\href {\doibase
  10.1088/1361-648x/aa7173} {\bibfield  {journal} {\bibinfo  {journal} {Journal
  of Physics: Condensed Matter}\ }\textbf {\bibinfo {volume} {29}},\ \bibinfo
  {pages} {255001} (\bibinfo {year} {2017})}\BibitemShut {NoStop}%
\bibitem [{\citenamefont {Haag}\ \emph {et~al.}(2020)\citenamefont {Haag},
  \citenamefont {L\"uftner}, \citenamefont {Haag}, \citenamefont {Seidel},
  \citenamefont {Kelly}, \citenamefont {Zamborlini}, \citenamefont {Jugovac},
  \citenamefont {Feyer}, \citenamefont {Aeschlimann}, \citenamefont {Puschnig},
  \citenamefont {Cinchetti},\ and\ \citenamefont {Stadtm\"uller}}]{Haag2020}%
  \BibitemOpen
  \bibfield  {author} {\bibinfo {author} {\bibfnamefont {N.}~\bibnamefont
  {Haag}}, \bibinfo {author} {\bibfnamefont {D.}~\bibnamefont {L\"uftner}},
  \bibinfo {author} {\bibfnamefont {F.}~\bibnamefont {Haag}}, \bibinfo {author}
  {\bibfnamefont {J.}~\bibnamefont {Seidel}}, \bibinfo {author} {\bibfnamefont
  {L.~L.}\ \bibnamefont {Kelly}}, \bibinfo {author} {\bibfnamefont
  {G.}~\bibnamefont {Zamborlini}}, \bibinfo {author} {\bibfnamefont
  {M.}~\bibnamefont {Jugovac}}, \bibinfo {author} {\bibfnamefont
  {V.}~\bibnamefont {Feyer}}, \bibinfo {author} {\bibfnamefont
  {M.}~\bibnamefont {Aeschlimann}}, \bibinfo {author} {\bibfnamefont
  {P.}~\bibnamefont {Puschnig}}, \bibinfo {author} {\bibfnamefont
  {M.}~\bibnamefont {Cinchetti}}, \ and\ \bibinfo {author} {\bibfnamefont
  {B.}~\bibnamefont {Stadtm\"uller}},\ }\bibfield  {title} {\enquote {\bibinfo
  {title} {Signatures of an atomic crystal in the band structure of a
  ${\mathrm{c}}_{60}$ thin film},}\ }\href {\doibase
  10.1103/PhysRevB.101.165422} {\bibfield  {journal} {\bibinfo  {journal}
  {Phys. Rev. B}\ }\textbf {\bibinfo {volume} {101}},\ \bibinfo {pages}
  {165422} (\bibinfo {year} {2020})}\BibitemShut {NoStop}%
\bibitem [{\citenamefont {Baumg\"artner}\ \emph {et~al.}(2022)\citenamefont
  {Baumg\"artner}, \citenamefont {Reuner}, \citenamefont {Metzger},
  \citenamefont {Kutnyakhov}, \citenamefont {Heber}, \citenamefont {Pressacco},
  \citenamefont {Min}, \citenamefont {Peixoto}, \citenamefont {Reiser},
  \citenamefont {Kim}, \citenamefont {Lu}, \citenamefont {Shayduk},
  \citenamefont {Izquierdo}, \citenamefont {Brenner}, \citenamefont {Roth},
  \citenamefont {Sch\"oll}, \citenamefont {Molodtsov}, \citenamefont {Wurth},
  \citenamefont {Reinert}, \citenamefont {Madsen}, \citenamefont
  {Popova-Gorelova},\ and\ \citenamefont {Scholz}}]{Baumgaertner2022}%
  \BibitemOpen
  \bibfield  {author} {\bibinfo {author} {\bibfnamefont {K.}~\bibnamefont
  {Baumg\"artner}}, \bibinfo {author} {\bibfnamefont {M.}~\bibnamefont
  {Reuner}}, \bibinfo {author} {\bibfnamefont {C.}~\bibnamefont {Metzger}},
  \bibinfo {author} {\bibfnamefont {D.}~\bibnamefont {Kutnyakhov}}, \bibinfo
  {author} {\bibfnamefont {M.}~\bibnamefont {Heber}}, \bibinfo {author}
  {\bibfnamefont {F.}~\bibnamefont {Pressacco}}, \bibinfo {author}
  {\bibfnamefont {C.-H.}\ \bibnamefont {Min}}, \bibinfo {author} {\bibfnamefont
  {T.~R.~F.}\ \bibnamefont {Peixoto}}, \bibinfo {author} {\bibfnamefont
  {M.}~\bibnamefont {Reiser}}, \bibinfo {author} {\bibfnamefont
  {C.}~\bibnamefont {Kim}}, \bibinfo {author} {\bibfnamefont {W.}~\bibnamefont
  {Lu}}, \bibinfo {author} {\bibfnamefont {R.}~\bibnamefont {Shayduk}},
  \bibinfo {author} {\bibfnamefont {M.}~\bibnamefont {Izquierdo}}, \bibinfo
  {author} {\bibfnamefont {G.}~\bibnamefont {Brenner}}, \bibinfo {author}
  {\bibfnamefont {F.}~\bibnamefont {Roth}}, \bibinfo {author} {\bibfnamefont
  {A.}~\bibnamefont {Sch\"oll}}, \bibinfo {author} {\bibfnamefont
  {S.}~\bibnamefont {Molodtsov}}, \bibinfo {author} {\bibfnamefont
  {W.}~\bibnamefont {Wurth}}, \bibinfo {author} {\bibfnamefont
  {F.}~\bibnamefont {Reinert}}, \bibinfo {author} {\bibfnamefont
  {A.}~\bibnamefont {Madsen}}, \bibinfo {author} {\bibfnamefont
  {D.}~\bibnamefont {Popova-Gorelova}}, \ and\ \bibinfo {author} {\bibfnamefont
  {M.}~\bibnamefont {Scholz}},\ }\bibfield  {title} {\enquote {\bibinfo {title}
  {Ultrafast orbital tomography of a pentacene film using time-resolved
  momentum microscopy at a fel},}\ }\href {\doibase 10.1038/s41467-022-30404-6}
  {\bibfield  {journal} {\bibinfo  {journal} {Nature Communications}\ }\textbf
  {\bibinfo {volume} {13}},\ \bibinfo {pages} {2741} (\bibinfo {year}
  {2022})}\BibitemShut {NoStop}%
\bibitem [{\citenamefont {Hellmann}\ \emph {et~al.}(2010)\citenamefont
  {Hellmann}, \citenamefont {Beye}, \citenamefont {Sohrt}, \citenamefont
  {Rohwer}, \citenamefont {Sorgenfrei}, \citenamefont {Redlin}, \citenamefont
  {Kall\"ane}, \citenamefont {Marczynski-B\"uhlow}, \citenamefont {Hennies},
  \citenamefont {Bauer}, \citenamefont {F\"ohlisch}, \citenamefont {Kipp},
  \citenamefont {Wurth},\ and\ \citenamefont {Rossnagel}}]{Hellmann2010}%
  \BibitemOpen
  \bibfield  {author} {\bibinfo {author} {\bibfnamefont {S.}~\bibnamefont
  {Hellmann}}, \bibinfo {author} {\bibfnamefont {M.}~\bibnamefont {Beye}},
  \bibinfo {author} {\bibfnamefont {C.}~\bibnamefont {Sohrt}}, \bibinfo
  {author} {\bibfnamefont {T.}~\bibnamefont {Rohwer}}, \bibinfo {author}
  {\bibfnamefont {F.}~\bibnamefont {Sorgenfrei}}, \bibinfo {author}
  {\bibfnamefont {H.}~\bibnamefont {Redlin}}, \bibinfo {author} {\bibfnamefont
  {M.}~\bibnamefont {Kall\"ane}}, \bibinfo {author} {\bibfnamefont
  {M.}~\bibnamefont {Marczynski-B\"uhlow}}, \bibinfo {author} {\bibfnamefont
  {F.}~\bibnamefont {Hennies}}, \bibinfo {author} {\bibfnamefont
  {M.}~\bibnamefont {Bauer}}, \bibinfo {author} {\bibfnamefont
  {A.}~\bibnamefont {F\"ohlisch}}, \bibinfo {author} {\bibfnamefont
  {L.}~\bibnamefont {Kipp}}, \bibinfo {author} {\bibfnamefont {W.}~\bibnamefont
  {Wurth}}, \ and\ \bibinfo {author} {\bibfnamefont {K.}~\bibnamefont
  {Rossnagel}},\ }\bibfield  {title} {\enquote {\bibinfo {title} {Ultrafast
  melting of a charge-density wave in the mott insulator
  $1t\mathrm{\text{\ensuremath{-}}}{\mathrm{tas}}_{2}$},}\ }\href {\doibase
  10.1103/PhysRevLett.105.187401} {\bibfield  {journal} {\bibinfo  {journal}
  {Phys. Rev. Lett.}\ }\textbf {\bibinfo {volume} {105}},\ \bibinfo {pages}
  {187401} (\bibinfo {year} {2010})}\BibitemShut {NoStop}%
\bibitem [{\citenamefont {Dendzik}\ \emph {et~al.}(2020)\citenamefont
  {Dendzik}, \citenamefont {Xian}, \citenamefont {Perfetto}, \citenamefont
  {Sangalli}, \citenamefont {Kutnyakhov}, \citenamefont {Dong}, \citenamefont
  {Beaulieu}, \citenamefont {Pincelli}, \citenamefont {Pressacco},
  \citenamefont {Curcio}, \citenamefont {Agustsson}, \citenamefont {Heber},
  \citenamefont {Hauer}, \citenamefont {Wurth}, \citenamefont {Brenner},
  \citenamefont {Acremann}, \citenamefont {Hofmann}, \citenamefont {Wolf},
  \citenamefont {Marini}, \citenamefont {Stefanucci}, \citenamefont {Rettig},\
  and\ \citenamefont {Ernstorfer}}]{Dendzik2020}%
  \BibitemOpen
  \bibfield  {author} {\bibinfo {author} {\bibfnamefont {M.}~\bibnamefont
  {Dendzik}}, \bibinfo {author} {\bibfnamefont {R.~P.}\ \bibnamefont {Xian}},
  \bibinfo {author} {\bibfnamefont {E.}~\bibnamefont {Perfetto}}, \bibinfo
  {author} {\bibfnamefont {D.}~\bibnamefont {Sangalli}}, \bibinfo {author}
  {\bibfnamefont {D.}~\bibnamefont {Kutnyakhov}}, \bibinfo {author}
  {\bibfnamefont {S.}~\bibnamefont {Dong}}, \bibinfo {author} {\bibfnamefont
  {S.}~\bibnamefont {Beaulieu}}, \bibinfo {author} {\bibfnamefont
  {T.}~\bibnamefont {Pincelli}}, \bibinfo {author} {\bibfnamefont
  {F.}~\bibnamefont {Pressacco}}, \bibinfo {author} {\bibfnamefont
  {D.}~\bibnamefont {Curcio}}, \bibinfo {author} {\bibfnamefont {S.~Y.}\
  \bibnamefont {Agustsson}}, \bibinfo {author} {\bibfnamefont {M.}~\bibnamefont
  {Heber}}, \bibinfo {author} {\bibfnamefont {J.}~\bibnamefont {Hauer}},
  \bibinfo {author} {\bibfnamefont {W.}~\bibnamefont {Wurth}}, \bibinfo
  {author} {\bibfnamefont {G.}~\bibnamefont {Brenner}}, \bibinfo {author}
  {\bibfnamefont {Y.}~\bibnamefont {Acremann}}, \bibinfo {author}
  {\bibfnamefont {P.}~\bibnamefont {Hofmann}}, \bibinfo {author} {\bibfnamefont
  {M.}~\bibnamefont {Wolf}}, \bibinfo {author} {\bibfnamefont {A.}~\bibnamefont
  {Marini}}, \bibinfo {author} {\bibfnamefont {G.}~\bibnamefont {Stefanucci}},
  \bibinfo {author} {\bibfnamefont {L.}~\bibnamefont {Rettig}}, \ and\ \bibinfo
  {author} {\bibfnamefont {R.}~\bibnamefont {Ernstorfer}},\ }\bibfield  {title}
  {\enquote {\bibinfo {title} {Observation of an excitonic mott transition
  through ultrafast core-cum-conduction photoemission spectroscopy},}\ }\href
  {\doibase 10.1103/PhysRevLett.125.096401} {\bibfield  {journal} {\bibinfo
  {journal} {Phys. Rev. Lett.}\ }\textbf {\bibinfo {volume} {125}},\ \bibinfo
  {pages} {096401} (\bibinfo {year} {2020})}\BibitemShut {NoStop}%
\bibitem [{\citenamefont {Pressacco}\ \emph {et~al.}(2021)\citenamefont
  {Pressacco}, \citenamefont {Sangalli}, \citenamefont {Uhlíř}, \citenamefont
  {Kutnyakhov}, \citenamefont {Arregi}, \citenamefont {Agustsson},
  \citenamefont {Brenner}, \citenamefont {Redlin}, \citenamefont {Heber},
  \citenamefont {Vasilyev}, \citenamefont {Demsar}, \citenamefont
  {Schönhense}, \citenamefont {Gatti}, \citenamefont {Marini}, \citenamefont
  {Wurth},\ and\ \citenamefont {Sirotti}}]{Pressacco2021}%
  \BibitemOpen
  \bibfield  {author} {\bibinfo {author} {\bibfnamefont {F.}~\bibnamefont
  {Pressacco}}, \bibinfo {author} {\bibfnamefont {D.}~\bibnamefont {Sangalli}},
  \bibinfo {author} {\bibfnamefont {V.}~\bibnamefont {Uhlíř}}, \bibinfo
  {author} {\bibfnamefont {D.}~\bibnamefont {Kutnyakhov}}, \bibinfo {author}
  {\bibfnamefont {J.~A.}\ \bibnamefont {Arregi}}, \bibinfo {author}
  {\bibfnamefont {S.~Y.}\ \bibnamefont {Agustsson}}, \bibinfo {author}
  {\bibfnamefont {G.}~\bibnamefont {Brenner}}, \bibinfo {author} {\bibfnamefont
  {H.}~\bibnamefont {Redlin}}, \bibinfo {author} {\bibfnamefont
  {M.}~\bibnamefont {Heber}}, \bibinfo {author} {\bibfnamefont
  {D.}~\bibnamefont {Vasilyev}}, \bibinfo {author} {\bibfnamefont
  {J.}~\bibnamefont {Demsar}}, \bibinfo {author} {\bibfnamefont
  {G.}~\bibnamefont {Schönhense}}, \bibinfo {author} {\bibfnamefont
  {M.}~\bibnamefont {Gatti}}, \bibinfo {author} {\bibfnamefont
  {A.}~\bibnamefont {Marini}}, \bibinfo {author} {\bibfnamefont
  {W.}~\bibnamefont {Wurth}}, \ and\ \bibinfo {author} {\bibfnamefont
  {F.}~\bibnamefont {Sirotti}},\ }\bibfield  {title} {\enquote {\bibinfo
  {title} {Subpicosecond metamagnetic phase transition in {FeRh} driven by
  non-equilibrium electron dynamics},}\ }\href {\doibase
  10.1038/s41467-021-25347-3} {\bibfield  {journal} {\bibinfo  {journal}
  {Nature Communications}\ }\textbf {\bibinfo {volume} {12}},\ \bibinfo {pages}
  {5088} (\bibinfo {year} {2021})}\BibitemShut {NoStop}%
\bibitem [{\citenamefont {Curcio}\ \emph {et~al.}(2021)\citenamefont {Curcio},
  \citenamefont {Pakdel}, \citenamefont {Volckaert}, \citenamefont {Miwa},
  \citenamefont {Ulstrup}, \citenamefont {Lanat\`a}, \citenamefont {Bianchi},
  \citenamefont {Kutnyakhov}, \citenamefont {Pressacco}, \citenamefont
  {Brenner}, \citenamefont {Dziarzhytski}, \citenamefont {Redlin},
  \citenamefont {Agustsson}, \citenamefont {Medjanik}, \citenamefont
  {Vasilyev}, \citenamefont {Elmers}, \citenamefont {Sch\"onhense},
  \citenamefont {Tusche}, \citenamefont {Chen}, \citenamefont {Speck},
  \citenamefont {Seyller}, \citenamefont {B\"uhlmann}, \citenamefont {Gort},
  \citenamefont {Diekmann}, \citenamefont {Rossnagel}, \citenamefont
  {Acremann}, \citenamefont {Demsar}, \citenamefont {Wurth}, \citenamefont
  {Lizzit}, \citenamefont {Bignardi}, \citenamefont {Lacovig}, \citenamefont
  {Lizzit}, \citenamefont {Sanders},\ and\ \citenamefont
  {Hofmann}}]{Curcio2021}%
  \BibitemOpen
  \bibfield  {author} {\bibinfo {author} {\bibfnamefont {D.}~\bibnamefont
  {Curcio}}, \bibinfo {author} {\bibfnamefont {S.}~\bibnamefont {Pakdel}},
  \bibinfo {author} {\bibfnamefont {K.}~\bibnamefont {Volckaert}}, \bibinfo
  {author} {\bibfnamefont {J.~A.}\ \bibnamefont {Miwa}}, \bibinfo {author}
  {\bibfnamefont {S.}~\bibnamefont {Ulstrup}}, \bibinfo {author} {\bibfnamefont
  {N.}~\bibnamefont {Lanat\`a}}, \bibinfo {author} {\bibfnamefont
  {M.}~\bibnamefont {Bianchi}}, \bibinfo {author} {\bibfnamefont
  {D.}~\bibnamefont {Kutnyakhov}}, \bibinfo {author} {\bibfnamefont
  {F.}~\bibnamefont {Pressacco}}, \bibinfo {author} {\bibfnamefont
  {G.}~\bibnamefont {Brenner}}, \bibinfo {author} {\bibfnamefont
  {S.}~\bibnamefont {Dziarzhytski}}, \bibinfo {author} {\bibfnamefont
  {H.}~\bibnamefont {Redlin}}, \bibinfo {author} {\bibfnamefont {S.~Y.}\
  \bibnamefont {Agustsson}}, \bibinfo {author} {\bibfnamefont {K.}~\bibnamefont
  {Medjanik}}, \bibinfo {author} {\bibfnamefont {D.}~\bibnamefont {Vasilyev}},
  \bibinfo {author} {\bibfnamefont {H.-J.}\ \bibnamefont {Elmers}}, \bibinfo
  {author} {\bibfnamefont {G.}~\bibnamefont {Sch\"onhense}}, \bibinfo {author}
  {\bibfnamefont {C.}~\bibnamefont {Tusche}}, \bibinfo {author} {\bibfnamefont
  {Y.-J.}\ \bibnamefont {Chen}}, \bibinfo {author} {\bibfnamefont
  {F.}~\bibnamefont {Speck}}, \bibinfo {author} {\bibfnamefont
  {T.}~\bibnamefont {Seyller}}, \bibinfo {author} {\bibfnamefont
  {K.}~\bibnamefont {B\"uhlmann}}, \bibinfo {author} {\bibfnamefont
  {R.}~\bibnamefont {Gort}}, \bibinfo {author} {\bibfnamefont {F.}~\bibnamefont
  {Diekmann}}, \bibinfo {author} {\bibfnamefont {K.}~\bibnamefont {Rossnagel}},
  \bibinfo {author} {\bibfnamefont {Y.}~\bibnamefont {Acremann}}, \bibinfo
  {author} {\bibfnamefont {J.}~\bibnamefont {Demsar}}, \bibinfo {author}
  {\bibfnamefont {W.}~\bibnamefont {Wurth}}, \bibinfo {author} {\bibfnamefont
  {D.}~\bibnamefont {Lizzit}}, \bibinfo {author} {\bibfnamefont
  {L.}~\bibnamefont {Bignardi}}, \bibinfo {author} {\bibfnamefont
  {P.}~\bibnamefont {Lacovig}}, \bibinfo {author} {\bibfnamefont
  {S.}~\bibnamefont {Lizzit}}, \bibinfo {author} {\bibfnamefont {C.~E.}\
  \bibnamefont {Sanders}}, \ and\ \bibinfo {author} {\bibfnamefont
  {P.}~\bibnamefont {Hofmann}},\ }\bibfield  {title} {\enquote {\bibinfo
  {title} {Ultrafast electronic linewidth broadening in the c $1s$ core level
  of graphene},}\ }\href {\doibase 10.1103/PhysRevB.104.L161104} {\bibfield
  {journal} {\bibinfo  {journal} {Phys. Rev. B}\ }\textbf {\bibinfo {volume}
  {104}},\ \bibinfo {pages} {L161104} (\bibinfo {year} {2021})}\BibitemShut
  {NoStop}%
\end{thebibliography}%

\end{document}